\newcommand{\aeff}{{\rm A}_{\lambda}^{\rm e}}
\newcommand{\aej}{{\rm A}_{\rm J}^{\rm e}}
\newcommand{\aeh}{{\rm A}_{\rm H}^{\rm e}}
\newcommand{\aeks}{{\rm A}_{\rm K_{\rm s}}^{\rm e}}
\newcommand{\ks}{{K_{\rm s}}}
\documentclass{emulateapj}
\usepackage{subfigure}
\usepackage{upgreek}
\usepackage{amsmath}
\usepackage{multirow}

\defcitealias{Schechtman-Rook13}{Paper 1}

\begin{document}
\title{Near-Infrared Structure of Fast and Slow Rotating Disk Galaxies}

\author{Andrew Schechtman-Rook\altaffilmark{1}, Matthew
  A. Bershady\altaffilmark{1}}

\altaffiltext{1}{University of Wisconsin, Department of Astronomy, 475
N. Charter St., Madison, WI 53706; andrew@astro.wisc.edu}


\begin{abstract}

We investigate the stellar disk structure of six nearby edge-on spiral
galaxies using high-resolution $JH\ks$-band images and 3D radiative
transfer models.  To explore how mass and environment shape spiral
disks, we selected galaxies with rotational velocities between
$69<V_{\rm rot}<245$ km sec$^{-1}$, and two with unusual
morphologies. We find a wide diversity of disk structure. Of the
fast-rotating (V$_{\rm rot}>150$ km sec$^{-1}$) galaxies, only
NGC~4013 has the super-thin+thin+thick nested disk structure seen in
NGC~891 and the Milky Way, albeit with decreased oblateness, while
NGC~1055, a disturbed massive spiral galaxy, contains disks with
$h_z\lesssim200$ pc. NGC~4565, another fast-rotator, contains a
prominent ring at a radius $\sim$5 kpc but no super-thin disk. Despite
these differences, all fast-rotating galaxies in our sample have inner
truncations in at least one of their disks. These truncations lead to
Freeman Type II profiles when projected face-on.  Slow-rotating
galaxies are less complex, lacking inner disk truncations and
requiring fewer disk components to reproduce their light
distributions. Super-thin disk components in undisturbed disks
contribute $\sim25$\% of the total $\ks$-band light, up to that of the
thin-disk contribution. The presence of super-thin disks correlates
with infrared flux ratios; galaxies with super-thin disks have
$f_{\ks}/f_{60\mu{\rm m}} \leq 0.12$ for integrated light, consistent
with super-thin disks being regions of on-going
star-formation. Attenuation-corrected vertical color gradients in
$(J-\ks)$ correlate with the observed disk structure and are
consistent with population gradients with young-to-intermediate ages
closer to the mid-plane, indicating that disk heating--or cooling--is
a ubiquitous phenomenon.

\end{abstract}

\section{Introduction}

Determining the vertical structure of spiral disks is central to
understanding how they form and evolve over time. The notion of galaxy
formation as a monolithic gravitational collapse \citep[][]{Eggen62}
has long since fallen by the wayside in favor of a hierarchical buildup
of mass that comes naturally in a cold dark-matter cosmological
scenario ($\Lambda$CDM). However, in the context of the disk galaxies
like the Milky Way that we see today, the two scenarios may not be too
dissimilar if the bulk of the mass assembly takes place early or very
slowly. What matters is the detailed interplay between the
initial mass build-up, angular momentum, and cooling of the gaseous disk;
and the subsequent kinematic heating of the stellar disk through
internal dynamical interactions, instabilities, and external mergers
\citep[][]{Walker96,Abadi03}. These conditions set the physical
parameters of the disks \citep[for discussion see][]{Bird13}.

Where $\Lambda$CDM is particularly helpful is in its statistical
prescription for the lumpiness and history of mass accretion; where it
is particularly uninformative concerns the coupling of the baryons to
the dark matter, and their processing into stars. Consequently, a
detailed picture of galaxy formation requires observational
constraints and independent (dynamical) arguments. For example, we
know that thick disks are forming at $z\sim2$
\citep{Forster_Schreiber09}, although it is less clear if these are
related to the fairly ubiquitous thick disk components that we see in
nearby galaxies like the Milky Way today (e.g., \citealt{Yoachim06},
\citealt{Comeron11}). Any very thin disks that we observe today may
have formed recently, since such disks are prone to disruption even in
minor satellite accretion events \citep{Purcell09}. These issues
highlight the uncertainties in our knowledge of the assembly and
dynamical evolution of disks, and point to the need to empirically
characterize the fossil record in today's disks.

The vertical distribution of mass and different stellar populations
are also critical variables in efforts to break the disk-halo
degeneracy and isolate the contributions of dark and luminous matter
to the gravitational potential of spiral galaxies
\citep{Bershady10a,Bershady10b}. While recent efforts have indicated
that galaxy disks contribute significantly less than the theoretical
maximum \citep[contradicting the hypothesis of `disk
  maximality';][]{Bershady11, Westfall11, Martinsson13} to the
gravitational potential, if significant luminosity is contained in
very thin disks it would serve to increase the estimate of disk
maximality from kinematic measurements of face-on disks.  This fact
was part of the initial motivation for the current study.

In the Milky Way (MW) the distribution of stars as a function of
distance from the disk midplane is relatively well known for stars of
different spectral types in the solar cylinder, e.g., \citet{Aumer09},
and more broadly within the disk for G-dwarfs of different abundance
and metallicity \citep[e.g.,][]{Bovy12}. However, resolved stellar
population studies are currently relegated to the MW and only the
brightest stars in very nearby galaxies. A study of nearby, edge-on
galaxies \citep{Seth05} also find evidence for vertical population
gradients, but their study does not contain any fast-rotating ($V_{\rm
  rot}>130$ km s$^{-1}$) disks. Because of the large optical midplane
dust attenuation found in all fast-rotating edge-on spiral galaxies,
the MW is the {\it only} galaxy of this type for which we have an
unambiguous (albeit radially limited) sampling of stellar population
gradients near the midplane.  Consequently our understanding of the
midplane structure of spiral galaxy disks is incomplete at best.  This
has necessarily led to simplified models of these systems which tend
to focus on the characterization of the more easily visible thicker
disk components.

We have undertaken a study of the vertical structure of nearby spiral
galaxies using high-resolution ($\le$1'' FWHM) near-infrared (NIR)
imaging, specifically focused on the light distribution near the
mid-plane. In order to probe the light distribution at these small
heights, we have employed state-of-the-art radiative transfer (RT)
models to estimate dust attenuation corrections.  In
\citet[][hereafter \citetalias{Schechtman-Rook13}]{Schechtman-Rook13}
we demonstrated our methods, constructing full spectral energy
distribution (SED) RT models of the well-studied MW-like galaxy
NGC~891. These models contained multiple stellar populations as well
as fractal dust clumps, and were used to compute a general
relationship between infrared color and attenuation. When applying
these corrections to NIR data for NGC~891, we found not only a
super-thin disk ($h_{z}\sim$100 pc) but also an inner disk truncation,
an exponential-like bar, and a nuclear disk.

In this work we apply an improved version of this procedure to a
small sample of 6 nearly edge-on spiral galaxies, including three
massive spirals with $V_{\rm rot}>150$ km s$^{-1}$ and morphological
similarity to NGC~891 (NGC 522, 4013, and 4565), one massive spiral
with disturbed morphology (NGC~1055), and two low-mass systems with
$V_{\rm rot}<100$ km s$^{-1}$ (NGC 4144 and 4244). The latter are
known to lack prominent dust lanes \citep{Dalcanton04} and therefore
may possess dramatically different disk structure than seen in more
massive spiral galaxies. While this is a modest sample, it does span
an intentionally broad range of parameters for star-forming disk
galaxies, and constitutes a large fraction of edge-on galaxies that
are close enough to resolve at or below the 70 pc level (the
super-thin disk scale-height of NGC~891 from
\citep{Schechtman-Rook13}) with extant wide-field NIR imaging
facilities.

The paper is organized as follows. In Section \ref{sec:data} we
describe the reduction process for our data. Section
\ref{sec:attencorr} contains significant revisions to our dust
attenuation estimation method based on (a) an improved coupling to
stellar population synthesis to our RT models; and (b) a new RT model
more suitable to the dust and stellar geometries of the slow-rotating
galaxies in our samples (detailed in the Appendix). This section also
presents the attenuation-corrected light profiles of our sample. In
Section \ref{sec:galaxycolors} we compare the apparent and corrected
near-infrared colors and their gradients in the context of
expectations from stellar population synthesis models for the purpose
of verifying our attenuation corrections.  The multi-component
modeling of the two-dimensional light profiles for each galaxy is
detailed in Section \ref{sec:model}.  In Section \ref{sec:discussion}
we revisit the inferred color gradients in the context of a
multi-component disk populations and disk heating, and characterize
the oblateness and luminosity contributions of different disk
components in the context of disk maximality estimates.  The face-on
appearance of the sample and indirect methods for inferring the
presence of super-thin disks in spiral galaxies are also presented in
this section.  The results of our study are summarized in Section
\ref{sec:conclusions}.  Note that all magnitudes used in this work are
in the Vega system unless otherwise specified.

\begin{figure*}
\plotone{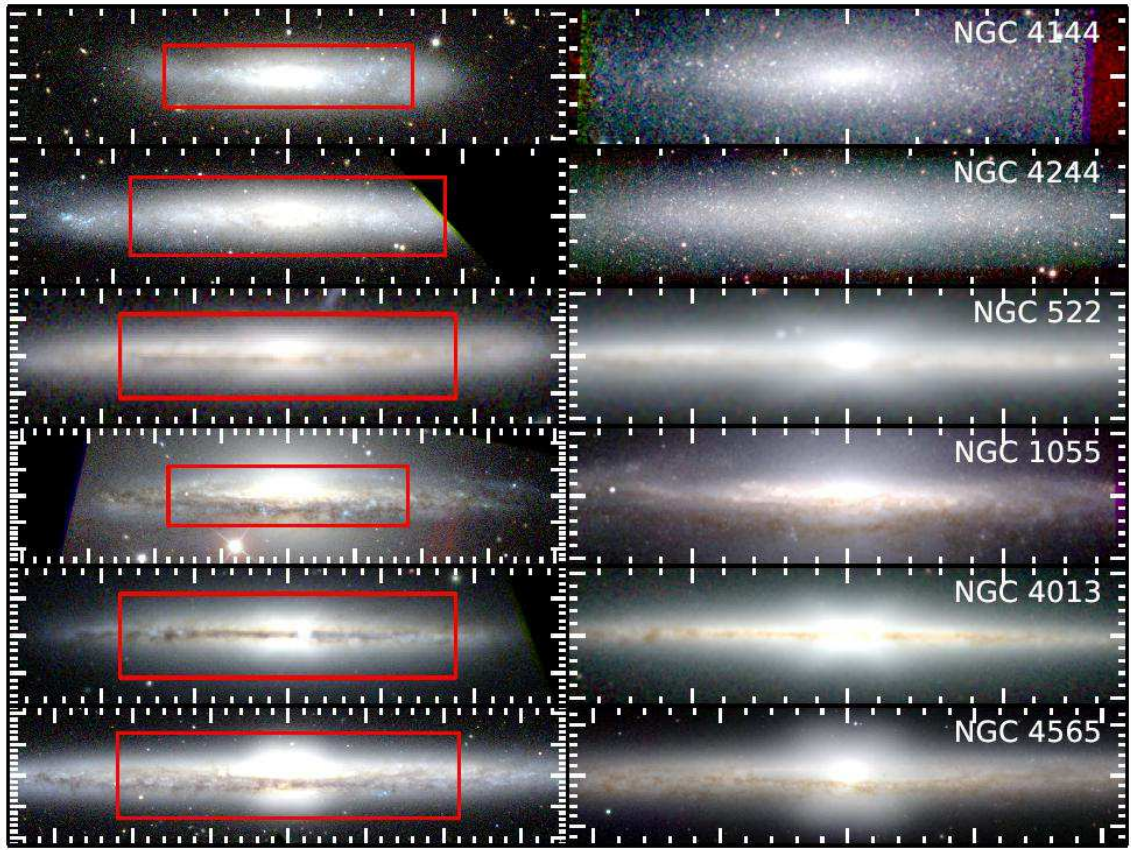}
\caption{SDSS gri' (left panel) and our full WHIRC JH$\ks$ (right
  panel) images of the galaxies in our sample, sorted by increasing
  V$_{\mathrm{rot}}$. Red boxes in the SDSS images indicate WHIRC
  FOV. Tickmarks show projected physical distance in increments of 1
  kpc (x axis) and 400 pc (y axis).}
\label{fig:gri_jhk_sample}
\end{figure*}

\begin{deluxetable*}{ccccccccccc}
\tablewidth{0pt}
\tablecaption{Summary of Observations and Image Mosaic Quality}
\tablehead{ 
\colhead{} & \colhead{Distance$^{a}$} &
\colhead{V$_{\mathrm{rot}}^{b}$} & \colhead{} &
\multicolumn{2}{c}{Exposure Time (s)} & \colhead{Seeing} &
\colhead{$\mu$(S/N=10)$^{f}$}& \colhead{$\mu({\rm lim})^{g}$}&&
\colhead{Dither}\\ \cline{5-6}
\colhead{Galaxy} & \colhead{(Mpc)} & \colhead{(km s$^{-1}$)} &
\colhead{Filter} & \colhead{Individual$^{c}$} & \colhead{Total$^{d}$}
& \colhead{ FWHM$^{e}$} & \multicolumn{2}{c}{(mag arcsec$^{-2}$)} &
\colhead{Pointings$^{h}$}
& \colhead{Pattern$^{i}$}
}
\startdata 
NGC 522 & 46.1 & 169 & J & 160 & 640 & $0\farcs5$ & 18.1$\pm$0.1& 23.1&1&A\\ 
& \nodata & \nodata & H & 100 & 800 & 0\farcs6 & 17.4$\pm$0.1& 22.2&1&A\\ 
 & \nodata & \nodata & $\ks$ & 40 & 6600 & 0\farcs6 & 16.5$\pm$01& 21.9&1&A\\ 
NGC 1055 & 16.6 & 181 & J & 100 & 800 & 0\farcs6 & 18.0$\pm$0.0& 22.8&1& S\\ 
& \nodata & \nodata & H & 100 & 600 & 0\farcs7 & 17.2$\pm$0.0& 22.1&1& S\\ 
& \nodata & \nodata & $\ks$ & 40 & 12400 & 0\farcs7 & 16.7$\pm$0.2& 20.7&2& S\\ 
NGC 4013 & 18.6 & 182 & J & 100 & 700 & 0\farcs7 & 18.1$\pm$0.1& 22.9&2& A\\ 
& \nodata & \nodata & H & 80 & 1040 & 0\farcs9 & 17.2$\pm$0.1& 22.4&2& A\\ 
& \nodata & \nodata & $\ks$ & 40 & 2640 & 0\farcs7 & 16.4$\pm$0.1& 21.8&2& A\\ 
NGC 4144 & 6.6 & 69 & J & 60 & 900 & 0\farcs7 & 17.7$\pm$0.5& 22.9&1& A\\ 
& \nodata & \nodata & H & 60 & 1380 & 0\farcs6 & 17.2$\pm$0.3& 23.0&1& A\\ 
& \nodata & \nodata & $\ks$ & 40 & 7240 & 0\farcs6 & 16.2$\pm$0.8& 21.9&1& A\\ 
NGC 4244 & 4.1 & 89 & J & 60 & 2100 & 0\farcs6 & 17.9$\pm$0.2& 22.2&3& A\\ 
& \nodata & \nodata & H & 60 & 3720 & 0\farcs6 & 17.4$\pm$0.3& 21.5&3& A\\ 
& \nodata & \nodata & $\ks$ & 40 & 11520 & 1\farcs0& 15.9$\pm$1.4& \nodata&3& S\\ 
NGC 4565 & 11.7 & 245 & J & 100 & 1000 & 0\farcs7 & 17.9$\pm$0.1& 23.0&3& S\\ 
& \nodata & \nodata & H & 100 & 1000 & 0\farcs8 & 17.1$\pm$0.1& 21.5&3& S\\ 
& \nodata & \nodata & $\ks$ & 40 & 3200 & 0\farcs6 & 16.4$\pm$0.1& 22.1&3& S\\ 
\enddata
\tablenotetext{a}{Distances come from the NASA/IPAC Extragalactic Database.}  
\tablenotetext{b}{From HyperLEDA \citep{Paturel03}, http://leda.univ=lyon1.fr.}
\tablenotetext{c}{Exposure time for a single exposure. These
  observations were taken during several observing runs that spanned
  the period of time between October 2010 and April 2012. The changing
  individual exposure times for the J and H filters are largely due to
  how our observing strategy evolved over that time.}
\tablenotetext{d}{Total exposure time over the entire galaxy, not
  including frames that were removed during reduction due to image
  artifacts or other issues.}
\tablenotetext{e}{Most galaxies required multiple pointings to obtain
  full coverage. Quoted seeing values are {\it worst case}; in
  several cases there are regions of the final image with
  $\ge$0\farcs1 better seeing than what is listed here.}
\tablenotetext{f}{The surface brightness of a pixel at S/N=10.}
\tablenotetext{g}{$\,$Limiting surface brightness due to large-scale
  image gradients. No value indicates images where gradients were not
  apparent. Limits used for masking, as noted in Section
  \ref{sec:imagemasking}, are 1.2 mag arcsec$^{-2}$ brighter than
  these values.}
\tablenotetext{h}{The number of observed positions along the major
  axis of the galaxy}
\tablenotetext{i}{Our sky subtraction methodology: `A' denotes the
  pattern which placed the galaxy on alternate sides of the detector,
  while `S' indicates that dedicated sky dithers were used.}
\label{tab:obslog}
\end{deluxetable*}

\section{Data}
\label{sec:data}

As in \citetalias{Schechtman-Rook13}, our data come from the
WIYN\footnote{The WIYN Observatory is a joint facility of the
  University of Wisconsin-Madison, Indiana University, Yale
  University, and the National Optical Astronomy Observatory.}
High-resolution InfraRed Camera \citep[WHIRC;][]{Meixner10} as well as
archival IRAC observations\footnote{AORs 31044096, 37342720, 42245120,
  14479104, 3626240, and 3628032.}. Our basic reduction procedure is
mostly unchanged from \citetalias{Schechtman-Rook13}, which can be
referred to for details.  In brief, we first trim, Fowler correct, and
linearity correct the images. Then we use temporally adjacent images
to compute and subtract a sky background. All sky-subtracted images
for a given galaxy in each filter are combined to produce a mosaic,
which is then used to create a mask of all foreground stars,
background galaxies, and the light from the target galaxy itself. The
mask is applied to the individual images and the background
subtraction is repeated, which produces a more accurate estimate for
the sky level. Finally, the mosaics are post-processed to obtain
astrometric and photometric calibrations. A log of our observations is
presented in Table \ref{tab:obslog}.  We have made some important
improvements to parts of our reduction scheme, specifically in our
foreground subtraction, cosmic-ray cleaning, error propagation, and
the masking of residual image gradients. These improvements are
detailed in the following sections.  False-color RGB images of our
final mosaics are compared to {\it Sloan Digital Sky Survey}
\citep[SDSS;][]{York00} \textit{gri}' color images in Figure
\ref{fig:gri_jhk_sample}.

\subsection{Foreground Subtraction}
\label{sec:data:backgroundsubtraction}

In \citetalias{Schechtman-Rook13}, we found that NGC~891 was large
enough on the sky to force the use of dedicated sky frames. These
provided very accurate sky subtraction at the cost of a $\sim$50\%
penalty in observing efficiency. Some galaxies in our sample, however,
are small enough that the vertical scales we wish to probe are smaller
than half the size of WHIRC. Therefore, using a careful dither pattern
which placed the galaxy on alternating sides of the chip
\citep[similar to the method used by][]{Dalcanton00} we avoided the
need for dedicated sky frames in these cases and doubled our
efficiency. Galaxies with which we were able to use this technique are
noted in Table \ref{tab:obslog}.

While the main purpose of this specialized dither pattern was to
``chop'' the target galaxy to different sides of the detector, we also
superimposed small ($\sim$3'') offsets along both dither axes. These
offsets are small enough to subtract extended emission from the
galaxy on the ``wrong'' half of the detector, but large enough to
preserve stellar sources (masked from the sky image).  Due to the
small size of the WHIRC field and the fact that most of our target
galaxies are in the North Galactic Cap, having fluxes for these stars
was crucial for computing our flux calibration.

Finally, the iterative relaxation algorithm used to interpolate over
regions with zero good foreground pixels in
\citetalias{Schechtman-Rook13} was replaced by a third order polynomial
fit to the pixels around each region where there was no overlap in the
sky images. This was done mainly as a computational speed boost; the
iterative relaxation algorithm slows down dramatically when
overlap-free regions are more than a few pixels in size.

\subsubsection{IRAC Foregrounds}

All of the IRAC images had non-zero foregrounds, and therefore also
needed foreground subtraction. Unlike in
\citetalias{Schechtman-Rook13}, the foreground levels appeared to be
generally constant offsets, and were removed by first computing a
clipped mean flux in a region well away from the galaxy using the
IRAF\footnote{IRAF is distributed by the National Optical Astronomy
  Observatory, which is operated by the Association for Research in
  Astronomy, Inc. under cooperative agreement with the National
  Science Foundation.} task IMSTAT and then subtracting this value
from the entire image.

\subsection{Cosmic Ray Cleaning}

In \citetalias{Schechtman-Rook13} we did not perform a dedicated
cosmic ray cleaning, relying on the process of making the mosaics to
average out any cosmic rays. In visual inspections of the final
mosaics of the other galaxies in our sample, however, it was apparent
that for at least some images there were still visible cosmic rays
and/or chip defects. Therefore, we added two algorithms we developed
for our post-processing steps to remove these features from our final
images. These algorithms are based on the IRAF CRCLEAN program, but
allow for iterative cleaning, better statistical control over
thresholding, interactive inspection, and control over pixel
replacement using simple image arithmetic and pixel masks to replace
bad pixels with linearly interpolated values from neighboring regions.

\subsection{Error Propagation}

Despite a careful and thorough propagation of different sources of
errors for NGC 891 in \citetalias{Schechtman-Rook13}, we neglected to
account for data covariance that naturally arises in registered,
coadded data. Here we have added the suitable corrrection for the
errors that come from the correlation of adjacent pixels in the
geometric transformation and mosaic-making steps of our
reduction. This addition results in slightly ($\sim$1.36 times) larger
errors, especially in regions of low surface brightness. While likely
not an issue for any of our fitting (considering that our fits were
clearly not biased by the low surface brightness data) in
\citetalias{Schechtman-Rook13}, it is something to be aware of when
inspecting those results.

\begin{figure*}
\epsscale{1.2}
\plotone{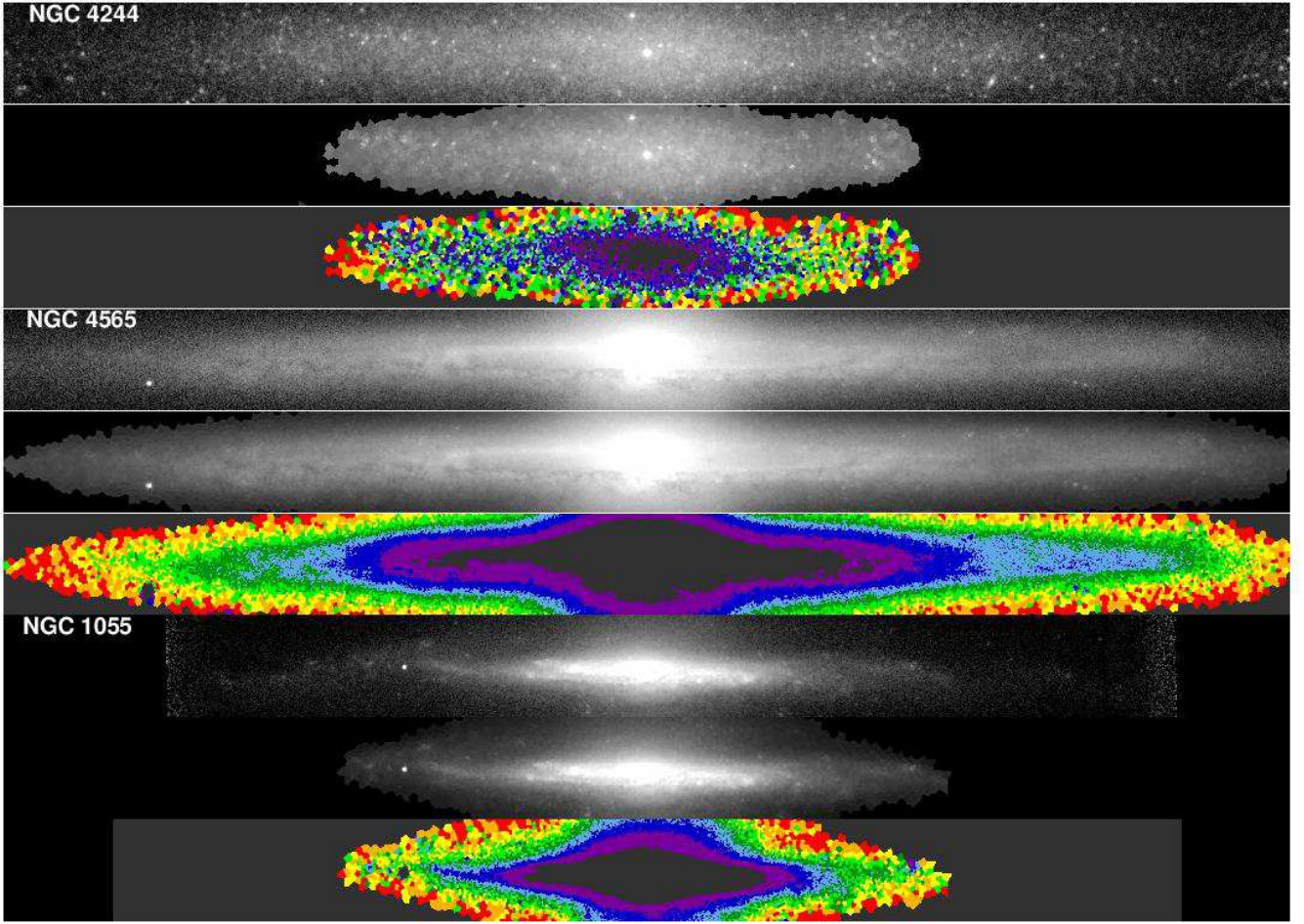}
\caption{Voronoi binning applied to final $\ks$-band mosaics to
  achieve S/N=20 per bin for three representative galaxies in our
  sample (the slow-rotator NGC~4244, the fast-rotator NGC~4565, and
  the disturbed NGC~1055). The top panel for each galaxy shows the
  original $\ks$-band mosaic, the middle panel shows the rebinned
    image chosen by the Voronoi binning algorithm and the same look-up
    table and stretch, and the bottom panel shows the bins
  color-coded by internal bin ID number to show the borders of the
  bins more clearly. In regions with very low S/N we truncate the
  binning process, and assume zero attenuation for our analysis. 
    The grayscale stretch limits are 16.6-20.9, 15.4-21.4, and
    15.4-21.2 mag arcsec$^{-2}$ for NGC 4244, NGC 4565, and NGC 1055,
    respectively.}
\label{fig:voronoi}
\end{figure*}

\subsection{Image Masking}
\label{sec:imagemasking}

All images are masked to remove both foreground and background
contaminants. This masking is done by hand, and aims to err on the
side of mask-inclusion when it was unclear whether an apparent source
was part of the galaxy or not. For NGC 4144 and 4244, which contain
many discrete star clusters (or possibly even individual super-giant
stars) across their disks, the false-color WHIRC images were consulted
to aid in identifying objects to mask. Specifically, background
galaxies visible through the disk had similar appearances to the star
clusters but were generally much redder. 

As a necessary compromise in order to subtract the foreground from
galaxies which we dithered between sides of the detector to eliminate
the need for dedicated sky frames, any faint extended galaxy light
contaminating the `sky' half of the WHIRC field was identically set to
zero by the sky-subtraction process. For these galaxies we therefore
masked out such regions of the final mosaics in order not to
introduce any artifical high-latitude truncations caused by our sky
subtraction.

Even with this effort, in most final mosaics there were still faint
brightness gradients over portions of the field. We measured the
magnitude of these gradients by hand over $\sim$100 arcsec$^{2}$
regions away from the galaxies, and found variations of $\sim$1 DN. We
then computed the corresponding surface-brightness values, which are
given in Table \ref{tab:obslog} as $\mu$(lim). For subsequent
analysis, we mask out any pixels below three times this value (i.e.,
$\mu>\mu({\rm lim})-1.2$) to avoid regions of these images which could
be dominated by these erroneous features.  Generally these gradients
were at much lower surface brightness than regions of interest in the
galaxy, and largely impact the vertical extent out to which we can
constrain thick, low-surface-brightness disk components.

\subsection{Image Quality and Depth}
\label{sec:imagequality}

Table \ref{tab:obslog} provides information about the overall quality
and depth of our sample images. Generally the resolution of
our mosaics is excellent, with 50\% at seeing $\le$0\farcs6 and
$>$80\% having seeing $\le$0\farcs7. NGC~4244, the only galaxy in our
sample with seeing at or above 1 arcsecond, happens to be the most
nearby galaxy (D=4.1 Mpc), making excellent resolution less of a
necessity. Also note that all of our data on NGC~522, our most distant
target, has seeing $\le$0\farcs6, some of the best in our sample.

For any given filter we achieve fairly uniform depth across each
mosaic and between sources. To provide a quantitative estimate on our
depth we compute the surface brightness corresponding to a
S/N$\approx$10 in a single pixel.  We compute statistics on all pixels
with 9.9$<$S/N$<$10.1 to estimate the mean and standard deviation of
this surface brightnesses value, given as $\mu$(S/N=10) in Table
\ref{tab:obslog}. We do the statistics in flux units before converting
to surface brightness, and report the mean because it was always
brighter than the median and is thus a more conservative
estimate. These values are relatively bright, but note that each WHIRC
pixel is only $\approx$0.01 arcsec$^{2}$. When producing the vertical
profiles we bin our data to ensure high S/N over our entire region of
interest.  As discussed in Section \ref{sec:imagemasking}, the
practical limit for accurate surface-photometry is governed by small
residual gradients in our images which persist even after attempting
to minimize them in our sky subtraction. Table \ref{tab:obslog}
indicates that these gradients are generally faint enough to pose little
problem for our analysis.

\section{Attenuation Correction}
\label{sec:attencorr}

\subsection{Slow Rotators}
\label{sec:slowrotatorattencorr}

We expect that the dust properties of fast-rotating galaxies will be
relatively similar to NGC~891 in their basic composition and geometry
(e.g., clumping fractions, dust density thresholds for embedding young
stars, scale-heights), although there will be differences in the
details. For example, NGC 891's extended dust tendrils at large
scale-heights may be unusual. Nonetheless, our attenuation correction
strategy is based on correlating the effective attenuation with
changes of near-to-mid-infrared colors that we find in our models,
and applying this {\it correlation} to the observations. Insofar
as the correlation is relatively immune to details of the models,
so too are our estimated effective attenuations.
 
For the two slow-rotating galaxies in our sample, however, it is not
clear that the assumed dust properties are even close to a good
representation, given the different dust morphologies and
metallicities of these systems compared to fast-rotators like NGC 891
and the MW \citep{Dalcanton04}.  Consequently, we redid the
attenuation correction modeling from \citetalias{Schechtman-Rook13}
for NGC~4244 (the better-studied of the two slow-rotators in our
sample) to test the sensitivity of method to the details of the dust
properties and geometry. A description of the procedure can be found
in \citetalias{Schechtman-Rook13}, with the application and results
for NGC~4244 presented in the Appendix \ref{sec:ngc4244attencorr}
here. The salient feature of the modeling is that it must reproduce
both the the attenuated star-light and the re-radiated thermal and
molecular dust emission.

We find that a slightly different attenuation correction is necessary
for slow-rotators. However, as will be shown in Section
\ref{sec:attencorrresults}, both slow-rotators appear to have very
little $\ks$-band attenuation. As a result, the exact form of the
slow-rotator attenuation correction is largely unimportant for our
results in these galaxies. Moreover, given the very large difference
in dust parameters and the rather modest change in the attenuation
correction compared to the model for NGC~891, this supports our claim
that our attenuation method, once properly calibrated by a relistic
model, is relatively insensitive to the detailed dust prescription
even for the fast rotating systems in our study.

\begin{deluxetable}{cc}
\tablewidth{0pt}
\tablecaption{Voronoi Binning Regions}
\tablehead{ \colhead{Galaxy} & \colhead{Region Shape (kpc)}}
\startdata 
NGC~522 & 15x1.6\\
NGC~891 & 11x1.5\\
NGC~1055 & 9.8x2.1\\
NGC~4013 & 9.8x1.5\\
NGC~4144 & 1.5x0.54\\
NGC~4244 & 2.4x0.44\\
NGC~4565 & 15x1.5
\enddata
\tablenotetext{a}{Regions are roughly elliptical, and are given as
  axb, where a and b are the semi-major and semi-minor axes.}
\label{tab:vbins}
\end{deluxetable}

\subsection{Voronoi Binning}

As in \citetalias{Schechtman-Rook13}, we focus our analysis on the
$\ks$-band data as it is least likely to be affected by errors in the
attenuation correction. Unlike NGC~891, however, many of these
galaxies are at lower surface brightness, and as a result low S/N
became an issue in computing the attenuation corrections at large
radii. This is especially true for NGCs~4144 and 4244, the slow
rotators, as well as NGC~1055, which as a disturbed galaxy at
$i\sim85^{\circ}$ has large regions of stellar emission at low surface
brightness.

As a result, we are forced to bin our data to ensure that our
attenuation correction is not dominated by pixel-to-pixel variations
in regions of low surface brightness. Since it is critical to maintain
our excellent WHIRC resolution near the midplane (where S/N is high),
we employ Voronoi binning to adaptively bin our data before computing
an attenuation correction. Our algorithm, written in Python, is based
on the procedure outlined by \citet{Cappellari03}, but with
significant numerical optimizations in order to run efficiently on
images with $>$1 million pixels. The most critical of these
  optimizations are the use of balltrees, which allow for the fast
  computation of nearest neighbors by partitioning the data into a
  binary tree \citep{Omohundro89}, and the use of fast native Numpy
  array operations wherever possible. Additionally, we choose a
maximum height (by visual inspection), above which the attenuation is
assumed to be zero. This keeps us from correcting low-SB data
  which we suspect to be uncontaminated by dust, and also serves to
  speed up the computation. While there are likely small amounts of
  high-latitude attenuation at UV/optical wavelengths in some of these
  galaxies (e.g. NGC~891; \citealt{Howk97,Seon14}), the average
  attenuation at large heights in the {\it NIR} is likely minimal. To
fully realize performance gains we are forced to relax the bin
circularity and adjacency criteria of \citet{Cappellari03}, but due to
the relative smoothness of the underlying surface brightness
distribution at low S/N we find no evidence of any issues arising as a
result of these decisions. We set our binning threshold at S/N = 20,
which we find empirically produces a smoothly varying attenuation map
while preserving as much spatial structure as possible. The
  source code for the Voronoi binning script is available for general
  use on Github\footnote{https://github.com/AndrewRook/astro/tree/master/voronoi}.

After compiling the initial bins, we perform 20 iterations of the
Centroidal Voronoi Tessellation algorithm \citep{Du99}. As a final
step, we reject very large bins, which we find tend to probe
background fluctuations rather than real features in the galaxy. An
example of the binning process is shown in Figure \ref{fig:voronoi}.
With this method the excellent WHIRC resolution is maintained in the
dusty midplane regions, while larger bins at large heights allow for a
robust attenuation correction even in areas where the S/N per pixel is
low. In regions with very low S/N (S/N per pixel $\lesssim$0.6) we
truncate the binning process and treat those pixels as having zero
attenuation. This assumption is well-justified by a visual inspection
of the optical and NIR images, which show no indication of dust in
these regions. Estimates of the physical size of the binned regions
are shown in Table \ref{tab:vbins}.

\begin{figure}
\epsscale{1.3}
\plotone{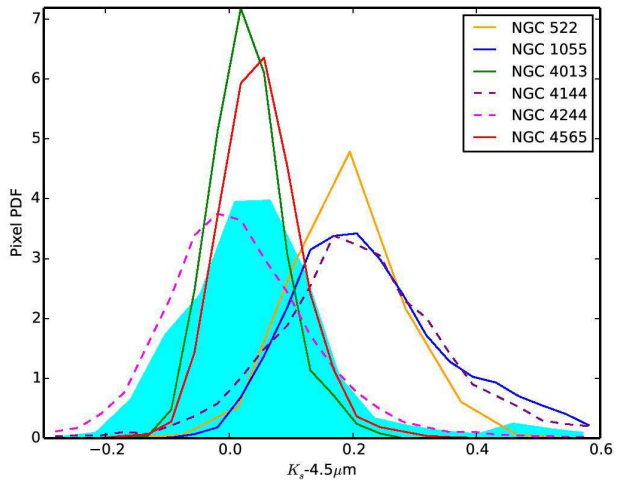}
\caption{Probability distribution function (normalized so the
  integrated y-value = 1) of $\ks$-4.5$\mu$m colors for Voronoi-binned
  pixels in unattenuated regions of the galaxies in our sample,
  color-coded by galaxy. For fast rotators (solid lines) we only use
  bins with $|z|\geq 1 kpc$, while due to their transparency we place
  no height restriction on the slow rotators (dashed lines). NGC~891,
  the galaxy on which we based our attenuation correction in
  \citetalias{Schechtman-Rook13}, is shown as the filled cyan curve.}

\label{fig:binnedcolor}
\end{figure}

\begin{figure*}
\epsscale{1.2}
\plotone{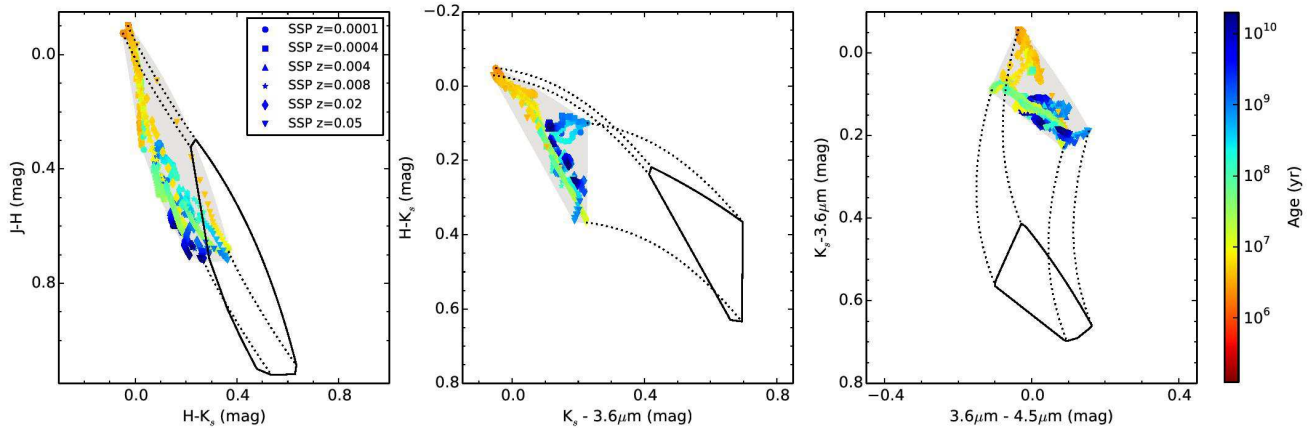}
\caption{IR color dependence on age and metallicity for SSP models
  from \cite{Bruzual03}. The different symbols denote the metallicity
  of the SSPs, with 0.02 being solar. Colors indicate the age of the
  SSPs. The gray shaded region shows the color space accessible by
  linear combinations of SSPs, while the dotted black lines and solid
  black shape show that color space for $\aeks$=0.5 mag. The
  attenuation correction results in a slight shift blueward in the
  3.6$\mu$m-4.5$\mu$m colors at small attenuations due to
  contributions from PAH emission.}
\label{fig:ssps}
\end{figure*}

\subsection{Improved Attenuation Correction Methodology}
\label{sec:attencorrresults}

\begin{figure}
\epsscale{1.3}
\plotone{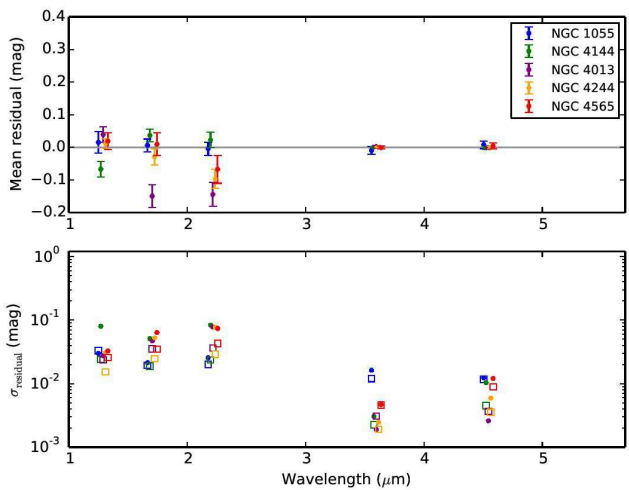}
\caption{Residuals from fitting SSPs to the multi-band near- and
  mid-infrared fluxes of individual sightlines for several galaxies in
  our sample in regions expected to have low attenuation. Top panel:
  mean residual (weighted by the errors on the data) are plotted as a
  function of wavelength (from left to right, J, H, $\ks$, IRAC
  3.6$\mu$m, and IRAC 4.5$\mu$m). Errors indicate the weighted
  standard deviation about the mean residual.  Bottom panel: weighted
  standard deviations (filled circles) are compared to the average
  uncertainty in the data (open squares). Points at each wavelength
  have been scattered slightly around the central wavelength of each
  filter to improve readability.}
\label{fig:sspjust}
\end{figure}

\begin{figure*}
\centering
\plotone{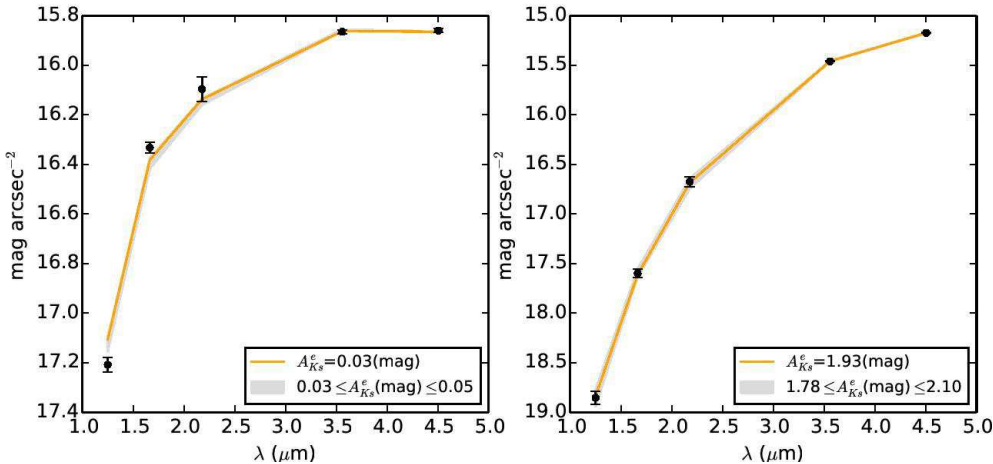}
\caption{ Demonstration of the attenuation correction method on
    individual binned pixels of NGC~891. Left panel: a low attenuation
    region at $R\sim2$kpc, $z\sim0.5$kpc, with a bin size of 0.25
    arcsec$^2$. Right panel: a high attenuation region at $R\sim2$kpc,
    $z\sim0$kpc, with a bin size of 0.1 arcsec$^2$. Black points show
    the data, the orange line indicates the best fit, while the gray
    shaded region indicates the fits within our 67\% confidence
    limits. The legend gives the best fitting attenuation, as well as
    the full range of $\aeks$ allowed by the models within the
    confidence limits.}
\label{fig:attencorrdemo}
\end{figure*}

\begin{figure}
\epsscale{1.3}
\centering
\plotone{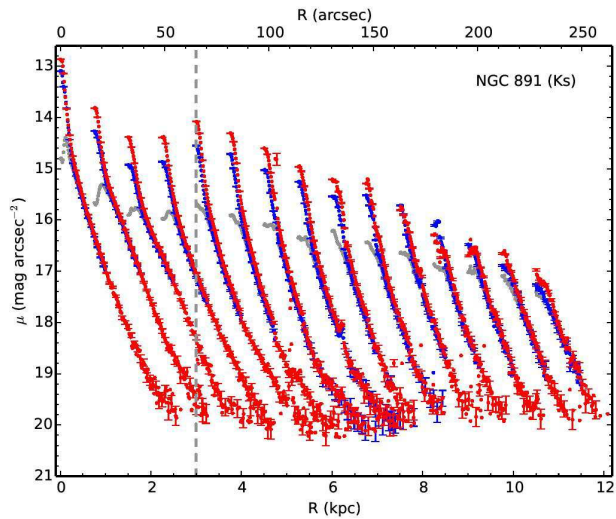}
\caption{ Comparison between the attenuation correction used in this work
    with that used on NGC~891 in \citetalias{Schechtman-Rook13}. Gray points
    show the uncorrected surface-brightness profiles, while the red and blue
    points denote profiles corrected with the \citetalias{Schechtman-Rook13}
    method and the method from this work (respectively).}
\label{fig:NGC891attencorrcompare}
\end{figure}

\begin{figure*}
\centering
\begin{tabular}{@{}c@{}c@{}}
\includegraphics[scale=0.4]{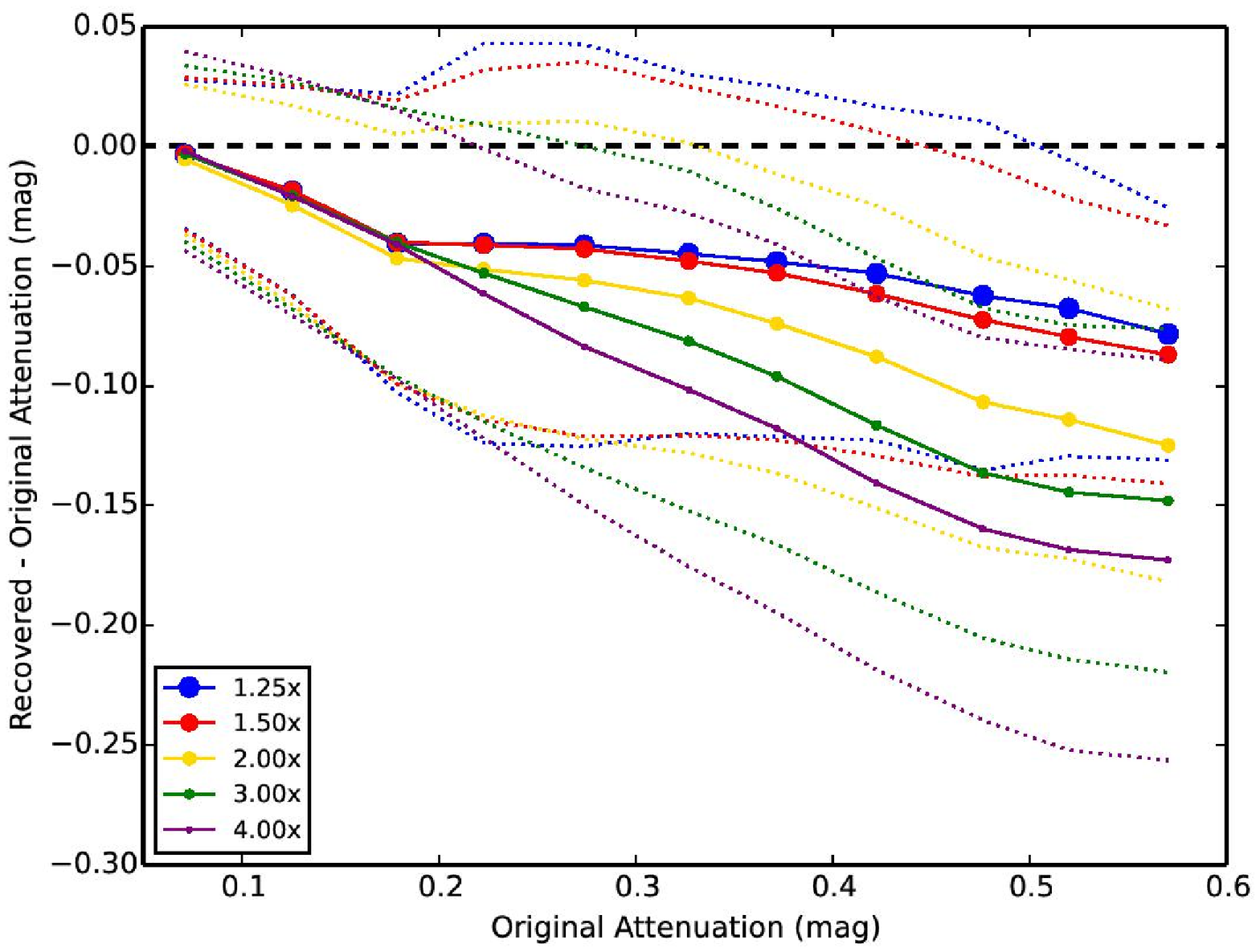}
&
\includegraphics[scale=0.4]{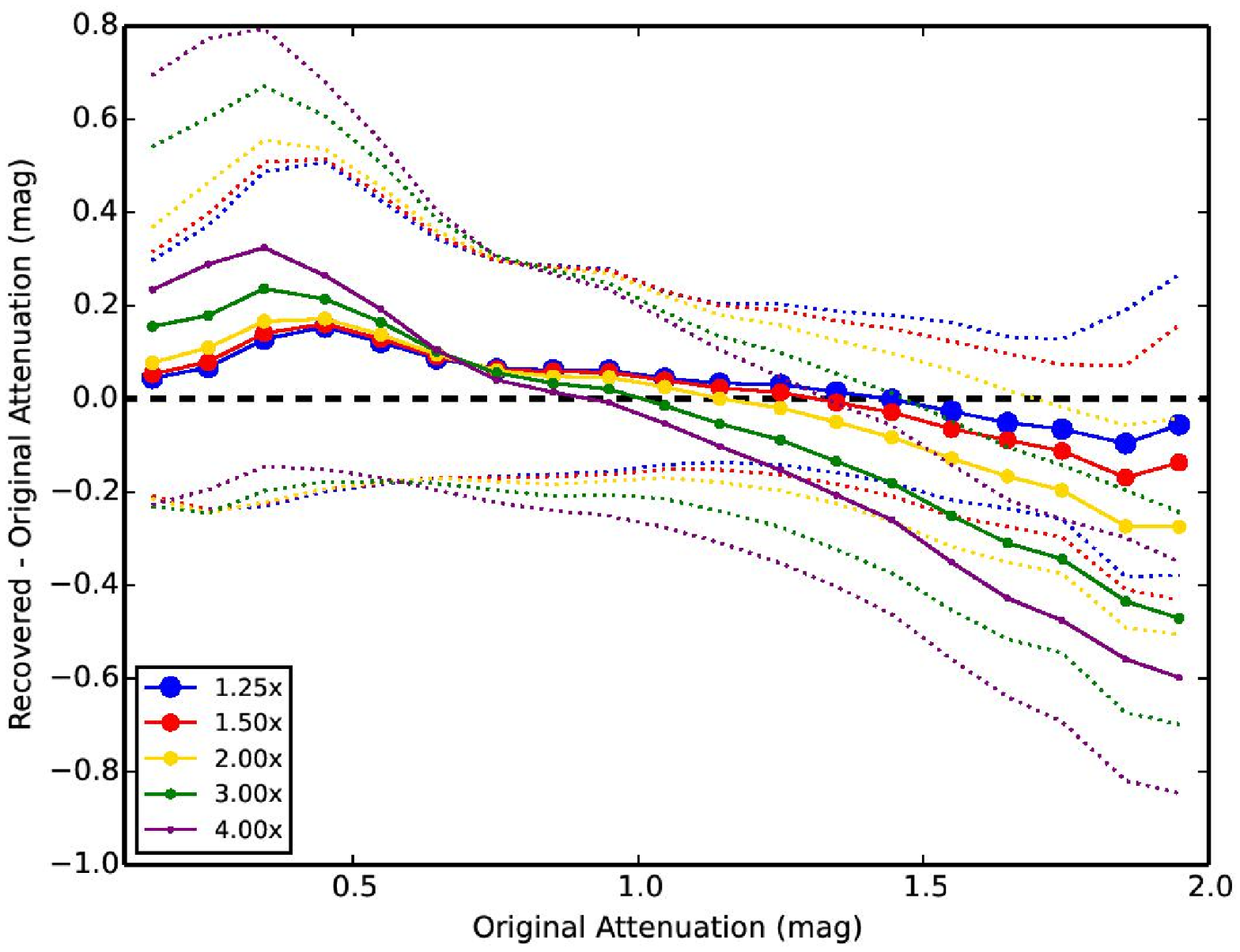}
\end{tabular}
\caption{Effect of increasing galaxy distance on our attenuation
  correction, for NGC~4565 (left panel) and NGC~891 (right
  panel). Each point shows the  residual between the recovered
  attenuation (averaged over individual pixels) at a given distance
  and the actual attenuation. Dotted lines show the standard deviation
  about the mean. Points are color coded by distance (shown as a
  multiplier on the actual distance to the galaxies, which is 11.7 and
  9.5 Mpc for NGC~4565 and NGC~891, respectively). The dashed black
  line shows zero residual as reference, i.e., the ideal
    case where going out to larger distance has no bias on the
  attenuation measurement.}
\label{fig:distattencorr}
\end{figure*}

\begin{figure*}
\centering
\begin{tabular}{@{}c@{}c@{}c@{}}
\multicolumn{3}{@{}c@{}}{\includegraphics[scale=0.8]{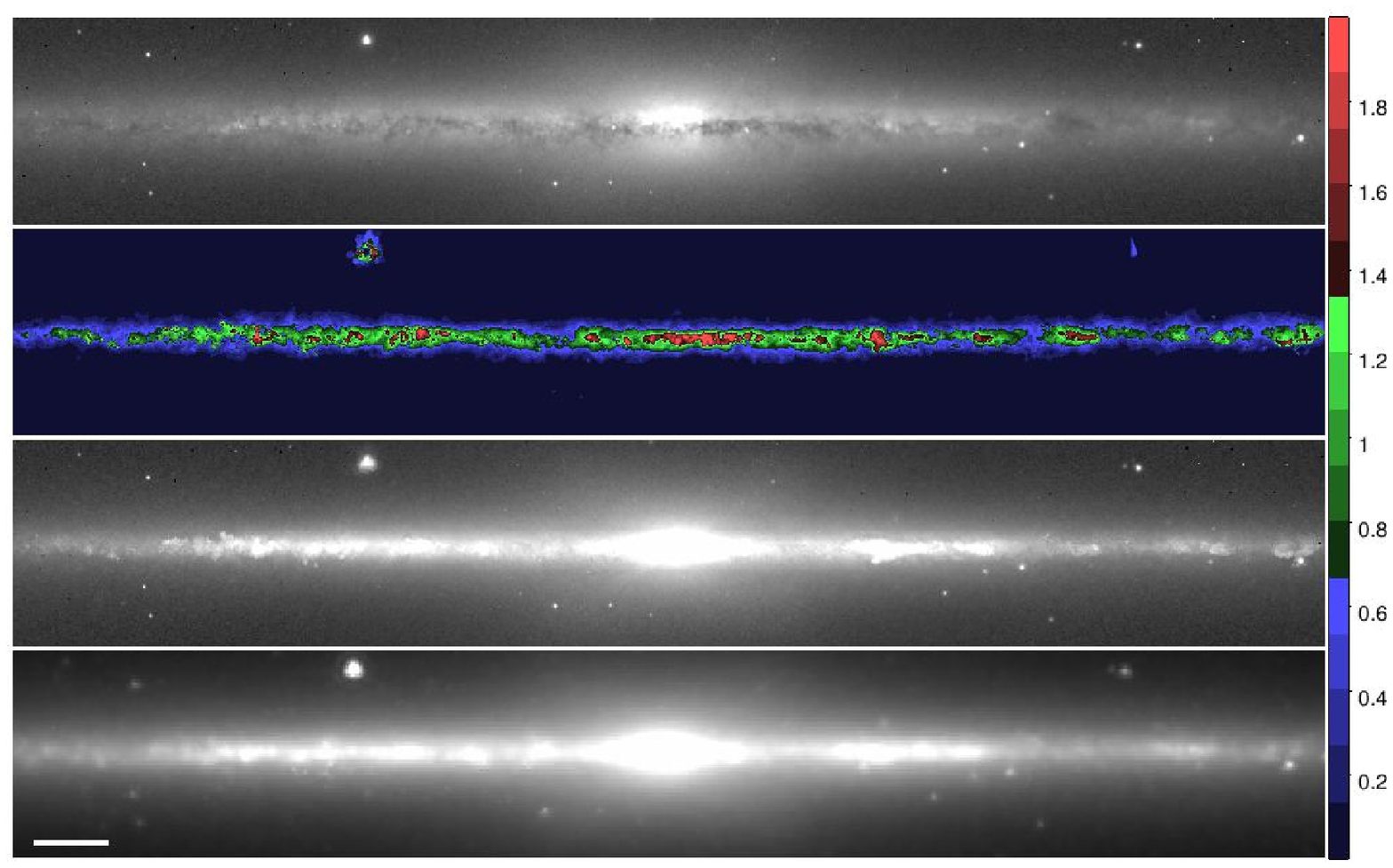}}
\\
\includegraphics[scale=0.5]{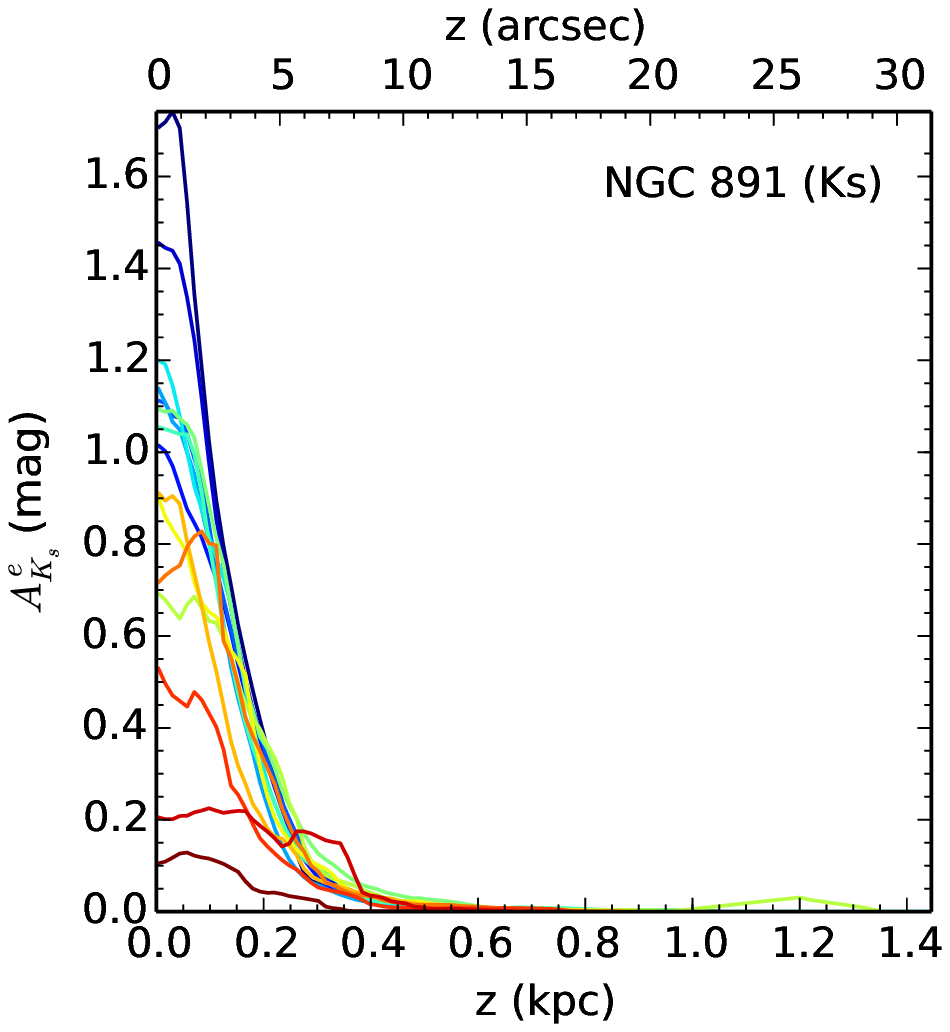}
&
\multicolumn{2}{@{}c@{}}{\includegraphics[scale=0.57]{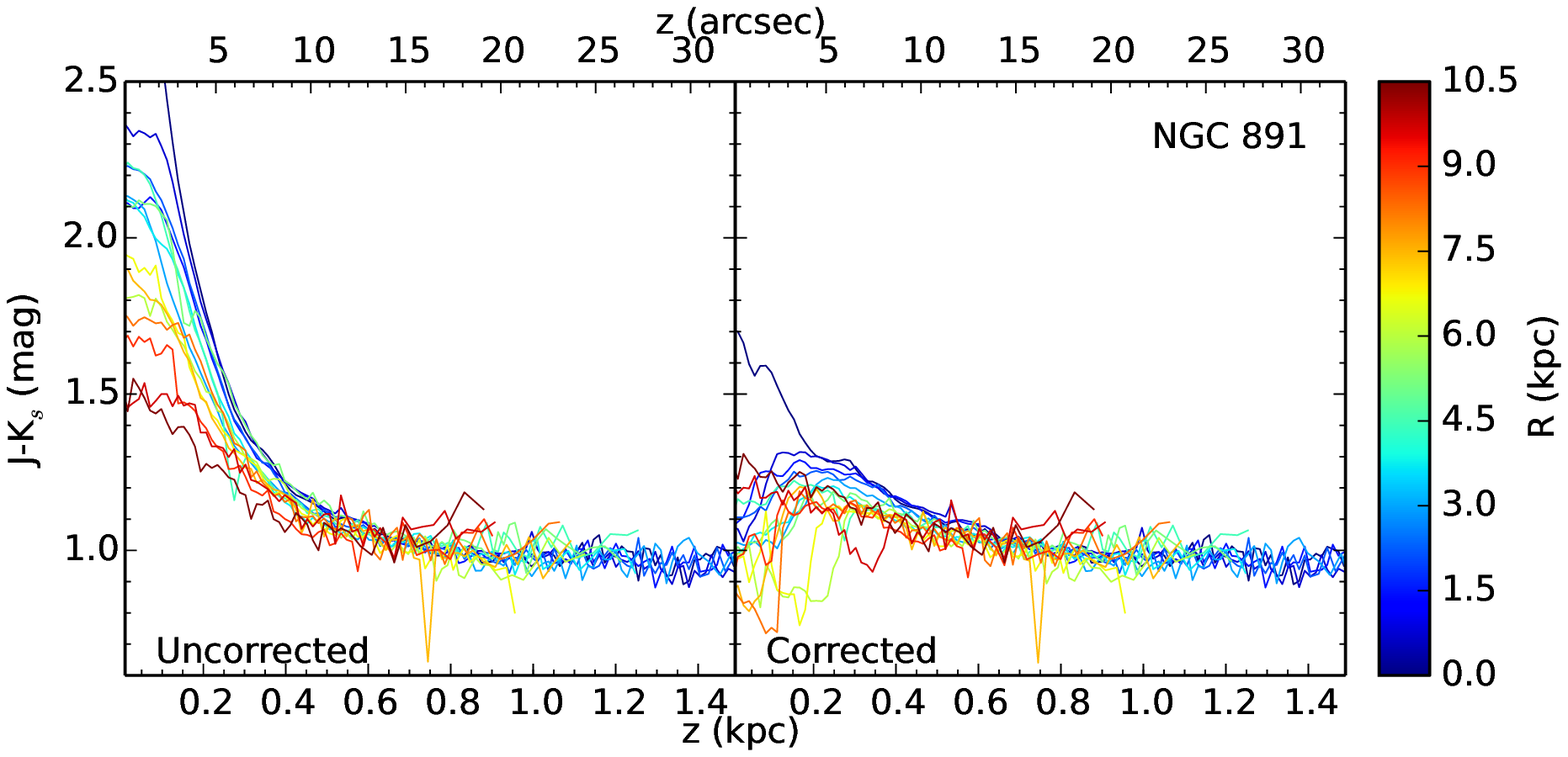}}
\end{tabular}
\caption{Images and vertical profiles illustrating attenuation effects
  and corrections for NGC~891. Images, top to bottom: central region
  of the $\ks$-band image (with a grayscale stretch from 14.7-21.2
    mag arcsec$^{-2}$); $\ks$-band attenuation map, in magnitude
  units; attenuation-corrected $\ks$-band image, with the same
  gray-scale as the top panel; IRAC 4.5$\mu$ image, for comparison.
  The white bar shows 1 kpc at the assumed distance to NGC~891. Bottom
  plots, left to right: vertical attenuation profiles; observed
  vertical J-$\ks$ profiles; and attenuation-corrected vertical
  J-$\ks$ profiles.  The color of each line shows the radius of that
  profile. The profiles are adaptively binned in the $z$ direction to
  maintain an error of $\le$0.05 mag, out to the limiting surface
  brightness of our data. The $z$ coordinate used here is the {\it
    projected} height, but this galaxy is estimated to be nearly
  perfectly edge on.}
\label{fig:ngc891attencorr}
\end{figure*}

\begin{figure*}
\centering
\begin{tabular}{@{}c@{}c@{}c@{}}
\multicolumn{3}{@{}c@{}}{\includegraphics[scale=0.8]{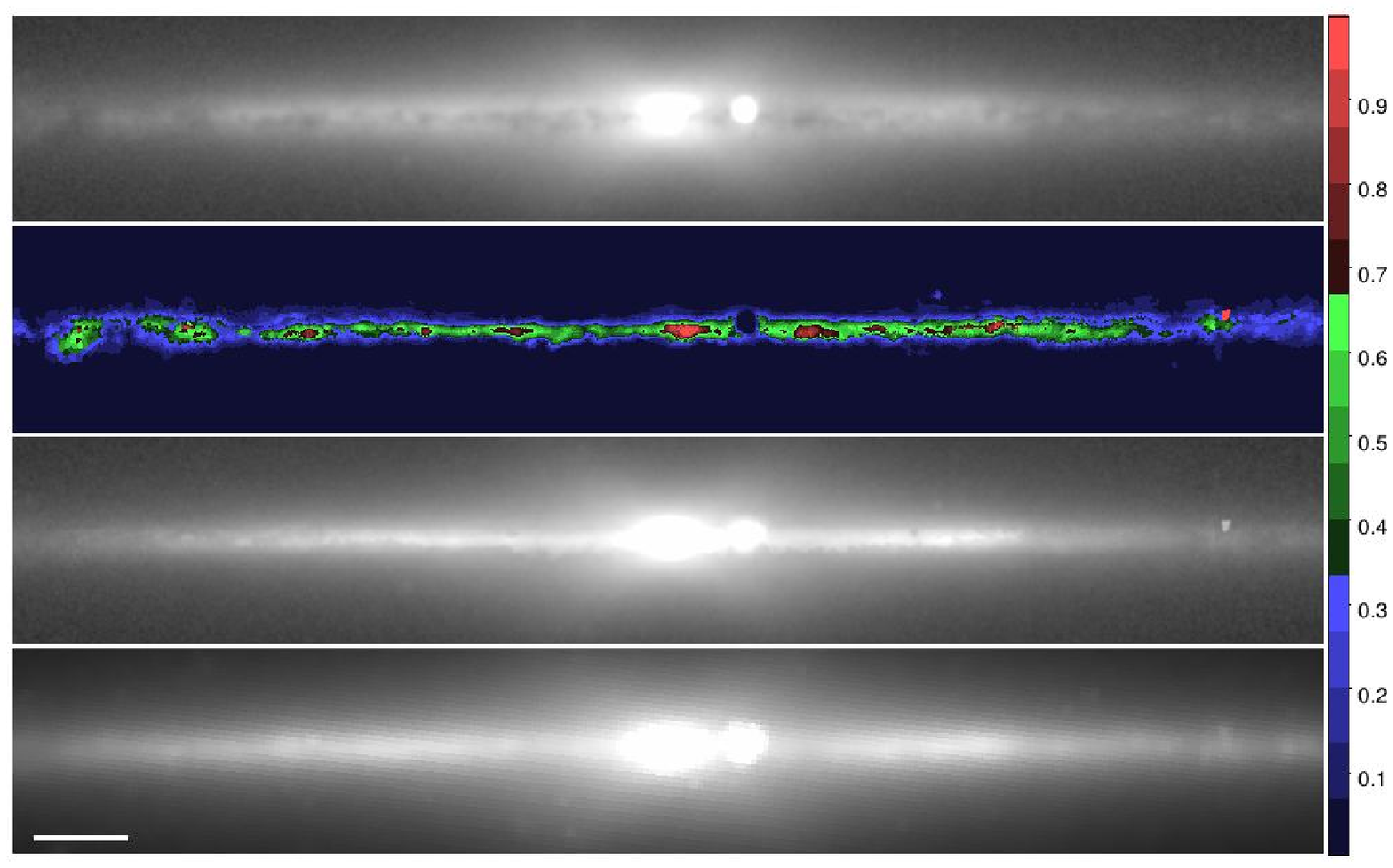}}
\\
\includegraphics[scale=0.5]{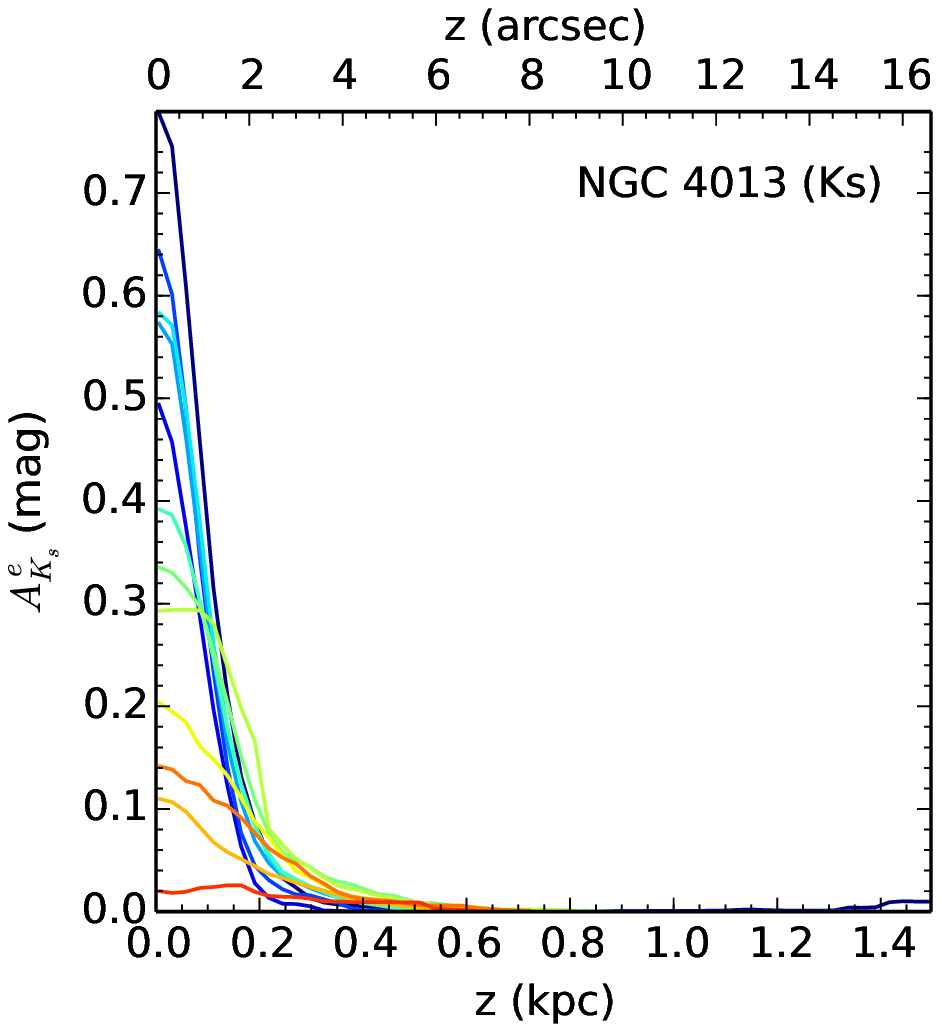}
&
\multicolumn{2}{@{}c@{}}{\includegraphics[scale=0.57]{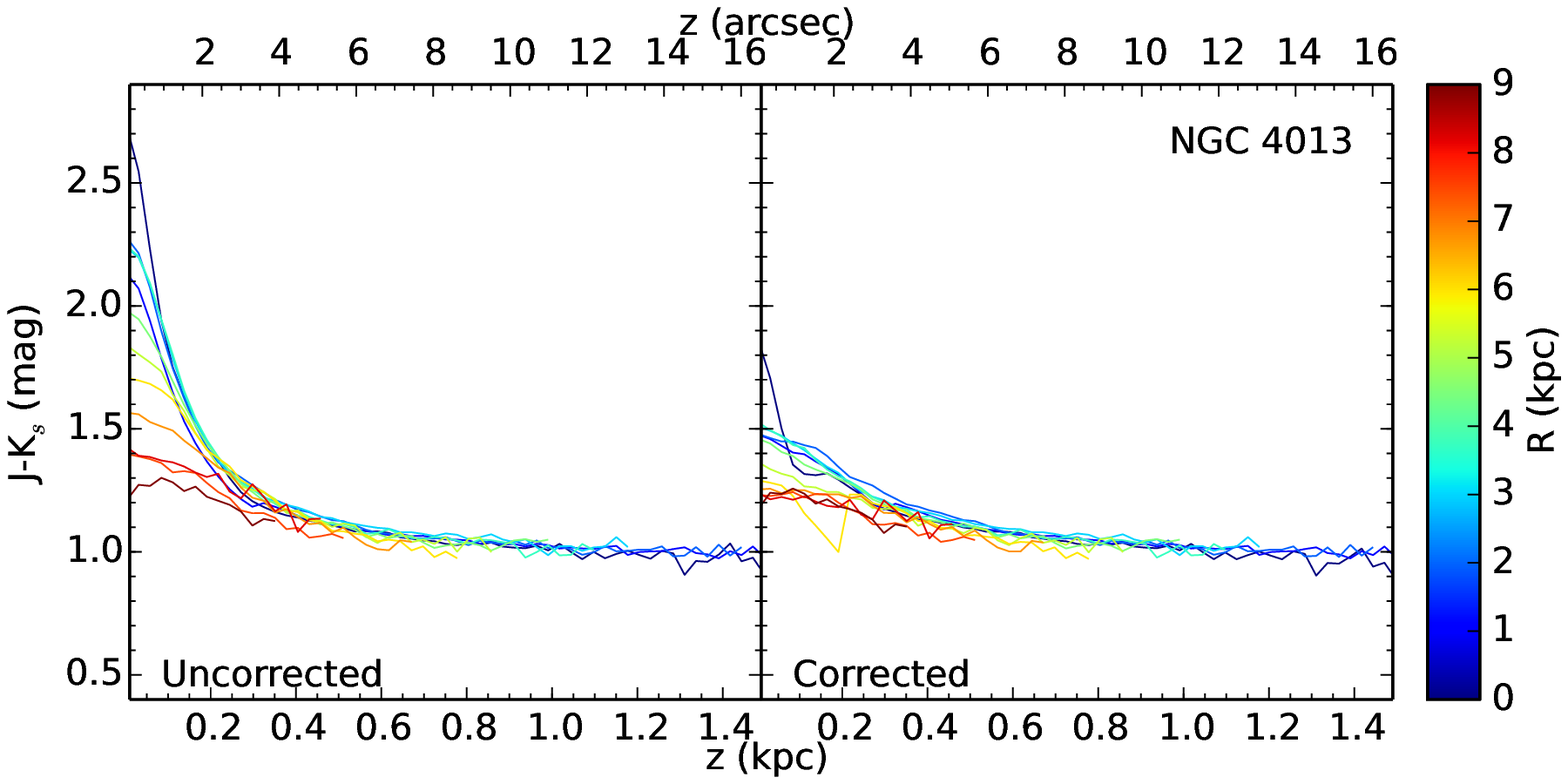}}
\end{tabular}
\caption{Same as Figure \ref{fig:ngc891attencorr}, but for NGC~4013
  (the $\ks$-band images grayscale range is from 15.0-21.2 mag
    arcsec$^{-2}$). Note the bright star just to the right of the
  center of the galaxy is masked out from all of our data after this
  step and so does not appear in any of our attenuation, color, or
  surface brightness profiles, nor does it affect our model fits.  The
  $z$ coordinate used here is the {\it projected} height, but this
  galaxy is estimated to be nearly perfectly edge on.}
\label{fig:ngc4013attencorr}
\end{figure*}

\begin{figure*}
\centering
\begin{tabular}{@{}c@{}c@{}c@{}}
\multicolumn{3}{@{}c@{}}{\includegraphics[scale=0.8]{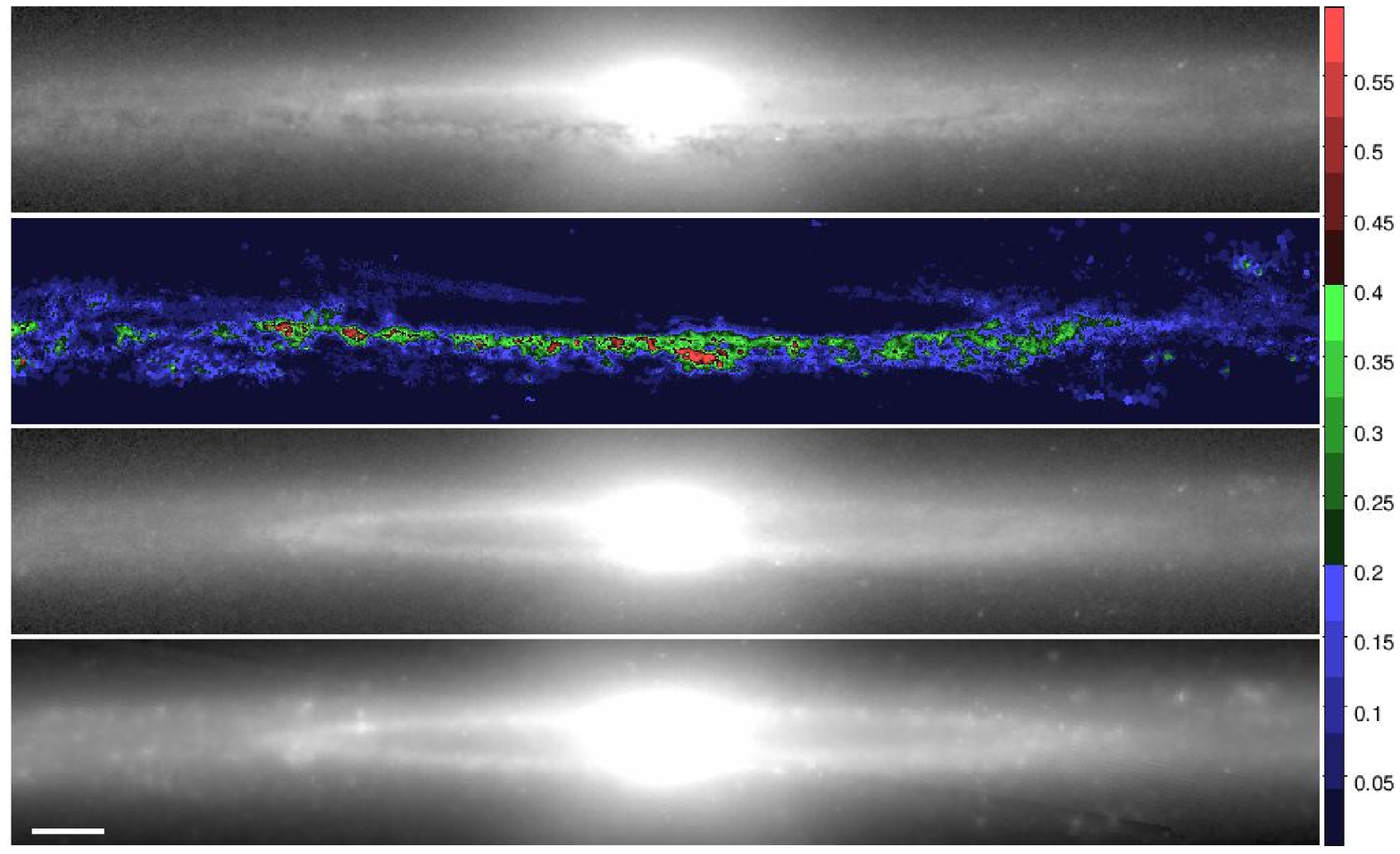}}
\\
\includegraphics[scale=0.5]{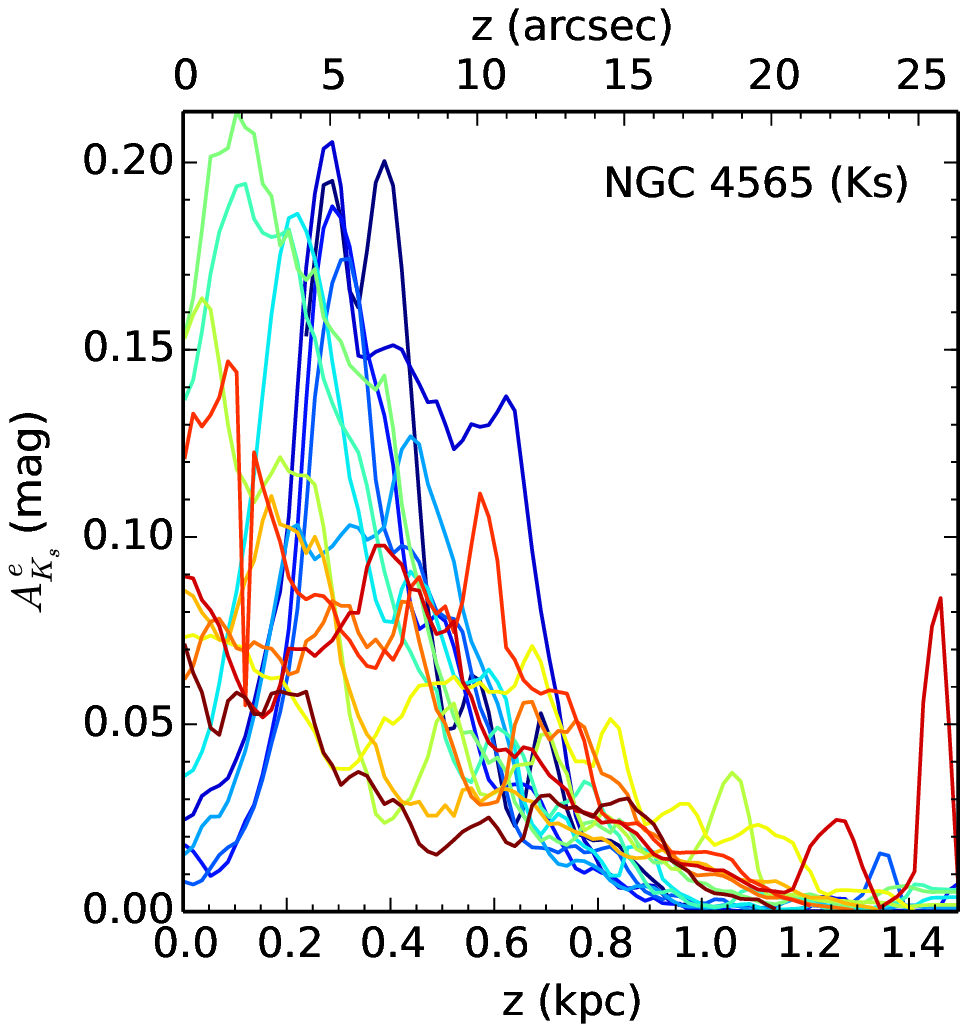}
&
\multicolumn{2}{@{}c@{}}{\includegraphics[scale=0.57]{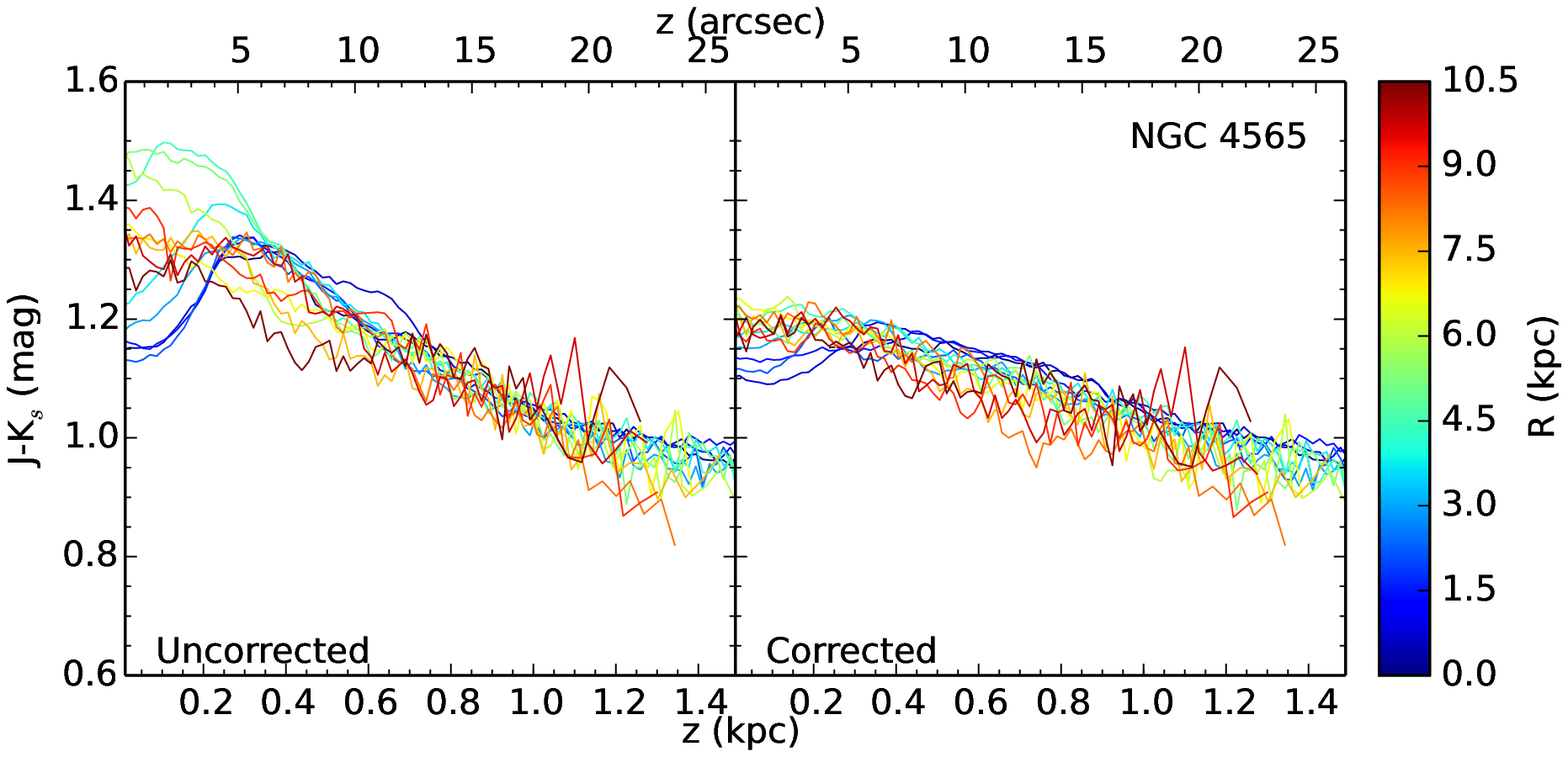}}
\end{tabular}
\caption{Same as Figure \ref{fig:ngc891attencorr}, but for NGC~4565
  (the $\ks$-band images grayscale range is from 15.4-21.4 mag
    arcsec$^{-2}$).  The $z$ coordinate used here is the {\it
    projected} height; this galaxy is estimated to be inclined at
  $87.6$ deg (see \ref{sec:ngc4565}).}
\label{fig:ngc4565attencorr}
\end{figure*}

\begin{figure*}
\centering
\begin{tabular}{@{}c@{}c@{}c@{}}
\multicolumn{3}{@{}c@{}}{\includegraphics[scale=0.8]{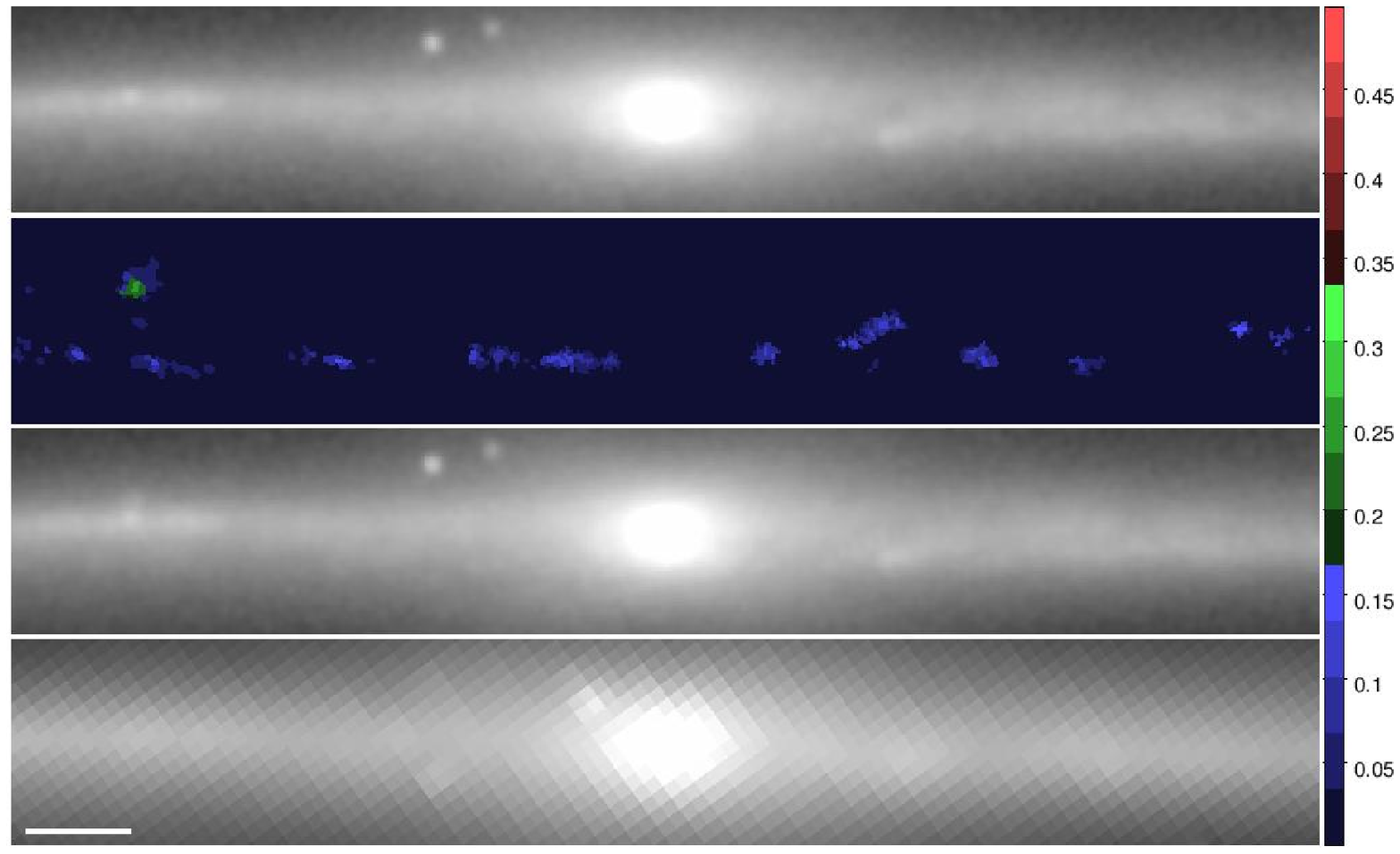}}
\\
\includegraphics[scale=0.5]{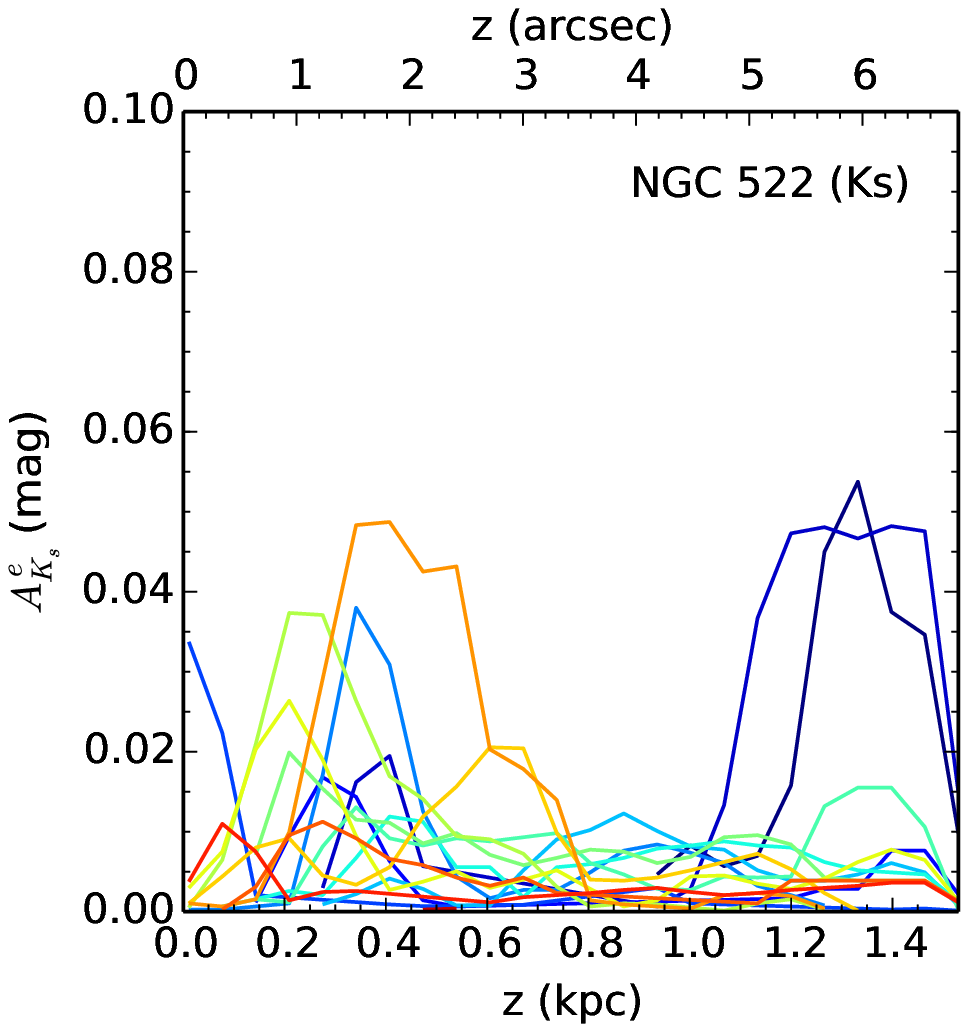}
&
\multicolumn{2}{@{}c@{}}{\includegraphics[scale=0.57]{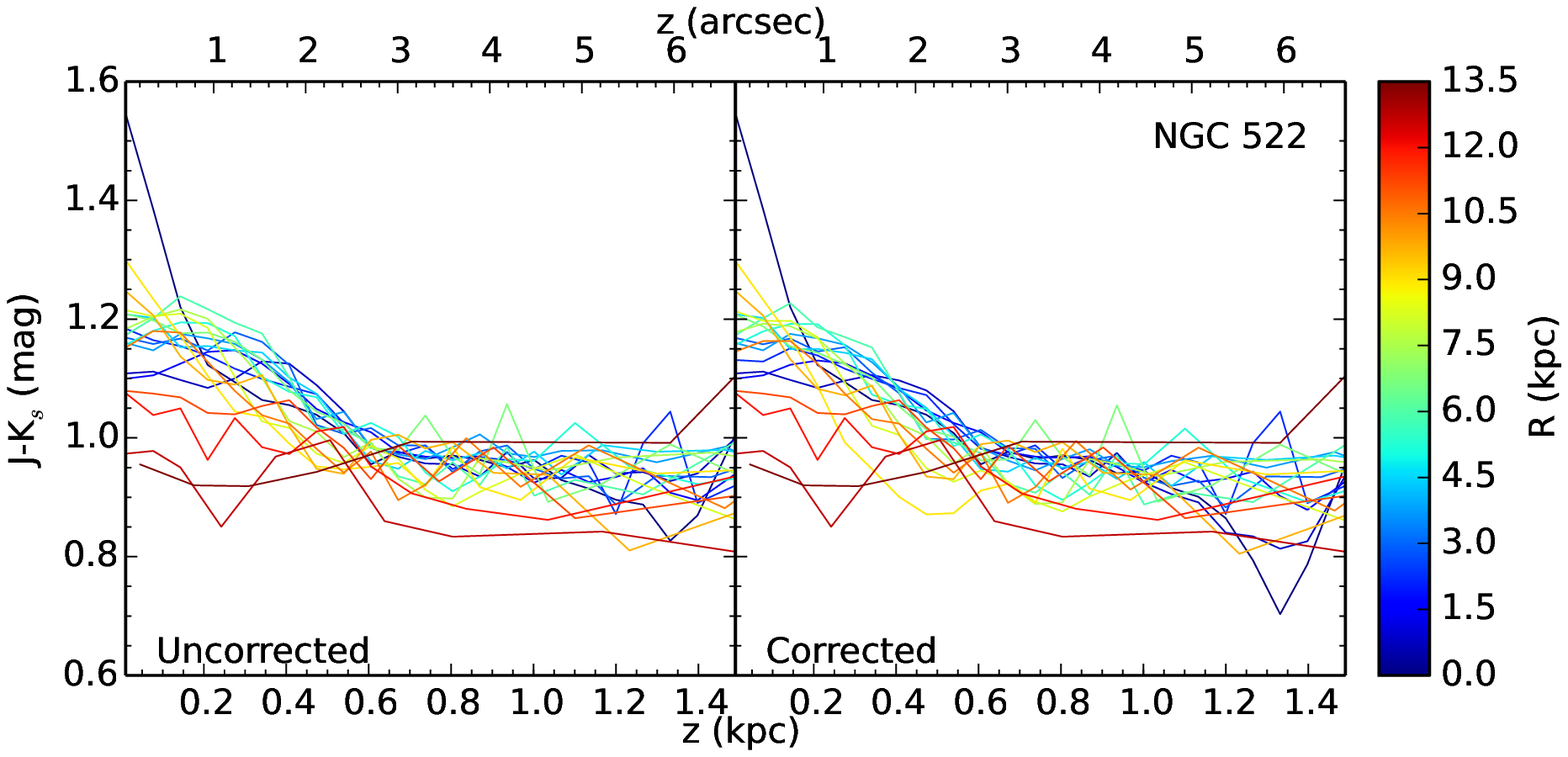}}
\end{tabular}
\caption{Same as Figure \ref{fig:ngc891attencorr}, but for NGC~522
  (the $\ks$-band images grayscale range is from 15.4-21.2 mag
    arcsec$^{-2}$).  The $z$ coordinate used here is the {\it
    projected} height; this galaxy is estimated to be inclined at
  $88.5$ deg (see \ref{sec:ngc522}).}
\label{fig:ngc522attencorr}
\end{figure*}

\begin{figure*}
\centering
\begin{tabular}{@{}c@{}c@{}c@{}}
\multicolumn{3}{@{}c@{}}{\includegraphics[scale=0.8]{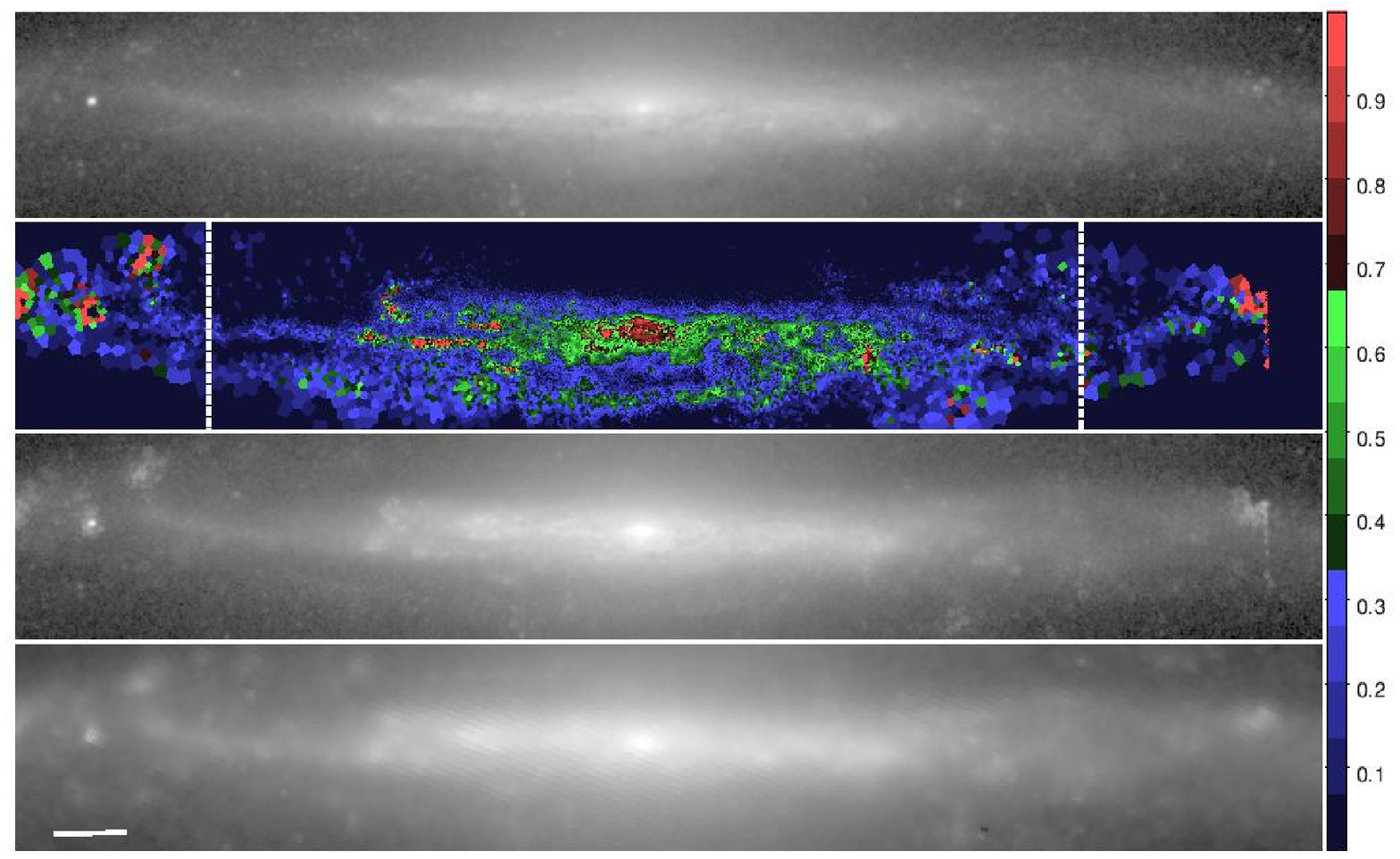}}
\\
\includegraphics[scale=0.5]{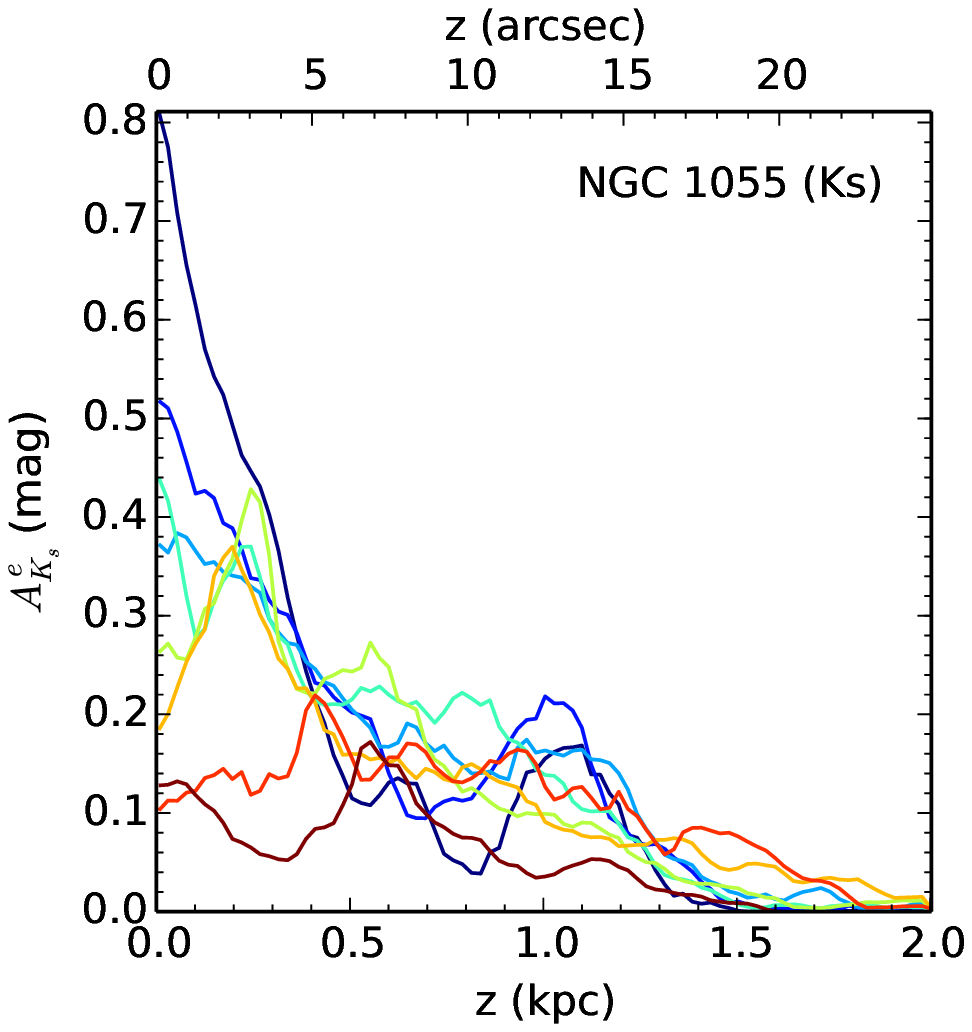}
&
\multicolumn{2}{@{}c@{}}{\includegraphics[scale=0.57]{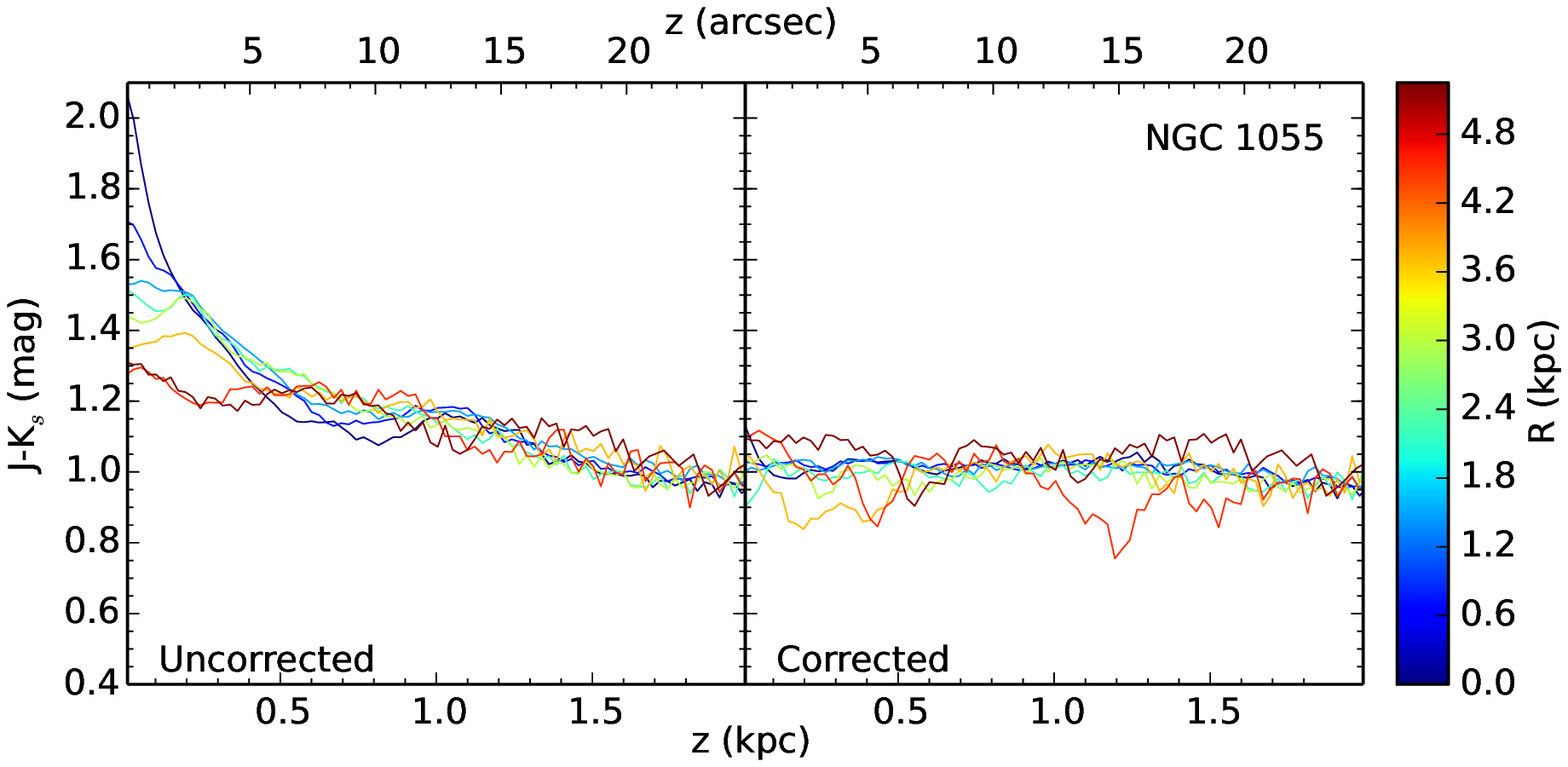}}
\end{tabular}
\caption{Same as Figure \ref{fig:ngc891attencorr}, but for NGC~1055
  (the $\ks$-band images grayscale range is from 13.7-21.2 mag
    arcsec$^{-2}$). The vertical dashed lines in the top panel
  indicate the outer limit of our model fitting for this galaxy, which
  avoids regions of low S/N where flux from the galaxy's spheroidal
  component begins to dominate over the disk. The $z$ coordinate used
  here is the {\it projected} height; this galaxy is estimated to be
  inclined at $85.5$ deg (see \ref{sec:ngc1055}).}
\label{fig:ngc1055attencorr}
\end{figure*}

\begin{figure*}
\centering
\begin{tabular}{@{}c@{}c@{}c@{}}
\multicolumn{3}{@{}c@{}}{\includegraphics[scale=0.8]{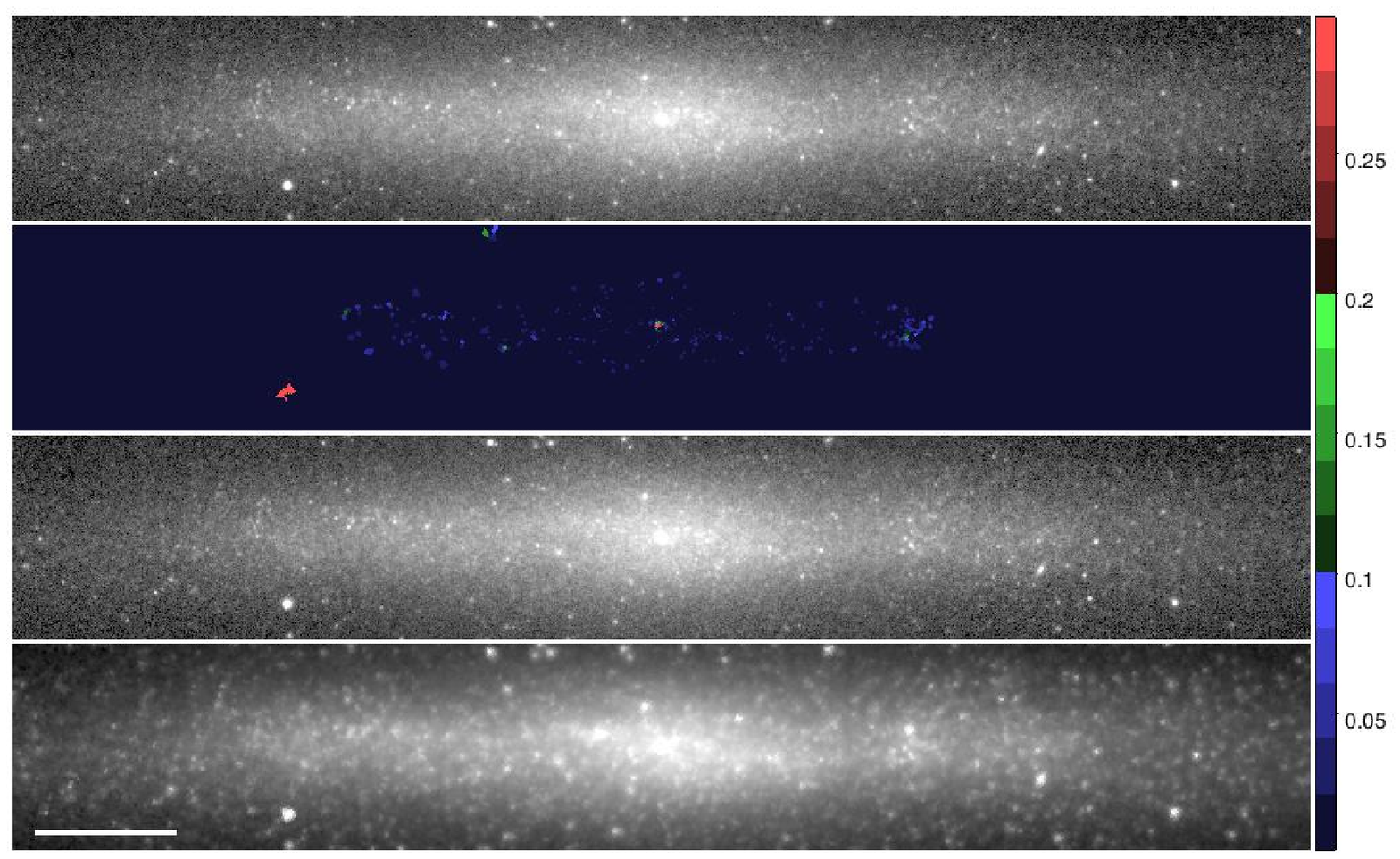}}
\\
\includegraphics[scale=0.5]{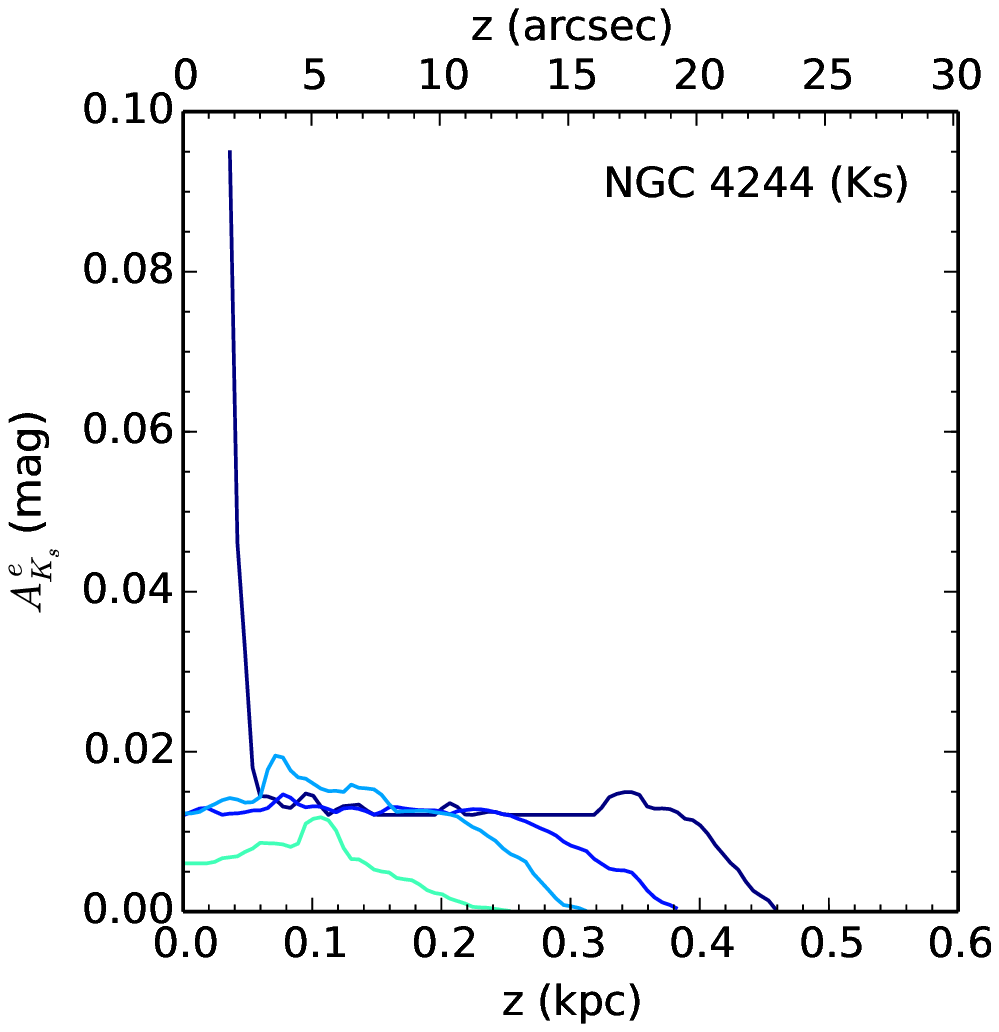}
&
\multicolumn{2}{@{}c@{}}{\includegraphics[scale=0.57]{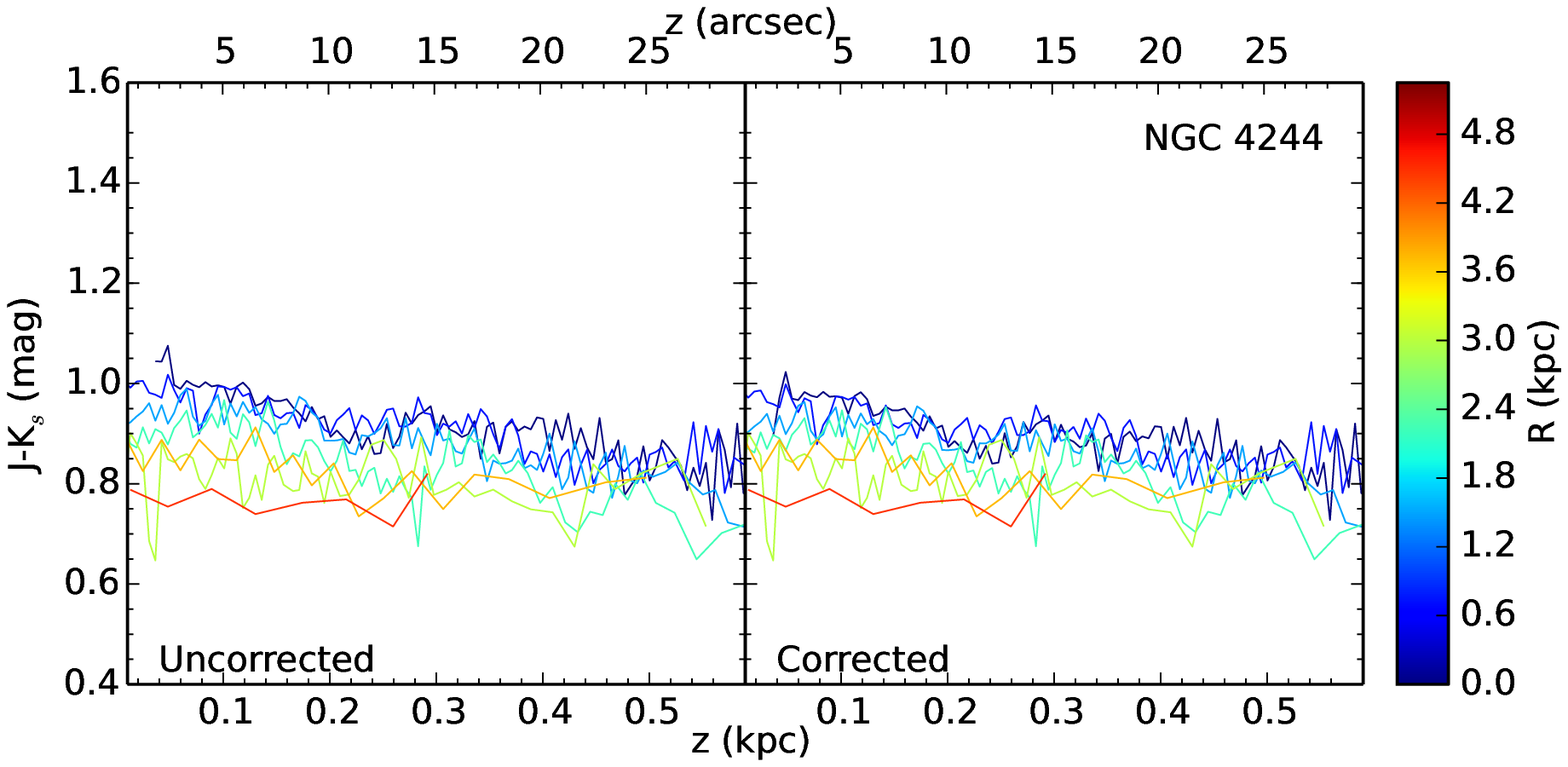}}
\end{tabular}
\caption{Same as Figure \ref{fig:ngc891attencorr}, but for NGC~4244
  (the $\ks$-band images grayscale range is from 17.0-20.7 mag
    arcsec$^{-2}$).  The $z$ coordinate used here is the {\it
    projected} height; this galaxy is estimated to be inclined at
  $85.8$ deg (see \ref{sec:ngc4244}).}
\label{fig:ngc4244attencorr}
\end{figure*}

\begin{figure*}
\centering
\begin{tabular}{@{}c@{}c@{}c@{}}
\multicolumn{3}{@{}c@{}}{\includegraphics[scale=0.8]{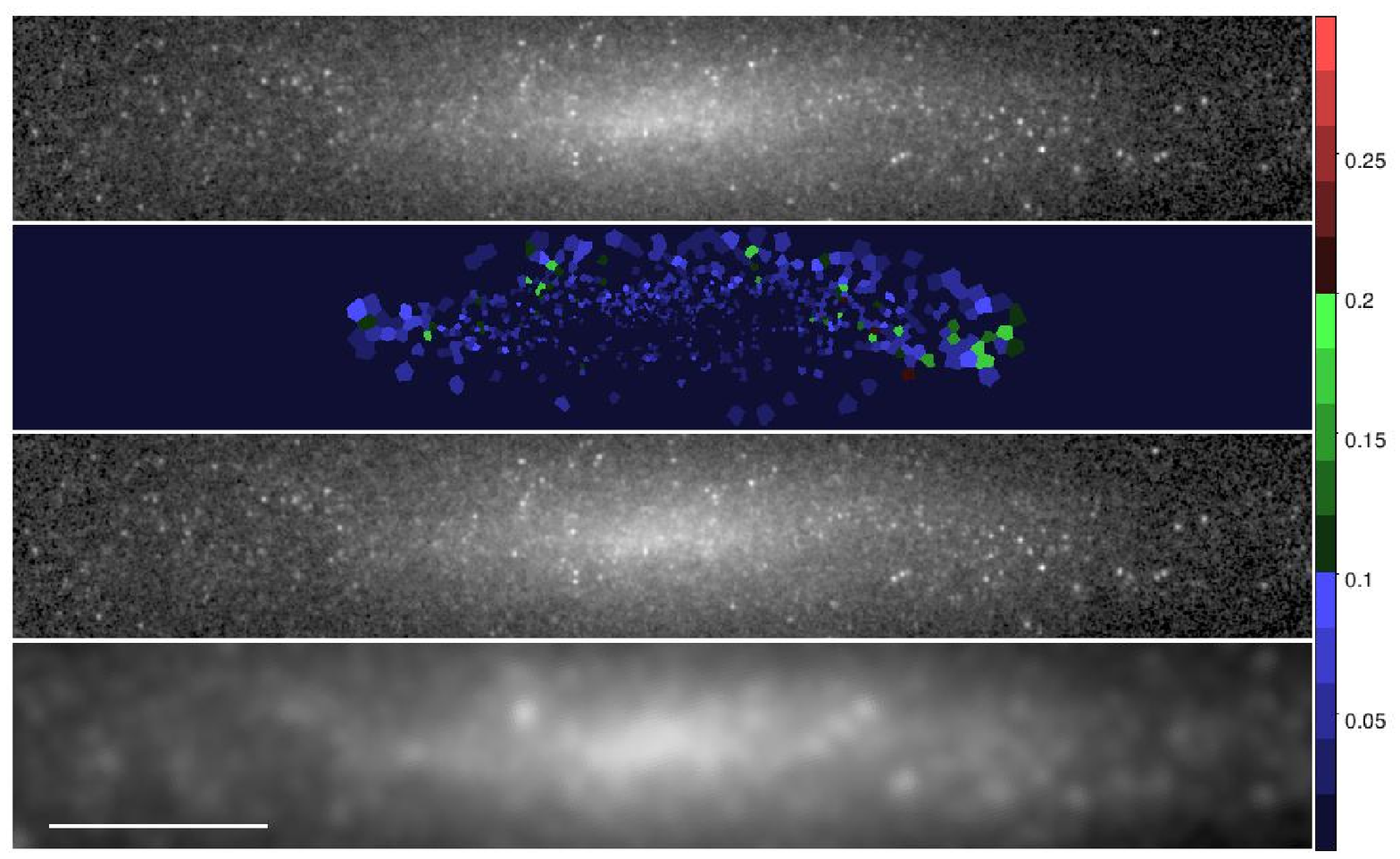}}
\\
\includegraphics[scale=0.5]{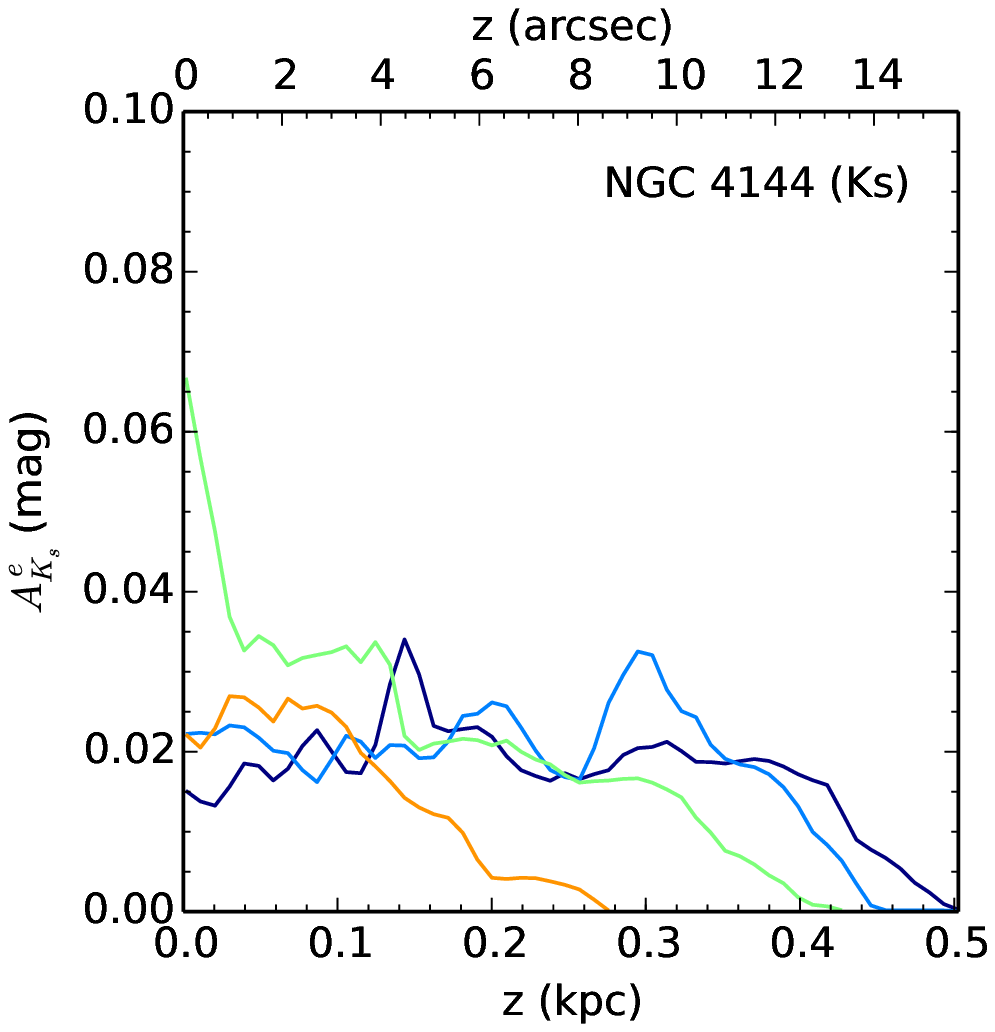}
&
\multicolumn{2}{@{}c@{}}{\includegraphics[scale=0.57]{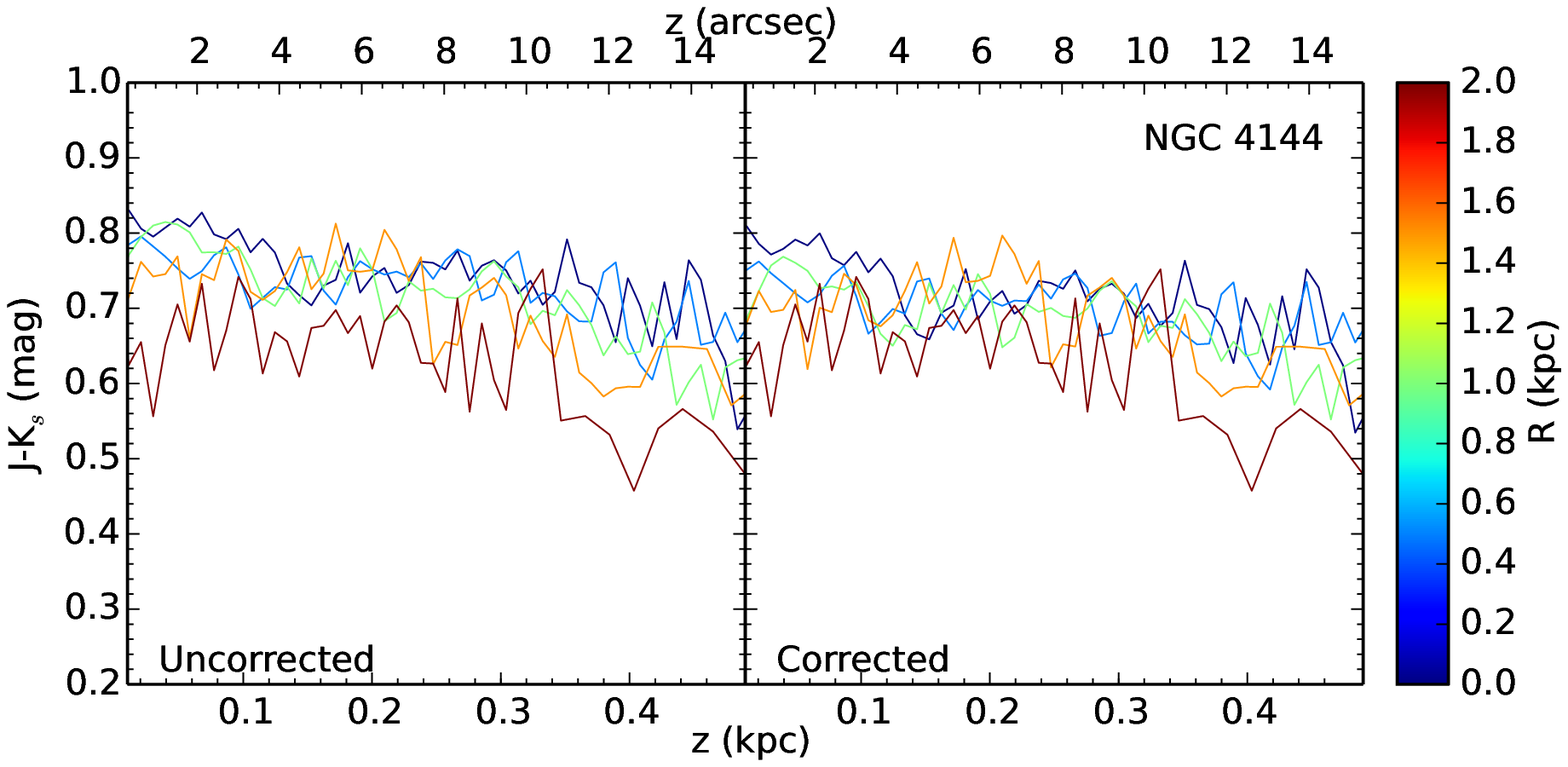}}
\end{tabular}
\caption{Same as Figure \ref{fig:ngc891attencorr}, but for NGC~4144
   (the $\ks$-band images grayscale range is from 16.9-21.1 mag
    arcsec$^{-2}$).  The $z$ coordinate used here is the {\it
    projected} height; this galaxy is estimated to be inclined at
  $85.2\pm0.09$ deg (see \ref{sec:ngc4144}).}
\label{fig:ngc4144attencorr}
\end{figure*}

Although our attenuation correction method was quite effective for
NGC~891, attempting to apply it in this work led to significant
problems due to variations in the unattenuated near- and mid-IR colors
of galaxies' stellar populations.  The attenuation correction from
\citetalias{Schechtman-Rook13} assumed an underlying color of
$\ks$-4.5$\mu{\rm m}\sim0$, which is plausible for a spectral energy
distribution dominated by the Rayleigh-Jeans tail from stellar
photospheric emission. As shown in Figure \ref{fig:binnedcolor},
however, there is a variation of $\sim$0.3 mag in mean color between
and {\it within} galaxies in our sample. This variation is
significantly larger than the photometric errors for our
Voronoi-binned data; with a bin S/N$\approx$20, the color errors are
only $\sim$0.07 mag, a factor of over four times smaller than the
observed range of $\ks$-4.5$\mu{\rm m}$. There is no correlation
between $\ks$-4.5$\mu$m color and rotation speed in this sample. The
apparent bimodality in the intrinsic color distribution between
galaxies see in Figure \ref{fig:binnedcolor} may simply be an artifact
of the small sample size, but this feature merits further exploration
with larger samples.

Even though NGC~891 does have a mean $\ks$-4.5$\mu{\rm m}\approx0$ 
in unattenuated regions, there appears to be some internal spread in
this color.  In general for this and the other galaxies in our sample,
the assumption of zero color in our previous attenuation correction
scheme leads to over or underpredictions of the actual attenuation
when applied to galaxies with different colors. However, due to our RT
modeling method, the {\it slope} of our derived attenuation correction
is independent of the intrinsic (unattenuated) colors of the stellar
population.  That is, the inferred effective attenuation can still be
determined by a {\it change} in, e.g., $\ks$-4.5$\mu$m color relative
to the unattenuated color.  Henceforth we formulate our attenuation
formulae in terms of the color excess $E(\ks-4.5\mu{\rm m}) \equiv
(\ks-4.5\mu{\rm m}) - (\ks-4.5\mu{\rm m})_0$, where $(\ks-4.5\mu{\rm
  m})_0$ is the unattenuated color, as we have done in the Appendix
here. The task at hand is to determine better estimates for
$(\ks-4.5\mu{\rm m})_0$.

It is tempting to account for the intrinsic variation of near- to
mid-IR colors by setting $(\ks-4.5\mu{\rm m})_0$ to the colors of the
galaxy at a point well above the midplane that is presumed to be 
dust-free. However, given that we know the $\ks$-4.5$\mu$m color
varies between galaxies, the assumption that it should be constant
{\it within} a single galaxy is also significantly weakened,
\textit{especially given the fact that we are specifically seeking
  evidence for different stellar disk components}.

An inspection of model simple stellar populations (SSPs) from
\citep{Bruzual03} indicates that there is a significant range of
allowed IR colors (a spread of $\sim$0.2 mag in the $\ks$-IRAC
colors), that depends both on the age of a population as well as its
metallicity (Figure \ref{fig:ssps}). This is simply a reflection of
departures from pure black-body emission due to absorption features
(mostly molecular) that arise in cool stars; the impact of these
features on the integrated light at these wavelengths depends on both
age and metallicity of the stellar population. However, the reddening
vectors in Figure \ref{fig:ssps} indicate that we may be able to
disentangle some of these effects from the attenuation by using our
full suite of WHIRC+IRAC colors.

We undertook a brute-force modeling approach, fitting the five
observed broad-band magnitudes ($J, H, \ks, 3.6\mu{\rm m}$, and $
4.5\mu{\rm m}$) for each Voronoi-binned pixel in each galaxy
independently to SSPs over a wide range of ages and
metallicities.\footnote{Each SSP was convolved with the WHIRC and IRAC
  filter curves and magnitudes were generated taking into account the
  IRAC aperture corrections necessary to perform extended-source
  photometry, which the IRAF Handbook
  (http://irsa.ipac.caltech.edu/data/SPITZER/docs/irac/\allowbreak
  iracinstrumenthandbook/) gives as 0.10 and 0.07 mag for 3.6 and
  4.5$\mu$m, respectively.} Our decision to use only NIR and MIR but
not optical bands was based on the desire to balance leverage on
reddening with sensitivity to changes in the underlying population
(e.g., age), while keeping the optical depth in different bands
comparable and modest. For every model, the attenuation was tuned to
optimize the fit (in a $\chi^{2}$ sense) for a particular binned
pixel:

\begin{equation}
\chi^{2} = \sum_{i=J,H,\ks,3.6,4.5}\frac{(f_{d,i}-\xi f_{m,i})^{2}}{\delta f_{d,i}^{2}},
\end{equation}
where $f_{d}$ and $f_{m}$ are the fluxes from the binned data pixel and the model,
respectively, $\xi$ is a constant offset set to minimize $\chi^{2}$, and
$\delta f_{d}$ is the error on the binned data pixel.
 Attenuated magnitudes were computed using the same
differential attenuation-color relationship found in
\citetalias{Schechtman-Rook13} or Appendix \ref{sec:ngc4244attencorr},
using the $E(\ks-4.5\mu{\rm m})$ formulation.  The distribution of
attenuation and $\chi^{2}$ from all of the models was saved for every
binned pixel.

As is well known, there is a significant degree of degeneracy in the
range of models with different age or metallicity that yield
comparably good fits to the data. However, there are always {\it some}
single SSPs that are able to fit every binned pixel well. As shown in
Figure \ref{fig:sspjust}, the standard deviation between best-fitting
model and data, averaged over many binned pixels, is comparable to the
measured errors, indicating that a more complex model would not result
in an improvement in the overall goodness-of-fit. It was for these
reasons that we did not consider a broader suite of models including
varying star-formation histories. No doubt it is possible to reduce
the number of models to a smaller number of principal components or
eigenspectra, e.g., \citet{Bershady95} or \citet{Connolly95}, but
this is a further refinement which essentially only improves
computational performance.

This model degeneracy is not a concern for our analysis here so long
as there is little covariance between different combinations of
intrinsic model colors and inferred attenuation for the set of models
that fit the data well. To account for this covariance, we compute the
uncertainty in the attenuation correction to be the full range of
attenuations found for all models with $\chi^{2}$ values within a 67\%
confidence limit relative to the best-fitting model. We find that
these uncertainties are generally small ($\lesssim\pm$10\% of
$\aeks$), and we treat this as a source of random error. We also
consider the impact of distance-dependent resolution effects in the
following section (see Figure \ref{fig:distattencorr}), but we do not
account for these effects in our calculation. Example fits from
  low and high attenuation regions of NGC~891 are shown in Figure
  \ref{fig:attencorrdemo} to illustrate the fidelity of our
  methodology. We compare the results of the new attenuation
  correction against the one used in \citetalias{Schechtman-Rook13} in
  Figure \ref{fig:NGC891attencorrcompare}.  The most dramatic changes
  are near the mid-plane in the sense that our attenuation corrections
  here are {\it smaller} than in our previous work. The impact of this
  change on our model fitting is discussed in \S 5.1.1.

Images showing the resulting attenuation maps, corrected data, and
vertical attenuation profiles for our sample are shown in in Figures
\ref{fig:ngc891attencorr} to \ref{fig:ngc4144attencorr}, in order of
the discussion of the surface brightness analysis in
Section~\ref{sec:model}.  This ordering roughly corresponds to the
complexity of the modeling analyisis (high to low), which
interestingly corresponds closely to decreasing
V$_{\mathrm{rot}}$. Given the significant changes made to our
attenuation correction technique, we re-analyzed our NGC~891 data from
\citetalias{Schechtman-Rook13}; the NGC~891 attenuation corrections in
Figure \ref{fig:ngc891attencorr} (and the other figures in this work)
show results using these new corrections.

\subsubsection{NGC~522: Distance-Dependence of our Attenuation Correction}
\label{sec:ngc522corr}

Despite the evidence for dust attenuation in NGC~522 (visible in the
both the optical and NIR images in Figure \ref{fig:gri_jhk_sample})
our correction method (Figure \ref{fig:ngc522attencorr}) finds very
little attenuation in any part of the galaxy. This model-based
conclusion is surprising, given the fact that this galaxy is a
fast-rotator and should therefore have a dense dust lane. However,
NGC~522 is $\sim$2 times more distant than any other galaxy in our
sample, so we must investigate the possibility that this galaxy is
beyond the range where our attenuation correction procedure is able to
resolve dust structure in super-thin layers with the data we have in
hand. If the spatial resolution of the IRAC is too low to resolve the
dust features in this galaxy, regions of high attenuation are
suppressed by mixing them with neighboring regions at lower
attenuation. The poor resolution at 4.5$\mu$m is readily apparent in
Figure \ref{fig:ngc522attencorr}.

In general, we do not expect to be able to resolve dust structure on
all physical scales, which makes investigating this potential issue
difficult. Creating a galaxy model with high enough spatial resolution
to perform a fully controlled test of this possibility is currently
computationally infeasible \citep[for an example of the
  state-of-the-art in this sort of modeling
  see][]{Schechtman-Rook12}. What is relevant for this study, however,
is that we are able to resolve the physical scales necessary for the
bulk characterization of disk structure. Based on what we know about
the Milky Way and NGC 891, the smallest scale of interest in the
edge-on perspective is the super-thin disk component with an expected
scale-height of 50-100 pc. Therefore a suitable approximation for
probing distance systematics is to rescale our data for NGC 891 and
4565, our two most nearby fast-rotators (both with resolution
comparable to or better than 50 pc at their assumed distances), to
simulate their appearance at larger distances. In this way we can test
whether we can resolve comparable attenuation distributions over the
full distance range of galaxies in our sample.

To simulate the appearance of NGC 891 and 4565 at larger distances, we
convolve them with a suitable gaussian kernel for each distance,
assuming the same seeing conditions.  We then rebin the data to
preserve the proper pixel scale.  We compute the attenuation in each
smoothed and rescaled pixel, and compare it to the attenuation
estimated for the matching pixels in the original galaxy images, as
shown in Figure \ref{fig:distattencorr}.  Since this Figure shows
averages and standard deviations binned by values of the original
attenuation value, each distance-shifted pixel is counted multiple
times for the corresponding set of pixels matched to the original
image.

As we expect, we find that increasing distance acts to flatten the
attenuation distribution: Pixels with larger attenuations are
suppressed while pixels with small attenuations are enhanced.  This is
simply a smoothing effect, the details of which depend on the
intrinsic distribution of regions with different levels of
attenuation. In edge-on disks (e.g., NGC~891), the regions of highest
attenuation are very thin, linear structures, and are spatially
promimate to more extended regions of lower attenuation (at larger
scale height). As disk projection moves away from purely edge on
(e.g., NGC~4565), the attenuation peak diminishes and the overall
distribution of attenuation is spatially broader; these attributes
tend to diminish the impact of spatial resolution on distance. This
qualitative description matches the observed behavior in Figure
\ref{fig:distattencorr}.  Overall, the result of moving either NGC~891
or NGC~4565 out to a distance comparable to NGC~522 (roughly four
times as distant, corresponding to the smallest points in Figure
\ref{fig:distattencorr}) is to bias estimates of the peak attenuation
downward by $\sim$15-60\% (depending on whether one uses the peak
attenuation difference in NGC~4565 or NGC~891, respectively).

Even after taking this correction into account, however, the peak
attenuation in NGC~522 would still be $\lesssim$0.1 mag. There are
some indications that NGC~522 is not exactly edge-on (most notably the
fact that the light from the central region of this galaxy is not
symmetrical above and below the dust lane in Figure
\ref{fig:gri_jhk_sample}), and may be more similar in gross morphology
to NGC~4565. For NGC~4565, this lower inclination acts to suppress the
peak attenuation since the super-thin mid-plane dust is projected over
a substantially thicker line of sight. However, even NGC~4565 has
attenuation $\gtrsim$0.2 mag. Applying a distance correction based on
the NGC~4565 simulation in Figure \ref{fig:distattencorr} to NGC~522
does not bring its attenuation level on par.

The lower dust attenuation for NGC~522 relative to NGC~4565 may be due
to the nearly 100 km sec$^{-1}$ difference in V$_{\mathrm{rot}}$
between these two galaxies, perhaps indicating that while NGC~522 has
sufficient mass to produce a dust lane \citep{Dalcanton04} it is not
capable of compressing it to the level needed to produce significant
$\ks$-band attenuation at its inclination. This explanation is
conjecture, and requires higher resolution imaging data to confirm.
We opt not to make any distance correction to NGC~522 in this work,
due to the large uncertainty on such a correction (a result of both the 
intrinsic scatter and the large difference in the magnitude of the
correction between NGCs 891 and 4565). Therefore we caution the reader
to be aware of these issues when examining our results for this
system.

Finally, while we were primarily interested in interpreting the
unusual attenuation correction results for NGC~522, understanding how
our attenuation estimates depend on distance is important for the
other fast-rotating galaxies in our sample. NGCs 1055 and 4013 are
$\sim$twice as distant as NGCs 891 and 4565, and Figure
\ref{fig:distattencorr} shows that we are likely underpredicting the
actual peak attenuation by $\sim$10-20\%. This will primarily affect
our measurements of very thin components, as the largest attenuations
are almost always found near the midplanes of spiral galaxies, and
would result in an {\it underestimate} of the central surface
brightness and an {\it overestimate} of the scale-heights of such
components, points which we return to below. The total light, however,
will be properly estimated. In future analysis, we suggest a forward
modeling approach might better account for the impact of spatial
resolution on the inferred attenuation and unattenuated light profiles
in a self-consistent manner; such an investigation is beyond the scope
of this work.

\section{Galaxy Colors}
\label{sec:galaxycolors}

We show the vertical $J-\ks$ color profiles for the galaxies in our
sample in the bottom middle and right panels of Figures
\ref{fig:ngc891attencorr}-\ref{fig:ngc4144attencorr}. These figures
include both the apparent and attenuation-corrected profiles. Several
interesting features relevant to an assessment of our attenuation
corrections, the distribution of dust, and gradients in stellar
populations in these galaxies are apparent in these color profiles.
The dust distribution and population gradients are interesting in
their own right as well as pertinent to the following analysis of the
photometric decomposition of these edge-on spiral galaxies into
multiple components. 

As a starting point for this discussion, we show the color excess
$E(J-\ks)$ as a function of attenuation in Figure~\ref{fig:modelejks}
based on our RT modeling in \citetalias{Schechtman-Rook13} for fast
rotators and in the Appendix here for slow rotators. For $\aeks < 0.5$
mag the relation between color excess and attenuation is nearly linear
with $E(J-\ks) / \aeks \sim 1.2$.  We also plot the $J-\ks$ colors
versus age and metallicity in Figure~\ref{fig:jkzage} for the same
models used in Figure \ref{fig:ssps} as well as models using a
different prescription for the late phases of intermediate-age stars
from \citet{Maraston05}.  From this it can be seen that simple stellar
populations predicted from stellar population synthesis models are no
redder than $J-\ks \sim 1.2$, corresponding to the stellar colors of
M4-5 III giants\footnote{We refer to \cite{Bessell88},
  \cite{Tokunaga00}, and \citet{Rayner09} for our estimates of the NIR
  colors of stars. We also note that we loosely refer to $J-\ks$
  colors on the 2MASS photometric system \citep{Carpenter01} for all
  studies discussed here, even though many of these studies predated
  2MASS and therefore adopted slightly different systems.  In the
  color range of our observations \citet{Frogel78}, \citet{Persson83},
  \citet{Frogel87}, and \citet{Terndrup94} calibrated to the
  CIT. \citet{Maraston05} adopted the Cousins-SAAO system. In the
  relevant color range their $J-K$ colors are, respectively, $\sim
  0.04$ redder and $\sim0.07$ bluer than $J-\ks$. The \cite{Bruzual03}
  colors were calculated by us directly from the spectra based on the
  correct filter transmission curves to place them in the 2MASS
  system.}.

Based on dynamical argument, at large distances from the disk
mid-plane we expect the stars to be primarily from old populations.
For stellar populations older than $\sim 1$ Gyr, different models
yield comparable predictions for $0.6 < J-\ks < 1.0$, varying weakly
with age and primarily with metallicity from about 1/20th to $\sim 3$
times solar. This variation reflects the position of the red giant
branch which varies (in a luminosity-weighted sense) between effective
spectral type K0 to M0. The colors of massive galaxies with little
dust and old stellar populations, such as E/S0s (e.g.,
\citealt{Frogel78}), are in good agreement with these models, having
$0.8 < J-\ks < 1.0$. This, of course, is for integrated light.  In the
bulge of the Milky Way, however, \citet{Frogel87} found a distribution
of variable and non-variable giants up to $J-\ks \lesssim 1.4$ in
Baade's window, with the integrated M-giant light dominated by
spectral type M6, or $J-\ks \sim 1.2$. In the context of recent
analysis of VVV and 2MASS data for the bulge by \citet{Gonzalez13}, who
also find such red giant colors, this is an extension of the
metallicity effect on the giant branch. We might expect, therefore,
that in specific locations of galaxies that are super metal rich that
the $1 < J-\ks < 1.2$ would occur even for older stellar populations.

The situation closer to the disk mid-plane is potentially more
complicated in the sense that a wider range of stellar population ages
likely co-exist, depending sensitively on the recent star-formation
history and the dynamical heating of the disk. In the age range from
0.2 to 2 Gyr the models in Figure~\ref{fig:jkzage} are discrepant
due primarily to the treatment of AGB stars. The seminal study by
\citet{Persson83} of the near-infrared properties of LMC and SMC star
clusters show some have very red $J-\ks$ colors, even after correcting
for reddening.  At intermediate ages there is a continuous
distribution of colors extending up to $J-\ks = 1.4$, with an outlier
(NGC 2209) at $J-\ks = 1.7$.  Given the strength of the $2.3\mu$m CO
feature in these red clusters, they conclude the colors are due to the
presence of carbon AGBs. Both the integrated colors and the
presence of carbon stars have been confirmed by more modern
measurements \citep{Mucciarelli06}. While using these results to
constrain the plausible range of near-infrared colors for more
massive, star-forming galaxies with possibly a wider range of
metallicities and ages is somewhat uncertain, it does inform us as to
what colors are plausible.

From these considerations, and accounting for an uncertainty of
$\pm0.1$ mag in our attenuation correction, we consider any $J-\ks >
1.3$ is likely to arise from reddening or unusual stellar populations.

\subsection{Apparent Colors}
\label{sec:galaxyapcolors}

\begin{figure}
\epsscale{1.3}
\plotone{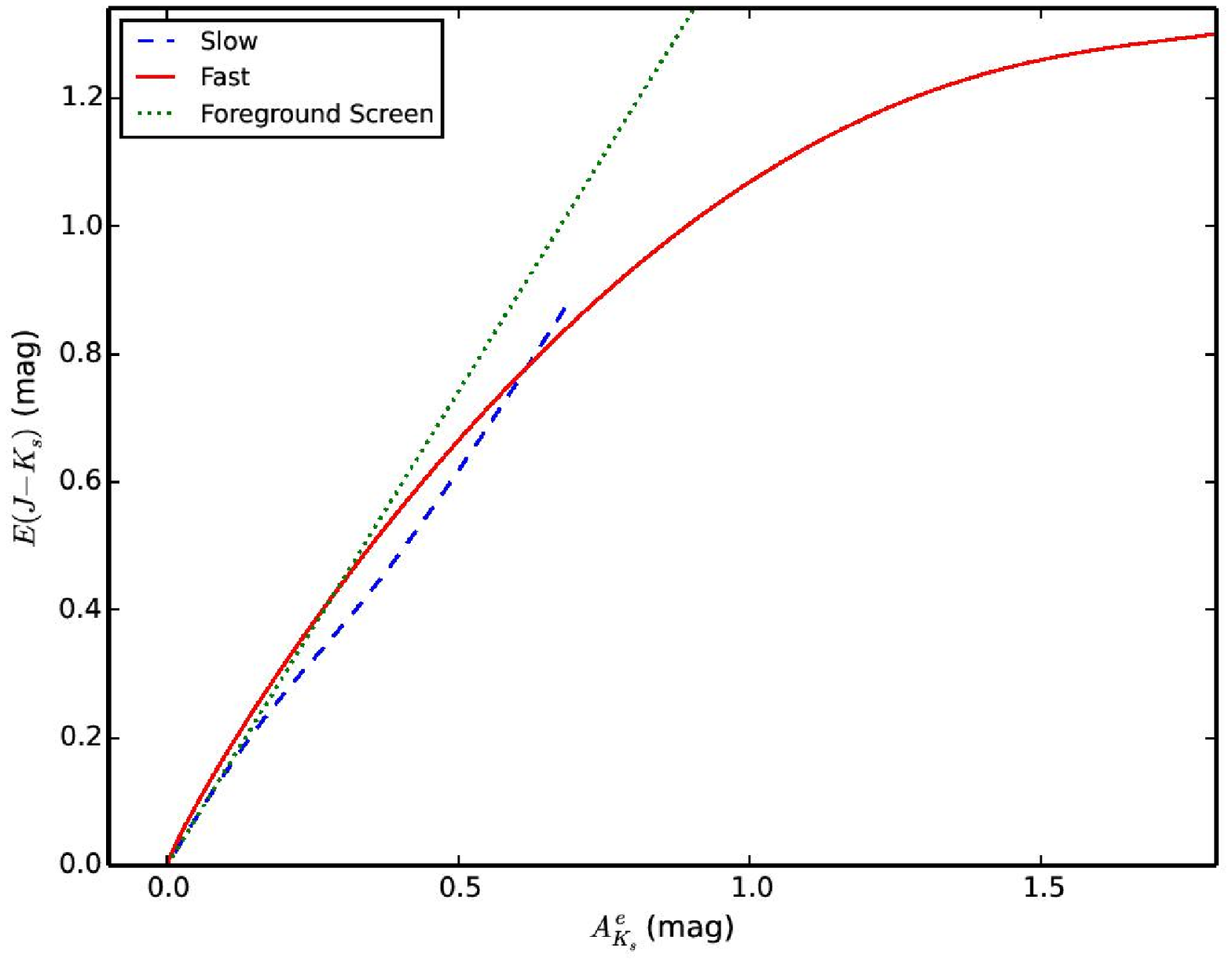}
\caption{Color excess, $E(J-\ks)$ as a function of $\aeks$
  predicted by our RT modeling for fast- and slow-rotators (red solid
  and blue dashed lines, respectively). For comparison we show this
  function for the foreground screen model of \citet{Cardelli89} as a
  green dotted line.}
\label{fig:modelejks}
\end{figure}

\begin{figure}
\epsscale{1.1}
\plotone{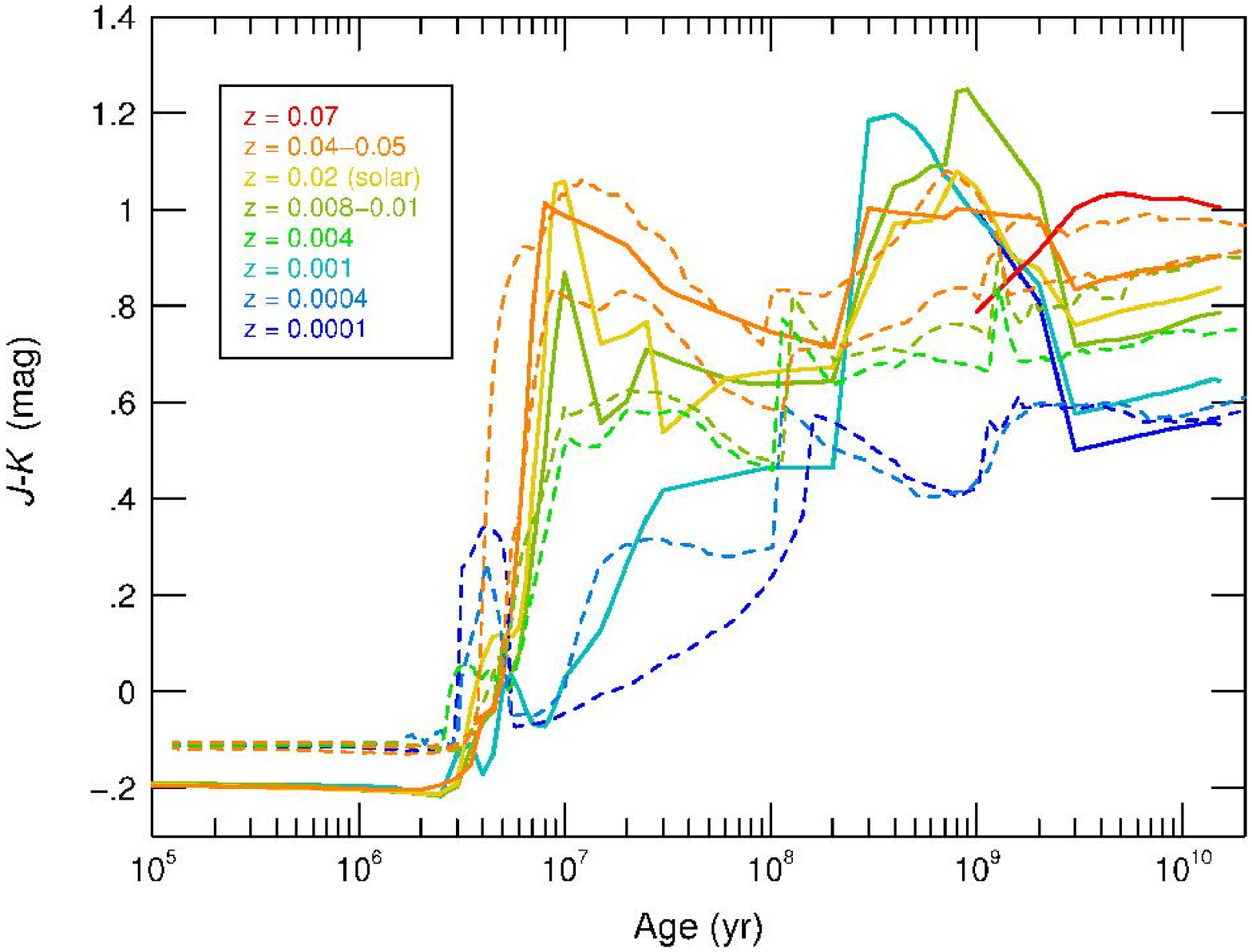}
\caption{Near-infrared colors for SSPs plotted as a function of age
  and metallicity for the same models used in Figure~\ref{fig:ssps}
  ($J-\ks$; dashed lines), as well as for the models of
\citet{Maraston05} ($J-K$, solid lines).}
\label{fig:jkzage}
\end{figure}

All of the galaxies in our sample become redder in {\it apparent}
$J-\ks$ color near the mid-plane and at smaller radius, with NGC~4565
being one exception. These reddening trends are likely a confluence of
dust attenuation {\it and} population gradients in radius and height
above the disk.

For NGC~4565, while generally following the trend in apparent $J-\ks$
color, there is significant drop in $J-\ks$ color near the mid-plane
at small radii.  This feature corresponds to the regions of
the galaxy inside the evacuated inner parts of this galaxy's disk
(readily visible in Figure \ref{fig:gri_jhk_sample}), and is a strong
indicator that this region has minimal attenuation. If NGC~4565 is not
aberrant in this regard and dust-free inner regions are common in
massive spiral galaxies, it would imply that much of the obscuring
dust in these galaxies is not smoothly distributed over all radii, but
instead preferentially at larger radii. This would explain, for
example, the relative prominence of dust lanes in many highly-inclined
spiral galaxies.

Only three galaxies have central apparent $J-\ks$ colors significantly
above 1.2 mag, namely NGCs 891, 4013, and 1055. All three are fast
rotators. The apparent red colors of NGC 891 and 4013 are easily
explained by their almost perfectly edge-on orientation.  Given the
amount of attenuation recovered by our correction (Figure
\ref{fig:ngc891attencorr}), it is unsurprising to find that NGC~891
has significantly redder apparent J-$\ks$ colors near the mid-plane
than any of the other galaxies in our sample.  Both NGCs 891 and 4013
have attenuation and $J-\ks$ color profiles which become
broader and shallower toward larger radii, possibly hinting at a
thickening of the dust layer and a commensurate reduction in the
mid-plane optical depth at larger radii. NGC~1055, however, is not
exactly edge-on; its red colors likely are the result of the
interaction that produced the other striking features of this galaxy
(discussed in more detail in Section \ref{sec:ngc1055}).

For the other fast-rotators, NGC~4565 has a peak color of $J-\ks \sim
1.4$ mag, because even though it probably contains a similar amount of
dust as NGC~4013 it has $i\sim87^{\circ}$ \citep{deLooze12}. NGC~522
also appears to have modest reddening near the mid-plane, consistent
with the little attenuation found for this system. Both of these
galaxies, and all of the fast rotators are distinguished from the two
slow-rotating disk systems that have apparent $J-\ks$ colors less than
1 mag. These blue $J-\ks$ colors are consistent with the small amount
of dust found in these systems, but also likely reflect differences in
their stellar populations.

Most of the galaxies converge to a $J-\ks$ color of $\sim$0.9-1 mag at
large heights where the dust attenuation is minimal.  This is
comparable to the color of a cool giant star in the spectral range of
late K to early M--not unreasonable for what should be an old,
metal-rich stellar population. The one exception is NGC~4144, which
asymptotes to J-$\ks \sim 0.65 \pm 0.1$, similar to that of an early K
giant. This is indicative of either an old, metal poor stellar
population or a much younger system, the latter possibly the result of
a recent global star-formation episode. NGC~4244, our other slow
rotator, has somewhat redder J-$\ks$ colors ($0.85\pm0.1$), but is
still slightly bluer than the fast rotators in our sample.

\subsection{Attenuation-Corrected Colors}
\label{sec:galaxycorcolors}

Once the color profiles have been corrected for attenuation, the range
of vertical and radial gradients in $J-\ks$ diminish considerably, as
expected. The two slow rotators NGC 4144 and 4244 exhibit shallow,
nearly linear gradients in height of -0.2 mag per kpc, independent of
radius, i.e., the disks appear to get bluer with height above the
mid-plane. These color profiles are largely unchanged by the
attenuation corrections because the level of attenuation is small.
With the exception of NGC~1055, all of the fast rotators show
pronounced vertical gradients in $J-\ks$. NGC~1055's lack of
discernable color gradients in radius or height is likely 
associated with its morphological peculiarities, discussed in Section
\ref{sec:ngc1055}. Given the large attenuation and attenuation
gradient in this galaxy, this indicates that the color gradients found
in the corrected profiles of the other galaxies in the sample (most
notably NGCs 891, 4013, and 4565) are not associated with systematics
in the attenuation corrections.

Above 300 pc, the remainder of the fast-rotating disks also become
bluer with height, like the slow-rotators, while closer to the
mid-plane there is more complex behavior. The color gradients for NGC
522, 891 and 4013 flatten at larger heights, but in all cases the
amplitude of the gradient is comparable to what is found for the
slow-rotators. The vertical gradients at large heights are plausibly
interpreted as decreasing metallicity in thicker (presumably older)
disk populations. If this is the case, as can be seen in
Figure~\ref{fig:jkzage} the change in metallicity would easily be
factors of 5-10 over the vertical range probed in these
galaxies. Regardless, the presence of color gradients indicates these
disks are not monolithic.

Near the disk mid-plane several features in the $J-\ks$ colors and
gradients for the fast rotators stand out.  First, the gradients tend
to flatten and in two cases (NGC 891 and 4565) reverse, leading to
bluer mid-plane colors.  The clear decrease in corrected $J-\ks$ color
near the midplane of NGC~891, especially for profiles at large
distances from NGC~891's center, is likely the product of young,
super-thin disk stars. Not only have we found evidence for such a disk
in \citetalias{Schechtman-Rook13} but NGC~891 is also known to have
young stars visible over the dust lane from optical imaging
\citep[likely due to the presence of a grand design spiral
  pattern;][]{Kamphuis07}. In NGC~4565, a small dip is still observed
even in the attenuation-corrected profiles, most noticably at small
radii; it may be the result of a relatively young nuclear star-forming
population.  The lack of such a large downturn in our other fast
rotators indicates they do not have as substantial a young stellar
population near their mid-plane compared to NGC 891 and 4565 {\it
  unless} their super-thin disk populations are preferentially
dominated by red super giants or intermediate-age asymptotic giant
branch stars.  On the other hand, the colors of all of the fast
rotators (besides NGC~1055) have $J-\ks>1$, indicating either a very
metal rich giant branch or intermediate age (AGB) population. Testing
this conjecture requires additional (spectral) diagnostics sensitive
to surface-gravity in cool stars, which we will explore in future
work.

For three galaxies in our sample (NGC 522, 891, and 4013), all fast
rotators, there are regions in projected radius and height which are
redder than $J-\ks \sim 1.3$.  For NGC 891 and 522 this only occurs in
the very central region of the galaxy. In the context of the
discussion in \ref{sec:galaxycolors}, we found it instructive to
compare these colors with nearby spiral galaxies, preferably at low
inclination, in three other samples. In this projection the effects of
dust attenuation in the near-infrared ought to be minimal. To
facilitate comparison, we integrated our attenuation-corrected colors
over height above the mid-plane, and plot these integrated colors as a
function of projected radius in Figure~\ref{fig:fon}. The range of $J-\ks$
colors decrease significantly in this projection, with only
the very central points for NGC 891 and 4013 having $J-\ks > 1.3$.

\subsubsection{Comparison to other samples}
\label{sec:colorcompare}

\begin{figure}
\epsscale{1.15}
\plotone{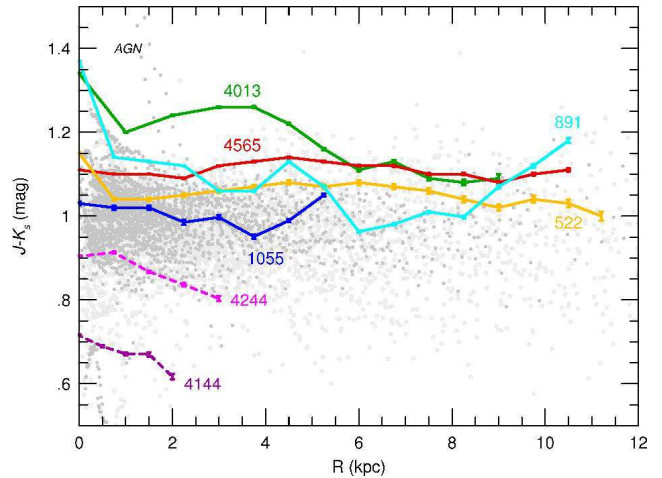}
\caption{Attenuation-corrected $J-\ks$ colors as a function of
  projected radius, integrated over $\pm 2$kpc in height for fast
  rotators and $\pm 1$kpc for slow rotators. Galaxies in our
  edge-on sample are represented by colored lines and are labeled by NGC.
  Points represent photometry of 231, low-inclination,
  intermediate-type spirals from the parent sample of the DiskMass
  Survey (see text). Filled (dark grey) points have color errors of
  $<0.1$ mag; open (light grey) points have color errors between 0.1
  and 0.2 mag. The inner color profile of two of these galaxies with
  known AGN are marked.}
\label{fig:fon}
\end{figure}

Two of the nearby spiral galaxy samples (\citealt{Terndrup94} and
\citealt{Jarrett03}) contain galaxies over a wide range of type and
inclination.  In \citet{Terndrup94}, 7 of the 43 galaxies are
convincingly below 45 deg, two of which are early-type spirals
(S0/Sa). None of the other 5 have $J-\ks>1$, but of the two early-type
galaxies, NGC~2681 (type Sa) has colors are between $1 < J-\ks <
1.35$, with a mean disk color of $J-\ks\sim 1.2$, while the other
(NGC~474) has a comparable nuclear color. \citet{Jarrett03} find that
that barred spirals tend to have redder nuclear colors, peaking at
$J-\ks>1.1$. Many of these galaxies, however, are considerably
inclined (e.g., NGC 253, which has a strong bar and nuclear starburst,
with colors of $J-\ks\sim2$ in this region).

To focus our comparison on purely low inclination systems, we
inspected surface-photometry performed on foreground-corrected 2MASS
images from a third study of 231 galaxies with nearly face-on
orientation. The sample preferentially selects intermediate type
galaxies similar to those we intended to select for the fast-rotators
in the current study of edge-on systems. These face-on systems formed
the Phase-A parent sample of the DiskMass Survey (DMS) described in
\citet{Bershady10a}; the photometry and foreground correction are
identical to that described in \citet{Martinsson13}. An important
point to note is that the DMS sample was intentionally biased {\it
  against} selecting galaxies with strong or large bars.

Figure~\ref{fig:fon} shows the DiskMass Survey photometry superimposed
on the vertically integrated colors of the galaxies in the current
study. In general, the 2MASS images have insufficient depth to provide
high-precision surface-photometry of spiral disks.  We have chosen to
limit the face-on sample data to those apertures with color errors
$<0.1$ mag. This clearly biases the comparison to high
surface-brightness disks. To show that this does not significantly
affect the resulting color distribution we have also plotted the
colors for apertures with errors between 0.1 and 0.2 mag. While there
are subtle effects that we will explore elsewhere, the primary change
is simply to broaden the color distribution by 0.1 mag due to larger
errors. Consequently, we focus our comparison here to the
highest-quality surface-photometry of this face-on sample.

In terms of the vertically integrated colors, all but NGC 4013 lie
within the observed color distribution of the DMS face-on sample.  In
the DMS, we find two cases (UGC 1727 and 9149) with $J-\ks >> 1.2$ at
small radius. Both galaxies have Seyfert nuclei. It is conceivable
that the very red $J-\ks \geq 1.5$ nuclear colors in NGCs 522, 891,
and 4013, when seen edge on, are due to small levels of AGN activity.
An additional 5 galaxies in the face-on sample have colors reaching
$1.3 < J-\ks < 1.4$ in their central 1-2 kpc (UGC 3701, 4542, 6157,
12270, 12893). This is a small fraction of the entire Phase A sample,
and the $J-\ks$ colors even in the face-on sample could still suffer
from some attenuation. However, we do note that NGC~4013 does have the
most pronounced and extended x-shaped morphology (see
Figure~\ref{fig:gri_jhk_sample}) of our sample, which is likely
indicitative of an extended, and possibly strong bar.  Such systems
are absent in the DMS sample, and as noted by \citet{Jarrett03} tend
to have redder $J-\ks$ colors.

We conclude that the majority of the attenuation estimates and
corrected colors are accurate, but that NGC~4013 may be too red by
0.15 mag in $J-\ks $ at projected radii below 4-5 kpc.  If this is
indeed a color excess, it implies a systematic error in $\aeks$ of at
most 0.1 mag in this one case.

\section{Multi-component Model Fitting}
\label{sec:model}

\begin{figure}
\epsscale{1.1}
\plotone{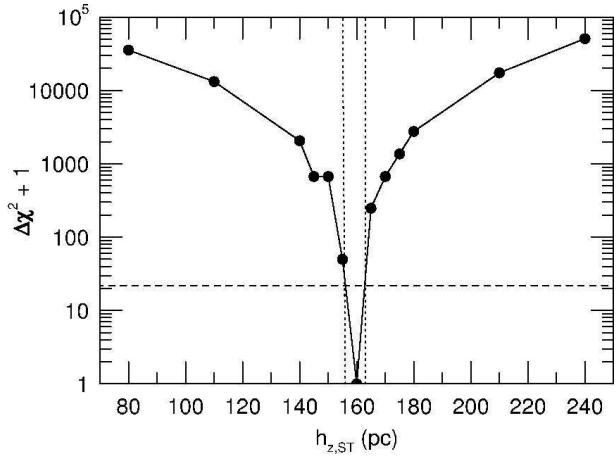}
\caption{ Difference between global minimum $\chi^2$ value for
  3-component disk model with 8 degress of freedom for NGC 891
  ($R>3$kpc), and minimum $\chi^2$ values for other values of the
  vertical scale-height of the super-thin disk component, $h_{z,ST}$.
  The 3$\sigma$ (99.73\%) confidence limit on $h_{z,ST}$ is estimated
  from the intersection with the dashed horizontal dashed line, given
  by the dotted vertical lines.}
\label{fig:goodness_of_fit}
\end{figure}

\begin{figure*}
\epsscale{0.85}
\plotone{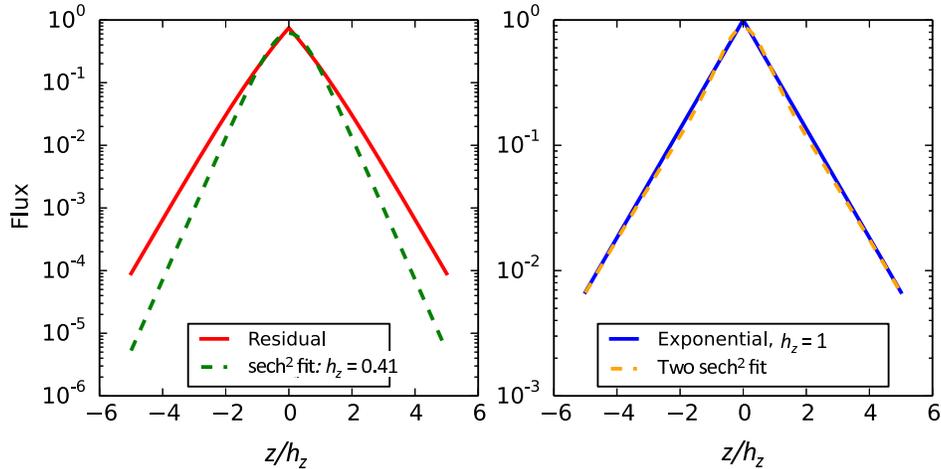}
\caption{Decomposition of exponential vertical fitting function into
  two sech$^2$ functions, as described in the text. The left-hand
  panel shows the fit of the second sech$^2$ function to the residual
  from the difference between the exponential and first sech$^2$
  functions, normalized at large heights, z. The right hand panel
  shows the resulting composite profile compared to the original
  exponential profile.}
\label{fig:exp_sechsq_fit}
\end{figure*}

We utilize the same basic fitting scheme as in
\citetalias{Schechtman-Rook13}, featuring a 2D Levenberg-Marquardt
non-linear least-squares algorithm run multiple times with randomized
initial parameter guesses to ensure a global minimum is
reached. However, unlike in \citetalias{Schechtman-Rook13}, the
assumption that our galaxies are viewed perfectly edge-on is incorrect
(especially obvious for NGC~4565). Therefore we add the inclination of
the galaxy as another free parameter, which means we can no longer
separate the $R$ and $z$ components of a galaxy. Following
\citet{Aoki91} the integrated luminosity of an exponential disk is
\begin{align}
I(R,z)
=\rho_{0}\int_{-\infty}^{\infty}\mathrm{exp}\bigg(&-\frac{\sqrt{R^{2}+t^{2}\mathrm{sin}^{2}\,i}}{h_{R}}
-\nonumber
\\ &\left.\frac{|z\mathrm{sin}\,i-t\mathrm{cos}\,i|}{h_{z}}\right)dt,
\label{eq:diskflux}
\end{align}
where $i$ is the inclination, $h_{R}$ and $h_{z}$ are the scale-length
and scale-height, and $\rho_{0}$ is a normalization factor.

Equation \ref{eq:diskflux} must be numerically integrated, which is
time-consuming. We replaced the core of our fitting algorithm with a
C++ function to do the actual integration, using a highly efficient
double-exponential integration
library\footnote{http://www.codeproject.com/Articles/31550/Fast-Numerical-Integration}
to ensure fast performance. Computation efficiency is important
because each model must be produced along millions of sight-lines in
our high resolution images. The load is exacerbated by the need to
oversample the data in order to perform an accurate seeing convolution
(which we now do in $z$ {\it and} $R$, as opposed to just in $z$, as
we had in \citetalias{Schechtman-Rook13}). As it will become clear
later, another extension of the present work is to include more
complicated structure than in our previous analysis. These additions
include multiple disk truncations, rings, and S\'{e}rsic profiles of
varying index, n.\footnote{Because even a S\'{e}rsic component with a
  large ellipticity will be much less flat than a disk and all of our
  galaxies are very close to edge-on, we treat all bulges as though
  they are viewed edge-on to save on computational time.} Based on our
inability to discern between disks with exponential, sech, or
sech$^{2}$ vertical light distributions in
\citetalias{Schechtman-Rook13} we only fit exponential models to
reduce the total compute time in our analysis.

Our fitting machinery is further augmented by a distributed network of
computers operated by the Center for High Throughput Computing at the
University of Wisconsin-Madison. This network uses the publicly
available HTCondor software \citep{Thain05} to manage job submission
and execution, and enables us to massively parallelize our
Levenberg-Marquardt fitting procedure\footnote{ Due to its
      complexity and the degree of customization required to make our
      code compatible with our distributed computing environment, this
      software is not as readily amenable to general use by other
      researchers. However, we have made it available at
      https://github.com/AndrewRook/astro/tree/master/galaxyfit, to
      serve at least as a resource for others seeking to add
      distributed computing capabilities to their software.}. As a
consequence we were able to run 1000 iterations per model--a factor of
10 more than in \citetalias{Schechtman-Rook13}--and to explore more
free parameters in our models on relatively short timescales. This
improvement led to more robust estimates of fitted parameters for
increasingly complex galaxy models, without which we would not have
been able to characterize the wide variety of morphologies in our
sample in a reasonable amount of time.  The model fits presented in
the following sections used over 25 {\it years} of computation time on
a single CPU.

\subsection{Robustness and Uniqueness}
\label{sec:results:robonique}

Multi-component models of one-dimensional vertical light profiles are
prone to degeneracies, manifest in covariance between component
vertical scales and normalizations. These degeneracies are ameliorated
(and the covariance reduced) by fitting the two-dimensional light
distribution instead of one-dimensional profiles (e.g., see discussion
in \citealt{Morrison97}). However, even with the two-dimensional
profile fitting we undertake here, it is difficult to anticipate the
degree of model parameter degeneracy without directly understanding
the uncertainties in their estimation. Given the complexity of our
models, we have made an assessment of the degree to which this is an
issue in our analysis.

In the subsequent analysis, we use the boot-strap method to estimate
uncertainties in our parameter estimates. Standard error estimates
from the nonlinear fitting are often too small (\citealt{Morrison94}),
while mapping out the full N-dimensional $\chi^2$ space of our models
is computationally prohibitive, even with the distributed processing
described in the previous section. The boot-strap errors are small
(often under 2\%), which provides an initial indication that parameter
covariance is not a significant concern.

To verify that boot-strap error estimates are reasonable, we explored
the model for NGC 891 that we will come to in \S \ref{sec:ngc891}, given in the first column of Table
\ref{tab:ngc891bestfits}. This model has three exponential disks fit
to the two-dimensional light distribution outside of a 3 kpc radius
from the center of the galaxy.  We chose this model because it is
representative of our model complexity for the one galaxy where we
have reanalyzed the attenuation correction; analysis of parameter
robustness is particularly relevant to later discussion. The model has
8 free parameters since the thick disk scale length and scale height
is fixed to literature values, and the inclination and center radial
position are fixed (the center vertical position is a free parameter,
but it is not tabulated.) We focus here on the super-thin
scale-height, $h_{z,ST}$, as this is a parameter of central interest
in this work, and because it might reasonably be considered to be
covariant with the scale-height of the thin-disk component $h_{z,T}$.

Our numerical experiment considered a series of fixed values of 
$h_{z,ST}$ between 80 and 240 pc, bracketing the best-fitting value of
160 pc.  By allowing the model to optimize the remaining 7 parameters
for each fixed value of $h_{z,ST}$, we find the minimum $\chi^2$ value
rapidly increases as $h_{z,ST}$ departs from the best fit value of
0.16 kpc. Based on a $\Delta\chi^2$ confidence interval, we find
$h_{z,ST}$ is constrainted to $\pm$ 3 pc at the 99.73\% (3$\sigma$),
consistent with our boot-strap estimates. Within these constraints the
variations in the remaining six parameters are equally small. In short,
the parameter estimates on scale-height, scale-length and
surface-brightness from our model fitting are remarkably well
constrained, with little covariance given the adopted model fitting
functions.

However, there is a second flavor of degeneracy which arises from our
choice of fitting functions. This was explored in
\citetalias{Schechtman-Rook13} for NGC 891 in the context of varying
the vertical luminosity profile from exponential to a sech$^2$
distribution. It is straightforward to show that it is possible to
decompose a single exponential function into two sech$^2$ with a high
degree of accuracy. One example is shown in Figure
\ref{fig:exp_sechsq_fit}. In this case, one of the sech$^2$ components
is required to have the same asymptotic scale-height and normalization
as the exponential.  The result is that the second sech$^2$ has half
the luminosity and $\sim$40\% of the scale-height of the original
exponential. In this one-dimensional case, it would imply that every exponential super-thin disk component we fit in our subsequent
analyisis could be reinterpreted as having a much thinner
sech$^{2}$ sub-component.

This interpretation is not entirely fair because it is based on a
one-dimensional analysis and it makes some a priori assumption of what
is the correct distribution function. 
In fitting two-dimensional light distribution with either exponential
or sech$^2$ vertical distribtion functions, we find we need the {\it
  same} number of components (\citealt{Schechtman-Rook13}). In other
words, there is no mathematical motivation for invoking an additional
component. For NGC 891, instead we found in
\citetalias{Schechtman-Rook13} that the best-fitting multi-component
disk models with sech$^2$ vertical profiles tend to have shorter
scale-heights and longer (exponential) scale-lengths than those models
with exponential vertical profiles. Both of these relative changes are
at the $\sim$25\% level such that the effect on disk oblateness is
roughly a factor of 1.5. The change in the scale-height is
qualitatively consistent with the illustration in Figure
\ref{fig:exp_sechsq_fit}, but the effect is smaller, and again there
is no need for extra components.

Concerning the point about what distribution function is correct,
\cite{Comeron11}, for example, take a physical approach of deriving
dynamically consistent fitting functions. While in some sense this is
desirable, it is a one-dimensional dynamical model and hence limited
to fitting one-dimensional light profiles.  Since we wish to mitigate
parameter covariance in multi-component models by fitting the
two-dimensional light distribution, and because the corresonding
dynamical model becomes considerably more complicated, in our analysis
we provide a phenomenological description of the light profile,
parameterized with a simple, analytic function. In any event, the
fitting function derived by \citealt{Comeron11} is intermediate
between an exponential and sech$^2$ function, and so it is bracketed
by the limiting phenomenological models considered in this discussion.

To summarize: We adopt an exponential vertical fitting function. Our resulting model parameters are well
constrained, with little parameter covariance. However, our resulting
model parameters deviate systematically from an identical analysis
adopting a different vertical fitting function, but in a well-defined
way. In the case of a sech$^2$ function, the conversion between
parameterizations can
be approximated given the above information based on the NGC 891 modelling
of \citetalias{Schechtman-Rook13}.

\begin{deluxetable*}{cccc}
\tablewidth{0pt}
\tablecaption{NGC~891 Best-fitting Models}
\tablehead{ & \multicolumn{2}{c}{Value} &  \\ \cline{2-3} 
& & \colhead{Three Disks+} & \\
  \colhead{Parameter$^{a}$} & \colhead{Three Disks$^{b}$} &
  \colhead{Bar+Nuclear Disk$^{c}$} &\colhead{Units}}
\startdata
$\mu_{0,ST}$ & 14.35$\pm$0.01& 14.41$\pm$0.03& mag arcsec$^{-2}$\\
$h_{R,ST}$ & 3.29$\pm$0.03& 3.36$\pm$0.05& kpc\\
$h_{z,ST}$ & 0.16$\pm$0.00& 0.16$\pm$0.00& kpc\\
$L_{ST}$ & 4.20 $\times$ 10$^{10}$& 2.99 $\times$ 10$^{10}$& $L_{\odot,K}$\\
$\mu_{0,T}$ & 15.93$\pm$0.02& 15.93$^{g}$& mag arcsec$^{-2}$\\
$h_{R,T}$ & 5.70$\pm$0.06& 5.70$^{g}$& kpc\\
$h_{z,T}$ & 0.47$\pm$0.00& 0.47$^{g}$& kpc\\
$L_{T}$ & 6.21 $\times$ 10$^{10}$& 5.67 $\times$ 10$^{10}$&$L_{\odot,K}$\\
$\mu_{0,Th}$ & 18.46$\pm$0.02 & $18.46^{g}$& mag arcsec$^{-2}$\\
$h_{R,Th}$ & 4.80$^{d}$ & 4.80$^{d}$& kpc\\
$h_{z,Th}$ & 1.44$^{d}$& 1.44$^{d}$& kpc\\
$L_{Th}$ & 1.56 $\times$ 10$^{10}$& 1.38 $\times$ 10$^{10}$& $L_{\odot,K}$\\
$\mu_{0,nuc}$ & \nodata & 12.89$\pm$0.02& mag arcsec$^{-2}$\\
$h_{R,nuc}$ & \nodata & 0.20$\pm$0.00& kpc\\
$h_{z,nuc}$ & \nodata & 0.10$\pm$0.00& kpc\\
$L_{nuc}$ & \nodata & 7.62 $\times$ 10$^{9}$& $L_{\odot,K}$\\
$\mu_{0,bar}$ & \nodata & 15.02$\pm$0.01& mag arcsec$^{-2}$\\
$h_{R,bar}$ & \nodata & 1.33$\pm$0.02& kpc\\
$h_{z,bar}$ & \nodata & 0.57$\pm$0.00& kpc\\
$L_{bar}$ & \nodata  & 2.52 $\times$ 10$^{10}$& $L_{\odot,K}$\\
$R_{trunc,1}^{e}$ & \nodata & 2.79$\pm$0.08& kpc\\
$R_{trunc,2}^{f}$ & 9.92$\pm$0.03 & 9.92$^{g}$ & kpc\\
$L_{tot}$ & 1.20 $\times$ 10$^{11}$ & 1.33 $\times$ 10$^{11}$ & $L_{\odot,K}$\\
$\chi_{\nu,inner}^{2}\,^{h}$& 18.62& 4.96&\nodata\\
$\chi_{\nu,outer}^{2}\,^{h}$& 3.47*& 3.47&\nodata\\
$\chi_{\nu,total}^{2}\,^{h}$& 7.39& 3.86*&\nodata
\enddata
\tablenotetext{a}{$ST$, $T$, $Th$, $nuc$, and $bar$ denote parameters that we associate with
  the super-thin disk, thin disk, thick disk, nuclear disk, and bar (respectively).}
\tablenotetext{b}{Only fit to regions of the galaxy with $R>3$ kpc.}
\tablenotetext{c}{Fit performed over entire radial range of data.}
\tablenotetext{d}{Fixed to literature value (see \citetalias{Schechtman-Rook13} for details).}
\tablenotetext{e}{Inner truncation for the super-thin, thin, and thick disks,
  outer truncation for the bar and nuclear disks.}
\tablenotetext{f}{Outer truncation for the super-thin disk only.}
\tablenotetext{g}{Fixed to the value found in the Three Disks model.}
\tablenotetext{h}{Reduced $\chi^{2}$. {\it inner} and {\it outer} correspond
  to the regions inside and outside of the 3 kpc fitting cutoff, while {\it
    total} is for both regions together. (*) denotes the $\chi_{\nu}^{2}$ that was
  minimized by each model.}
\label{tab:ngc891bestfits}
\end{deluxetable*}

\begin{figure*}
\centering
\includegraphics[scale=0.7]{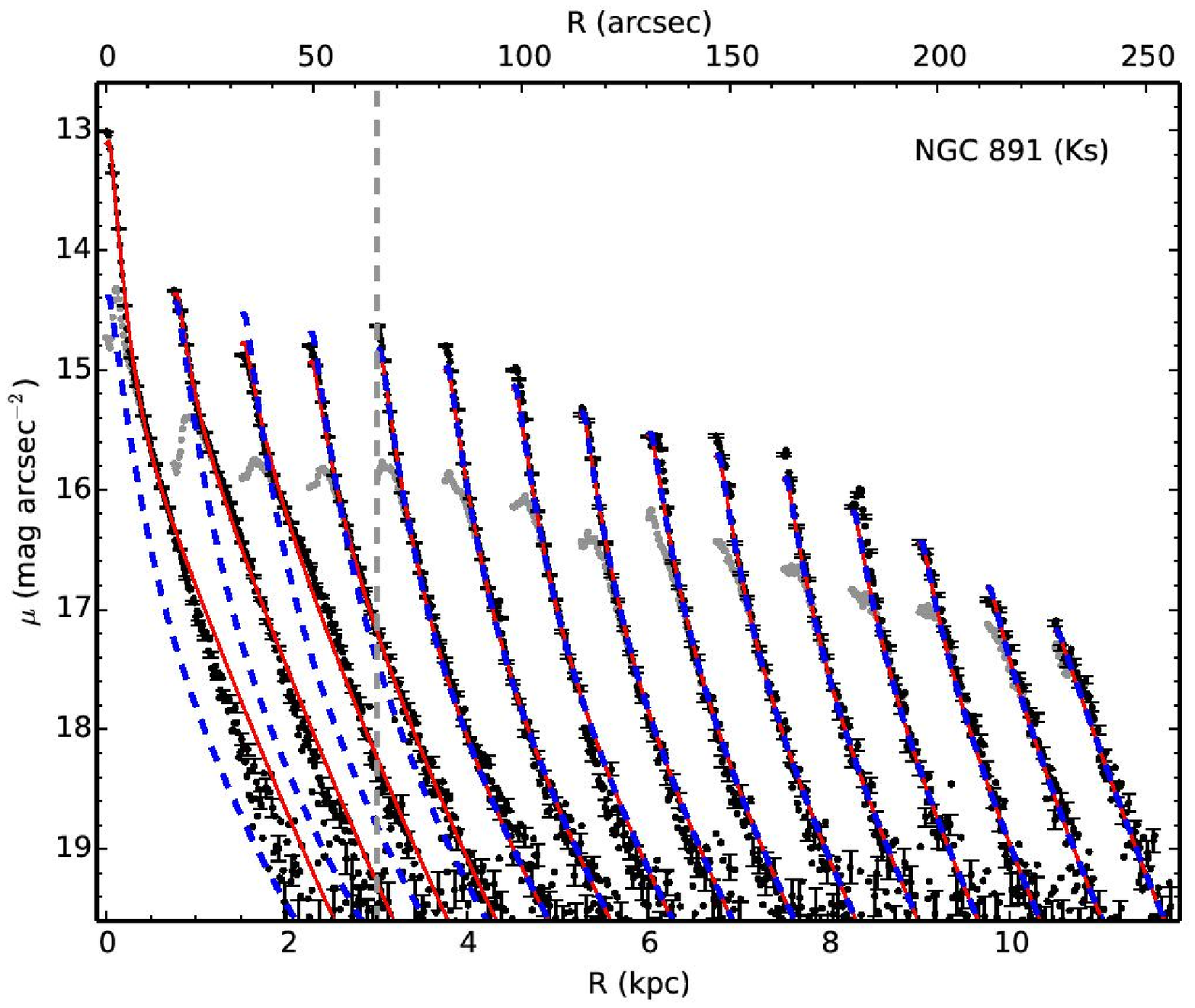}\\
\includegraphics[scale=0.7]{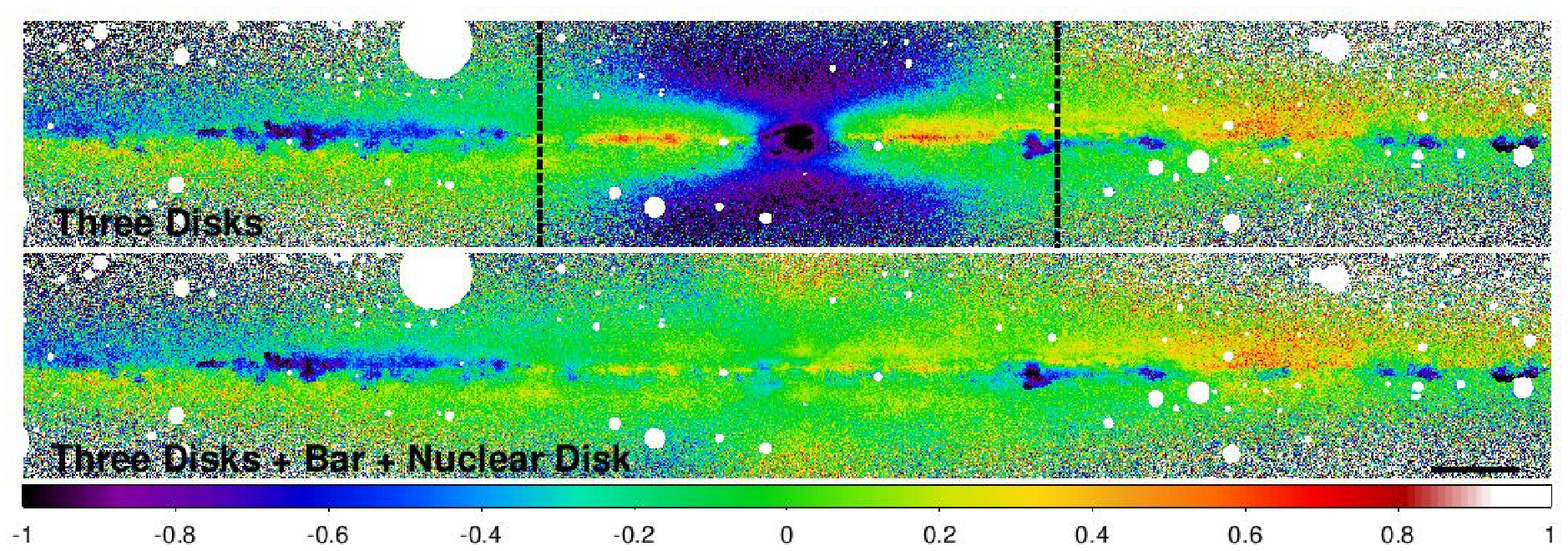}
\caption{Top: best-fitting models for NGC~891. Gray and black points
  represent un-corrected and attenuation corrected data
  (respectively), where each set of points shows a vertical profile
  starting at the radius of the innermost point.The blue dashed lines
  show the best fitting three disk model (with only an outer
  truncation for the super-thin disk), fit to all regions outside of 3
  kpc in radius (denoted by the vertical gray dashed line). The solid
  red lines show the best fitting three disk + exponential bar +
  nuclear disk model, fit to the entire profile. This latter model
  also has an inner truncations for all three disks and and an outer
  truncation for the thinnest disk, and were fit at all radii. To
  improve S/N the data in the 1D profiles have all been binned to 5''
  x 0\farcs3 (R x z). Bottom: magnitude residual (data - model)
    images. Dashed black lines show the radial limits of our fits,
    where appropriate, and the solid black line indicates 1 kpc at our
    assumed distance. Details of the model parameters are in Table
  \ref{tab:ngc891bestfits}}.
\label{fig:ngc891fitplot}
\end{figure*}

\subsection{Individual Galaxy Fits}
\label{sec:results:indivgals}

While we use the same general framework to find our best fits, each
galaxy is unique; there is no `one-size-fits-all' fitting
procedure. Therefore, while some elements may be similar, we discuss
our fits to each galaxy individually. We order the galaxies in this
section not by NGC number but rather by the complexity of the system,
starting with our four fast-rotators and then moving to our two
slow-rotating galaxies. Additionally, for NGCs 891, 4244 ,and 4565,
the three most spatially extended galaxies in our sample, we
boxcar-smoothed the images with a kernel 0\farcs3 on a side. By
performing this smoothing (roughly 3x3 pixels on WHIRC) we were able
to dramatically decrease our fitting times without significantly
impacting our ability to recover very thin galaxy structure, as the
resolution of these images was more than twice the size of the
smoothing kernel.

It is also worth discussing the nomenclature we will use, which has
not always been consistent in the literature.  This is especially
relevant because the light distribution of the galaxies in our sample
does not always follow the nested super-thin + thin + thick disk
paradigm of the MW and NGC~891. (We prefer to use the term `super-thin'
instead of `young' or 'star-forming' because with our data alone the
true age of these disks is unknown, although in some cases, as
discussed in the previous section, the $J-\ks$ indicates the mid-plane
light is indeed dominated by young stars.) Generally, when confronted
with a galaxy which requires three disks with scale-heights all
significantly different from each other, we use super-thin, thin, and
thick to describe the three disks as a function of increasing
scale-height. For galaxies with two disks we use `thin' and `thick',
and only `thin' for apparently single-disk systems; generally we find
that these disks have similar values of scale-height as the thin and
thick disks in full three disk systems like the MW. In one case
(NGC~1055) we find that the two most luminous disks have comparable
scale-heights and total luminosities, but very different
scale-lengths, and so we refer to them simply as disks 1 and 2.

It is important to note that our naming convention does not, nor is it
intended to, convey any information about the astrophysics of these
galaxies. It is possible, for instance, that two disks which we give
the same designation to are the result of very different formation
processes. Our naming scheme only provides a convenient reference for
the different components of a galaxy. The model $\ks$-band
luminosities given in the following tables are in solar units,
adopting $L_{\odot,K}= 3.909\times10^{18}$ erg s$^{-1}$ Hz$^{-1}$, as
we did in \citetalias{Schechtman-Rook13}.

\subsubsection{NGC~891}
\label{sec:ngc891}

Despite the changes to our attenuation correction methodology, we do
not expect to find significant changes in the intrinsic disk
morphologies. Therefore we begin in much the same way as
\citetalias{Schechtman-Rook13}, although we skip the one and two disk
models and first constrain a model with three disks at radii larger
than 3 kpc. Additionally, after careful inspection of NGC~891's
surface brightness profiles, we added an outer truncation of the
super-thin disk to this model; as we will show in subsequent sections
such a feature is fairly common in the fast rotating galaxies in our
sample.

The results of this model are shown as blue dashed lines in Figure
\ref{fig:ngc891fitplot} and in the left column of Table
\ref{tab:ngc891bestfits}. The new outer disk truncation occurs around
10 kpc, and can clearly be seen on the right side of Figure
\ref{fig:ngc891fitplot}. The new attenuation correction results in two
main changes to our results from \citetalias{Schechtman-Rook13}: the
{\it total} luminosity of the galaxy is $\sim$25\% fainter, and
  the luminosity profile is shallower. The luminosity change is
  entirely due to the change in attenuation correction, and is
  independent of the model fitting.  The profile shape change is
  reflected in larger super-thin and thin disks scale-heights (100\%
  and 62\% increase, respectively) and scale-lengths (68\% and 47\%
  increase, respectively).\footnote{These changes are relative to the
    results using exponential vertical fitting functions in
    \citealt{Schechtman-Rook13}; this work also considers sech$^2$
    vertical fitting functions.}  Given our discussion and analysis in
  \S \ref{sec:results:robonique}, these changes are robust, and
  reflect the change in the profile shape, and are not due to
  parameter covariance.  Despite the decrease in the total luminosity
  as well as the increase in the scale-height of the super-thin disk,
  the luminosity ratios between all of NGC~891's stellar components
  are essentially unchanged between \citetalias{Schechtman-Rook13} and
  this work.

Using this model to fix most of the disk parameters, we then extended
our fits into the central portions of NGC~891 by adding an inner
truncation for all three disks, a bar (mimicked by an exponential
disk), as well as a small nuclear disk--the same scheme as in
\citetalias{Schechtman-Rook13}. The best fitting instance of this
model is shown as the solid red line in Figure \ref{fig:ngc891fitplot}
and in the right column of Table \ref{tab:ngc891bestfits}.  This model
generally does a good job of fitting NGC~891's surface brightness
profile, roughly comparable to the fit quality of the equivalent model
in \citetalias{Schechtman-Rook13} (See Figure 21 of that work for the
most direct comparison).  The change in the $\chi^2$ values
  between the two models in Table \ref{tab:ngc891bestfits} is
  significant, as can be seen in the residual images in Figure
  \ref{fig:ngc891fitplot}.

\begin{deluxetable*}{c c c c }
\tablewidth{0pt}
\tablecaption{NGC~4013 Best-fitting Models}
\tablehead{ & \multicolumn{2}{c}{Value} &  \\ \cline{2-3}
&&Three Disks+&\\
  \colhead{Parameter$^{a}$} & \colhead{Three Disks$^{b}$} &
  \colhead{Bar+Nuclear Disk$^{c}$} &\colhead{Units}}
\startdata
$\mu_{0,ST}$ & 14.27$\pm$0.00& 14.53$\pm$0.01& mag arcsec$^{-2}$\\
$h_{R,ST}$ & 1.94$\pm$0.01& 2.25$\pm$0.01& kpc\\
$h_{z,ST}$ & 0.21$\pm$0.00& 0.21$\pm$0.00& kpc\\
$L_{ST}$ & 4.36 $\times$ 10$^{10}$& 2.25 $\times$ 10$^{10}$& $L_{\odot,K}$\\
$\mu_{0,T}$ & 16.05$\pm$0.02& 16.05$^{e}$& mag arcsec$^{-2}$\\
$h_{R,T}$ & 2.80$\pm$0.01& 2.80$^{e}$& kpc\\
$h_{z,T}$ & 0.60$\pm$0.01& 0.60$^{e}$& kpc\\
$L_{T}$ & 3.49 $\times$ 10$^{10}$& 2.50 $\times$ 10$^{10}$&$L_{\odot,K}$\\
$\mu_{0,Th}$ & 19.04$\pm$0.02& 19.04$^{e}$& mag arcsec$^{-2}$\\
$h_{R,Th}$ & 3.95$\pm$0.01& 3.95$^{e}$& kpc\\
$h_{z,Th}$ & 2.96$^{d}$& 2.96$^{d}$& kpc\\
$L_{Th}$ & 1.55 $\times$ 10$^{10}$& 1.28 $\times$ 10$^{10}$& $L_{\odot,K}$\\
$\mu_{0,nuc}$ & \nodata & 13.81$\pm$0.03& mag arcsec$^{-2}$\\
$h_{R,nuc}$ & \nodata & 0.13$\pm$0.00& kpc\\
$h_{z,nuc}$ & \nodata & 0.11$\pm$0.00& kpc\\
$L_{nuc}$ & \nodata &  2.34 $\times$ 10$^{9}$& $L_{\odot,K}$\\
$\mu_{0,bar}$ & \nodata & 14.91$\pm$0.00& mag arcsec$^{-2}$\\
$h_{R,bar}$ & \nodata & 1.46$\pm$0.01& kpc\\
$h_{z,bar}$ & \nodata & 0.49$\pm$0.00& kpc\\
$L_{bar}$ & \nodata  & 2.54 $\times$ 10$^{10}$& $L_{\odot,K}$\\
$R_{trunc,1}^{f}$ & \nodata & 2.94$\pm$0.01& kpc\\
$R_{trunc,2}^{g}$ & \nodata & 10.29$\pm$0.01& kpc\\
$L_{tot}$ & 9.40 $\times$ 10$^{10}$& 8.80 $\times$ 10$^{10}$& $L_{\odot,K}$\\
$\chi_{\nu,inner}^{2}\,^{h}$& 3.21& 0.97&\nodata\\
$\chi_{\nu,outer}^{2}\,^{h}$& 0.25*& 0.24&\nodata\\
$\chi_{\nu,total}^{2}\,^{h}$& 1.57& 0.57*&\nodata
\enddata
\tablenotetext{a}{$ST$, $T$, $Th$, $nuc$, and $bar$ denote parameters
  that we associate with the super-thin disk, thin disk, thick disk,
  nuclear disk, and bar (respectively).}
\tablenotetext{b}{Only fit to regions of the galaxy with $R>4$ kpc.}
\tablenotetext{c}{Fit performed over entire radial range of data.}
\tablenotetext{d}{Fixed to literature value (see text for details).}
\tablenotetext{e}{Fixed to the value found in the Three Disks model.}
\tablenotetext{f}{Inner truncation for the super-thin, thin, and thick disks,
  outer truncation for the nuclear disk and bar.}
\tablenotetext{g}{Outer truncation for the super-thin disk.}
\tablenotetext{h}{Reduced $\chi^{2}$. {\it inner} and {\it outer} correspond
  to the regions inside and outside of the 4 kpc fitting cutoff, while {\it
    total} is for both regions together. (*) denotes the $\chi_{\nu}^{2}$ that was
  minimized by each model.}
\label{tab:ngc4013bestfits}
\end{deluxetable*}

\subsubsection{NGC~4013}
\label{sec:ngc4013}

\begin{figure*}
\centering
\includegraphics[scale=0.7]{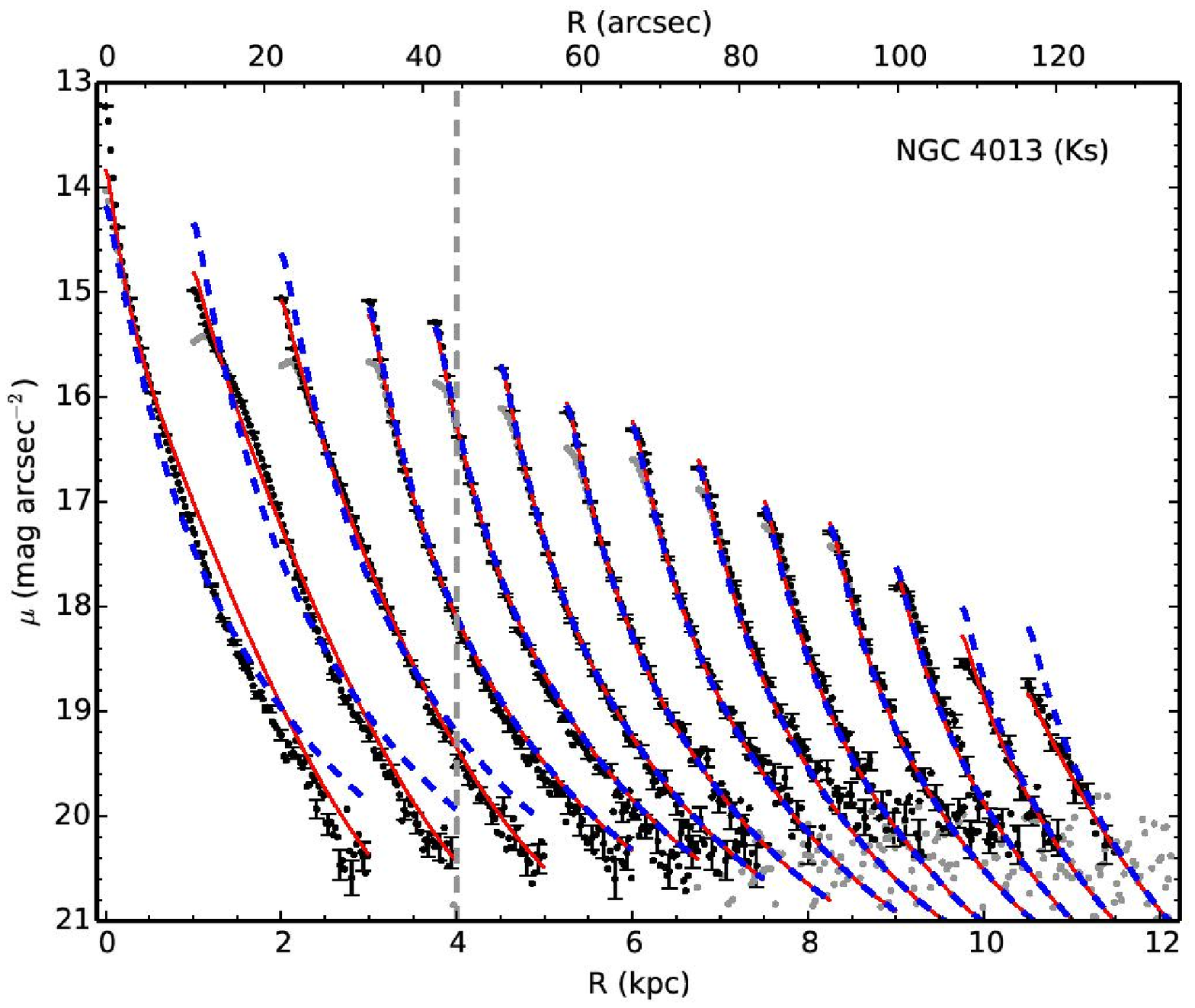}\\
\includegraphics[scale=0.7]{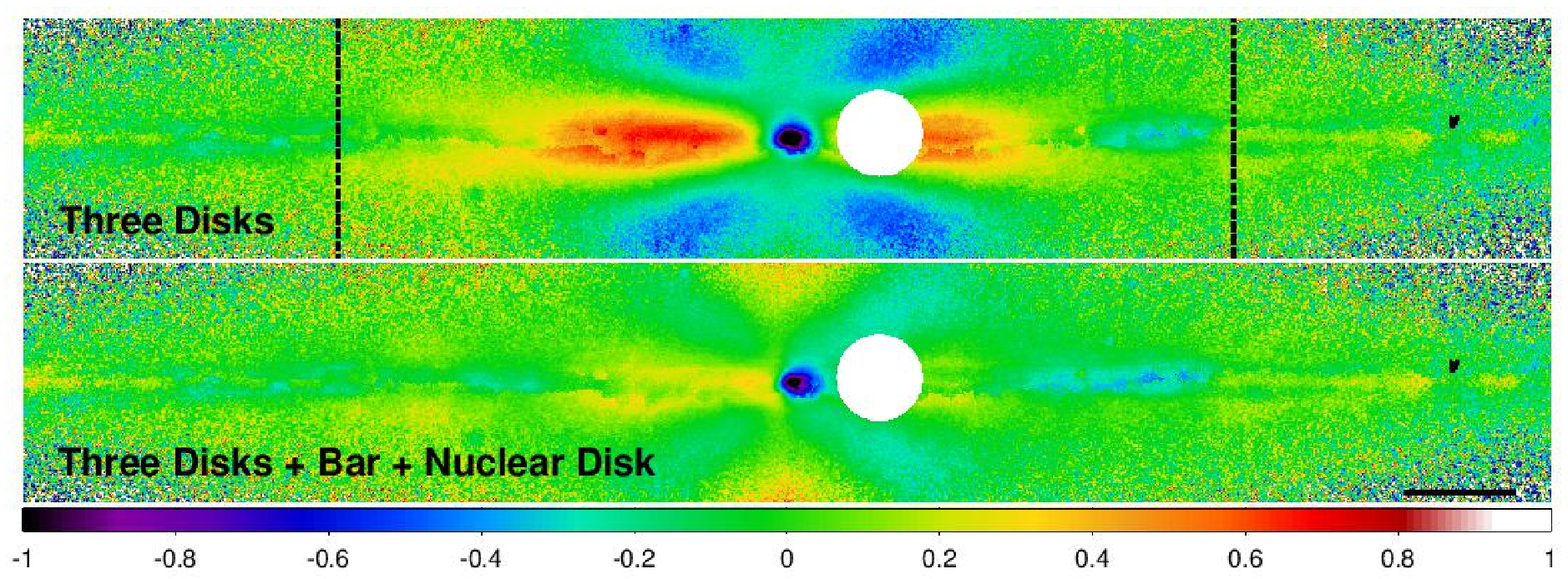}
\caption{Similar to Figure \ref{fig:ngc891fitplot}, but for
  NGC~4013. The blue dashed lines show the best fitting three disk
  model (with no truncations), fit to all regions outside of 4 kpc in
  radius (denoted by the vertical gray dashed line). The solid red
  lines show the best fitting three disk + exponential bar + nuclear
  disk model. This latter model also has an inner truncations for all
  three disks and and an outer truncation for the thinnest disk, and
  were fit at all radii. To improve S/N the data in the 1D profiles have all been binned
  to 5'' x 0\farcs3 ($R \times z$). Details of the model parameters are in
  Table \ref{tab:ngc4013bestfits}.}
\label{fig:ngc4013fitplot}
\end{figure*}

\citet{Comeron11b} performed 1D disk fits to this galaxy and found
that three stellar disks were required to fit the IRAC 3.6 $\mu$m
vertical light profile. They adopt the same distance for NGC~4013 as
we do, and their thinnest disk has $h_{z}\sim$125 pc while their
intermediate-height disk has $h_{z}\sim$565 pc, comparable to the
super-thin and thin disks in the MW and NGC~891. Furthermore, they
found that the thickest of these three disks had a scaleheight of
$\sim$3 kpc, about two times larger than `normal' for a traditional
thick disk.

Similar to \citetalias{Schechtman-Rook13}, we first exclude the
central 4 kpc of the galaxy, chosen by eye to avoid contamination from
NGC~4013's X-shaped orbits, and focus our fit on the disk
structure. Because of NGC~4013's almost exactly edge-on inclination,
we fix the inclination at 90$^{\circ}$. Our data go deep enough to
easily recover the two thinner disks in NGC~4013, but we are not able
to probe large enough heights to easily fit the extended $h_{z}$=3 kpc
disk. However, by fixing the scaleheight of a third disk component to
the scaleheight found for this disk by \citet{Comeron11b}, we find
that our models do benefit from its inclusion. A plot showing our
best-fitting 3-disk model is shown in Figure \ref{fig:ngc4013fitplot},
with the individual parameters detailed in Table
\ref{tab:ngc4013bestfits}.

It is clear from Figure \ref{fig:ngc4013fitplot} that NGC~4013
requires an inner truncation for all three disks as well as an outer
truncation for the thinnest disk only. Therefore, as for NGC~891 we
add an abrupt inner truncation to all three disk components, as well
as an outer truncation for just the super-thin disk. To improve
computation time, we fix the two thicker disks at the best-fitting
values from the original un-truncated disks-only model; we leave the
super-thin disk parameters free to ensure that the outer truncation
(which was not masked out in our disk-only fits) is not biasing the
super-thin disk fit.

Following our procedure for NGC~891, to fit the inner portions of
NGC~4013 we added an exponential disk with an outer truncation at the
radius of the inner disk truncation to mimic a bar component, as well
as a nuclear star-forming disk. The best-fitting bar has a
scale-height of 490 pc and scale-length of 1.46 kpc. This is roughly
comparable to the scale-height of the intermediate-thickness disk
component, but with only half the scale-length. The results for this
more complex model are shown in Figure \ref{fig:ngc4013fitplot} and
Table \ref{tab:ngc4013bestfits}. The $\chi^2$ values are reduced
  significantly compared to the original model, as seen in the
  residual images.

Like NGC~891, the exponential bar model for NGC~4013 is an imperfect
fit at small radii. The model overpredicts the flux at small radii and
medium heights, and underpredicts the light at the very center of the
galaxy. These discrepancies are likely due to the fact that our
exponential disk assumption, while a fairly close match to NGC~891's
bar, is inadequate for the X-orbits visible in NGC~4013 (the strongest
such feature in our sample). A model with a S\'{e}rsic bulge instead
of an exponential bar was also tried, but did not improve the
correspondence between data and model (and in fact made the fit
worse).

However, given the fact that this mismatch is only an issue at very
small radii, it should not impact the disk parameters we are most
interested in measuring. It is clear we are accurately fitting
NGC~4013's disks, and obtain reasonable results for our fitted
parameters. We find a scale-height of 540 pc for our thin disk, almost
identical to that of the intermediate disk in \citet{Comeron11b}. Our
super-thin disk has a scale-height $\sim$1.5 times as large as
Comeron's thinnest disk. This discrepancy could arise because of our
different choices of fitting function (in
\citetalias{Schechtman-Rook13} we found that more rounded vertical
profiles produced thinner disks, especially for thinner components),
or from our use of 2D fits rather than multiple 1D analyses.

The addition of the outer super-thin disk truncation is quite
important for accurately determining the scale-length of the
super-thin disk, which increases by $\sim$15\% in the model with a
bar. The super-thin disk has a scale-length only 80\% as long as the
thin disk, but is 2.3 times flatter than the thin disk. The thick disk
scale-length is 1.4 times as long as the thin disk, but due to its
extremely large scale-height this still results in a decrease of
$\frac{h_{R}}{h_{z}}$ to about 1.3.

\begin{deluxetable*}{ccccc}
\tablewidth{0pt}
\tablecaption{NGC~4565 Best-fitting Models}
\tablehead{ & \multicolumn{3}{c}{Value} &  \\\cline{2-4}
& & \colhead{Two Disks+Ring+} & \colhead{Two Disks+Ring+Bulge+} & \\
  \colhead{Parameter$^{a}$} &  \colhead{Two Disks+Ring$^{b}$} & \colhead{Bulge+Nuclear Disk$^{c}$} & \colhead{Nuclear Disk+Truncation$^{c}$} & \colhead{Units}}
\startdata
$\mu_{0,T}$ &  15.38$\pm$0.01& 15.38$^{d}$&15.38$^{d}$& mag arcsec$^{-2}$\\
$h_{R,T}$ &  5.96$\pm$0.01& 5.96$^{d}$&5.96$^{d}$& kpc\\
$h_{z,T}$ &  0.35$\pm$0.00& 0.35$^{d}$&0.35$^{d}$& kpc\\
$L_{T}$ &  8.02 $\times$ 10$^{10}$& 8.02 $\times$ 10$^{10}$&6.04 $\times$ 10$^{10}$& $L_{\odot,K}$\\
$\mu_{0,Th}$ &  18.86$\pm$0.02& 18.86$^{d}$&18.86$^{d}$& mag arcsec$^{-2}$\\
$h_{R,Th}$ &  5.44$\pm$0.06& 5.44$^{d}$&5.44$^{d}$& kpc\\
$h_{z,Th}$ &  2.23$^{e}$& 2.23$^{e}$&2.23$^{e}$& kpc\\
$L_{Th}$ & 1.89 $\times$ 10$^{10}$& 1.89 $\times$ 10$^{10}$&1.36 $\times$ 10$^{10}$& $L_{\odot,K}$\\
$\mu_{0,nuc}$ &  \nodata& 13.45$\pm$0.02&13.49$\pm$0.02& mag arcsec$^{-2}$\\
$h_{R,nuc}$ &  \nodata& 0.16$\pm$0.00&0.18$\pm$0.00& kpc\\
$h_{z,nuc}$ &  \nodata& 0.18$\pm$0.00&0.17$\pm$0.00& kpc\\
$L_{nuc}$ &  \nodata& 6.55 $\times$10$^{9}$&6.71 $\times$ 10$^{9}$& $L_{\odot,K}$\\
$\mu_{0,ring}$ &  18.07$\pm$0.04& 18.33$\pm$0.01&16.11$\pm$0.08& mag arcsec$^{-2}$\\
$r_{0_{ring}}$ &  5.34$\pm$0.02& 5.39$\pm$0.01&4.55$\pm$0.23& kpc\\
$\sigma_{0_{ring}}$ &  0.42$\pm$0.01& 0.41$\pm$0.01&0.94$\pm$0.03& kpc\\
$h_{z_{ring}}$ &  0.27$\pm$0.00& 0.36$\pm$0.01&0.30$\pm$0.01& kpc\\
$L_{ring}$ & 4.67 $\times$ 10$^{9}$& 4.94 $\times$ 10$^{9}$&2.69 $\times$ 10$^{10}$& $L_{\odot,K}$\\
$\mu_{0,bulge}$ &  \nodata& 15.48$\pm$0.01&15.20$\pm$0.01& mag arcsec$^{-2}$\\
$R_{e}$ & \nodata& 1.50$\pm$0.00&1.77$\pm$0.03& kpc\\
$a/b$ &  \nodata& 1.36$\pm$0.00&1.73$\pm$0.03& kpc\\
$n^{f}$ &  \nodata& 0.69$\pm$0.01&0.71$\pm$0.01& \\
$L_{bulge}$ &  \nodata& 3.26 $\times$ 10$^{10}$&4.50 $\times$ 10$^{10}$& $L_{\odot,K}$\\
$i$ &  87.59$\pm$0.01& 87.59$^{d}$&87.59$^{d}$& degrees\\
$R_{trunc,1}^{g}$ &  \nodata& \nodata&5.68$\pm$0.35& kpc\\
$L_{tot}$ & 1.04 $\times$ 10$^{11}$& 1.43 $\times$ 10$^{11}$& 1.53 $\times$ 10$^{11}$& $L_{\odot,K}$\\
$\chi_{\nu,inner}^{2}\,^{h}$& 122.16& 7.19& 6.14&\nodata\\
$\chi_{\nu,outer}^{2}\,^{h}$& 3.23*& 3.24& 3.33&\nodata\\
$\chi_{\nu,total}^{2}\,^{h}$& 52.15& 4.87*& 4.49*&\nodata
\enddata
\tablenotetext{a}{$T$, $Th$, $nuc$, and $ring$ denote parameters that we associate with
  the thin disk, thick disk, nuclear disk, and ring (respectively).}
\tablenotetext{b}{Only fit to regions of the galaxy with $R>5$ kpc.}
\tablenotetext{c}{Fit performed over entire radial range of data.}
\tablenotetext{d}{Fixed to Two Disks+Ring model value.}
\tablenotetext{e}{Fixed to literature value (see text for details).}
\tablenotetext{f}{Sersic index.}
\tablenotetext{g}{Inner truncation for the thin and thick disks, outer
  truncation for the nuclear disk.}
\tablenotetext{h}{Reduced $\chi^{2}$. {\it inner} and {\it outer} correspond
  to the regions inside and outside of the 5 kpc fitting cutoff, while {\it
    total} is for both regions together. (*) denotes the $\chi_{\nu}^{2}$ that was
  minimized by each model.}
\label{tab:ngc4565bestfits}
\end{deluxetable*}

\begin{figure*}
\centering
\includegraphics[scale=0.7]{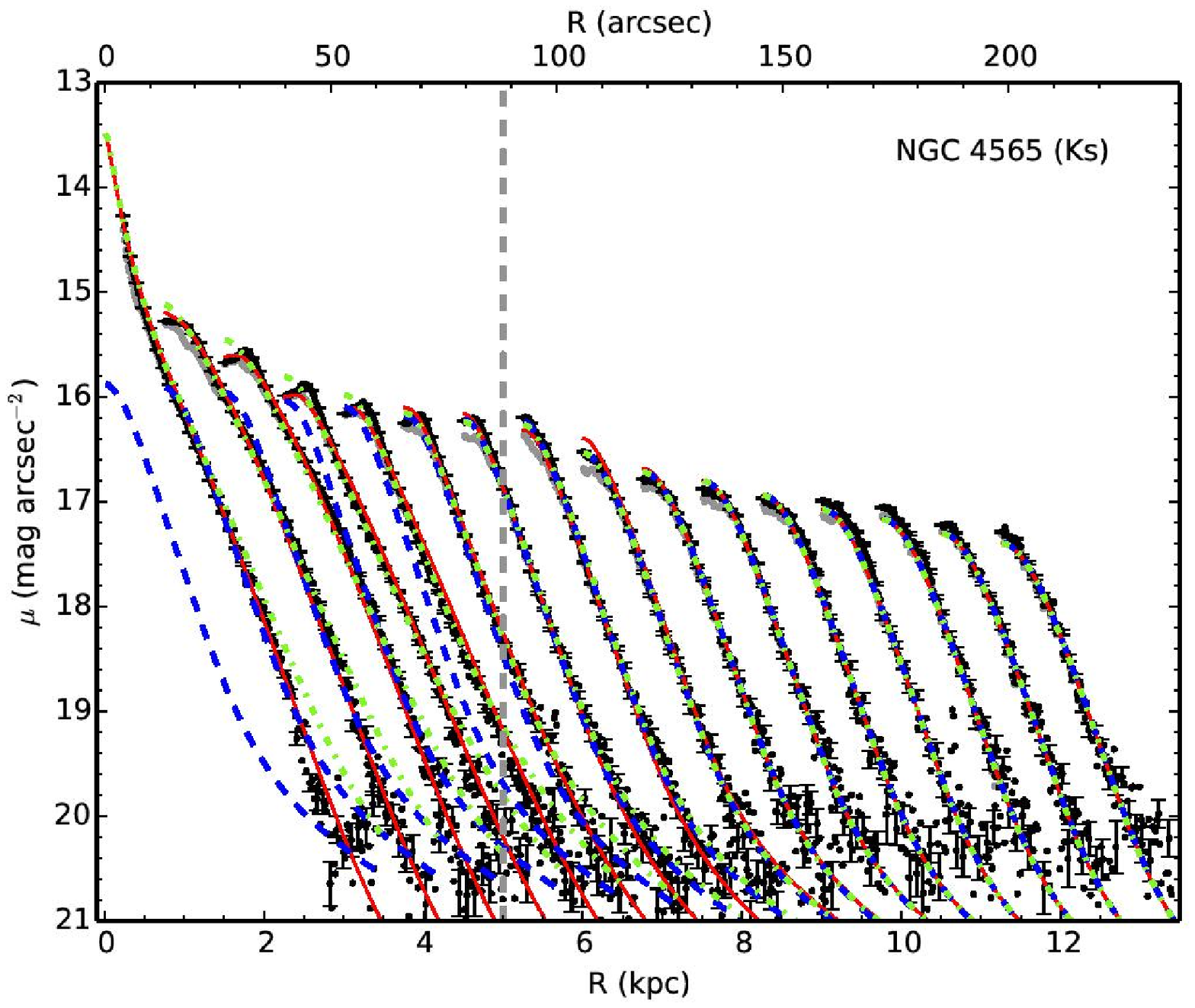}\\
\includegraphics[scale=0.7]{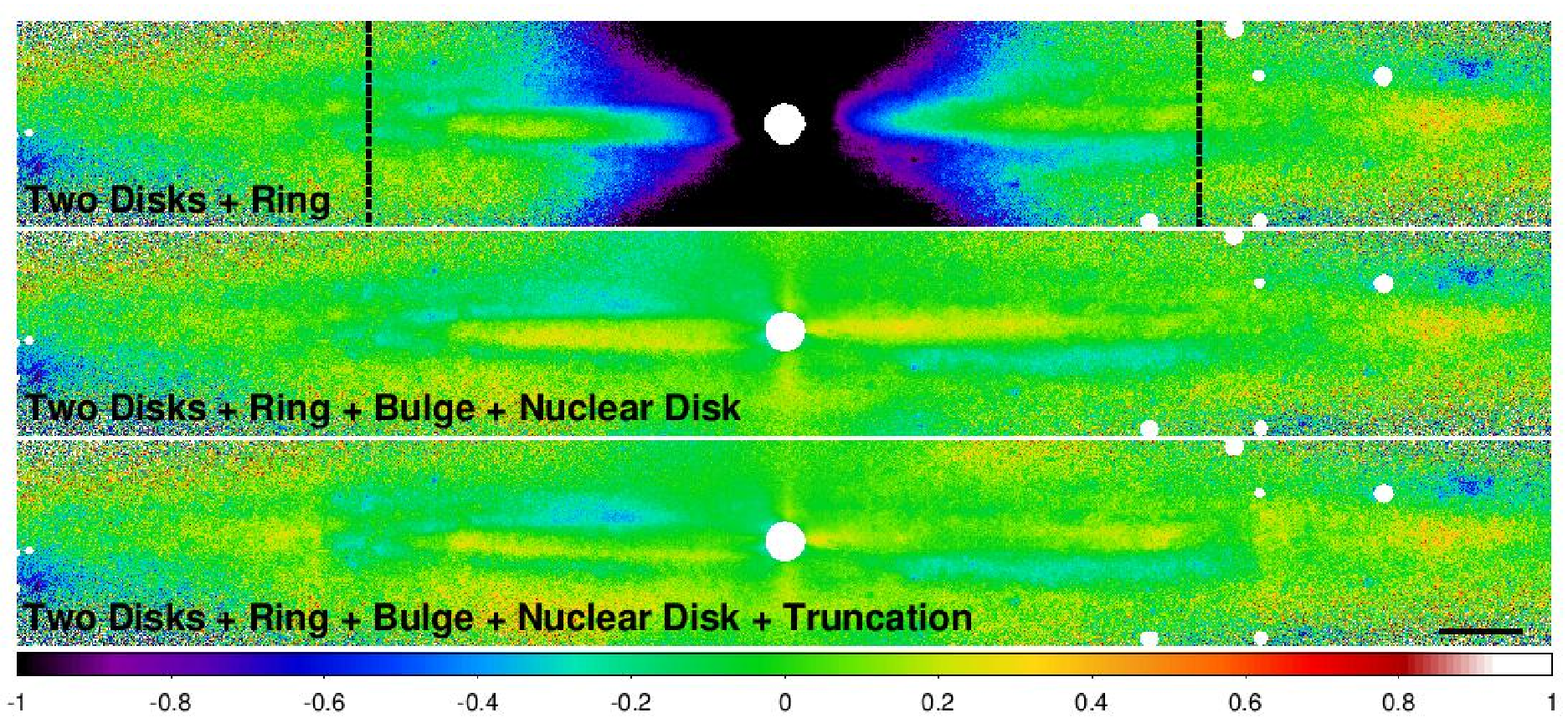}
\caption{Similar to Figure \ref{fig:ngc891fitplot}, but for
  NGC~4565. Here the blue dashed lines show the best fitting two disk
  + ring model (the model shown in the left column of Table
  \ref{tab:ngc4565bestfits}), fit to all regions outside of 5 kpc in
  radius (denoted by the vertical gray dashed line). The green
  dot-dashed lines show the best fitting two disk + ring + S\'{e}rsic
  bulge + nuclear disk model without a truncation (center column of
  Table \ref{tab:ngc4565bestfits}), while the solid red lines show the
  best fit to that same model with an added disk truncation (right
  column of Table \ref{tab:ngc4565bestfits}). To improve S/N the data
  in the 1D profiles have all been binned to 5'' x 0\farcs3 (R x z).}
\label{fig:ngc4565fitplot}
\end{figure*}

\subsubsection{NGC~4565}
\label{sec:ngc4565}

\begin{deluxetable*}{c c c c c}
\tabletypesize{\footnotesize}
\tablewidth{0pt}
\tablecaption{NGC~522 Best-fitting Models}
\tablehead{ & \multicolumn{3}{c}{Value} &  \\\cline{2-4}
  \colhead{Parameter$^{a}$} &  \colhead{Two Disks$^{b}$} &
  \colhead{Two Disks+Bulge$^{c}$} & \colhead{Two Disks+Bulge+Bar$^{c}$} & \colhead{Units}}
\startdata
$\mu_{0,T}$ & 15.55$\pm$0.00& 15.62$\pm$0.00&15.54$\pm$0.00& mag arcsec$^{-2}$\\
$h_{R,T}$ & 7.93$\pm$0.08& 10.00$^{d}$&7.66$\pm$0.07& kpc\\
$h_{z,T}$ & 0.38$\pm$0.00& 0.41$\pm$0.00&0.38$\pm$0.00& kpc\\
$L_{T}$ & 4.90 $\times$ 10$^{10}$& 4.71 $\times$ 10$^{10}$&3.84 $\times$ 10$^{10}$& $L_{\odot,K}$\\
$\mu_{0,Th}$ & 18.01$\pm$0.03& 18.01$^{e}$&18.01$^{e}$& mag arcsec$^{-2}$\\
$h_{R,Th}$ & 9.55$\pm$0.16& 9.55$^{e}$&9.55$^{e}$& kpc\\
$h_{z,Th}$ & 1.22$\pm$0.02& 1.22$^{e}$& 1.22$^{e}$& kpc\\
$L_{Th}$ & 3.98 $\times$ 10$^{10}$& 3.94 $\times$ 10$^{10}$&3.66 $\times$ 10$^{10}$& $L_{\odot,K}$\\
$\mu_{0,bar}$ &  \nodata& \nodata&16.23$\pm$0.01& mag arcsec$^{-2}$\\
$h_{R,bar}$ &  \nodata& \nodata&5.00$^{d}$& kpc\\
$h_{z,bar}$ &  \nodata& \nodata&0.64$\pm$0.00& kpc\\
$L_{bar}$ &  \nodata& \nodata&1.26 $\times$ 10$^{10}$& $L_{\odot,K}$\\
$\mu_{0,bulge}$ &  \nodata& 14.88$\pm$0.02&14.84$\pm$0.03& mag arcsec$^{-2}$\\
$R_{e}$ & \nodata& 1.10$\pm$0.01&1.09$\pm$0.01& kpc\\
$a/b$ &  \nodata& 1.45$\pm$0.01&1.57$\pm$0.01& kpc\\
$n^f$ &  \nodata& 1.30$\pm$0.01&1.39$\pm$0.02& kpc\\
$L_{bulge}$ &  \nodata& 1.11 $\times$ 10$^{10}$& 9.03 $\times$ 10$^{9}$& $L_{\odot,K}$\\
$i$ & 88.49$\pm$0.02& 88.49$^{e}$&88.49$^{e}$& degrees\\
$R_{trunc,1}^{g}$ &  \nodata& 1.25$\pm$0.03&4.45$\pm$0.05& kpc\\
$R_{trunc,2}^{h}$ & 13.18$\pm$0.03& 13.18$^{e}$&13.18$^{e}$& kpc\\ 
$L_{tot}$ & 8.88 $\times$ 10$^{10}$& 9.76 $\times$ 10$^{10}$& 9.66 $\times$ 10$^{10}$& $L_{\odot,K}$\\
$\chi_{\nu,inner}^{2}\,^{i}$& 4.10& 0.57& 0.54&\nodata\\
$\chi_{\nu,outer}^{2}\,^{i}$& 0.65*& 0.66& 0.65&\nodata\\
$\chi_{\nu,total}^{2}\,^{i}$& 1.68& 0.64*& 0.61*&\nodata
\enddata
\tablenotetext{a}{$T$, $Th$, and $nuc$, denote parameters that we
  associate with the thin disk, thick disk, and nuclear disk
  (respectively).}
\tablenotetext{b}{Only fit to regions of the galaxy with $R>4$ kpc.}
\tablenotetext{c}{Fit performed over entire radial range of data.}
\tablenotetext{d}{Parameter minimized to the upper boundary of allowed
  parameter space.}
\tablenotetext{e}{Fixed to Two Disks model value.}
\tablenotetext{f}{Sersic index.}
\tablenotetext{g}{Inner truncation for the thin and thick disks, outer
  truncation for the bar.}
\tablenotetext{h}{Outer truncation for the thin disk only.}
\tablenotetext{i}{Reduced $\chi^{2}$. {\it inner} and {\it outer} correspond
  to the regions inside and outside of the 4 kpc fitting cutoff, while {\it
    total} is for both regions together. (*) denotes the $\chi_{\nu}^{2}$ that was
  minimized by each model.}
\label{tab:ngc522bestfits}
\end{deluxetable*}

\begin{figure*}
\centering
\includegraphics[scale=0.7]{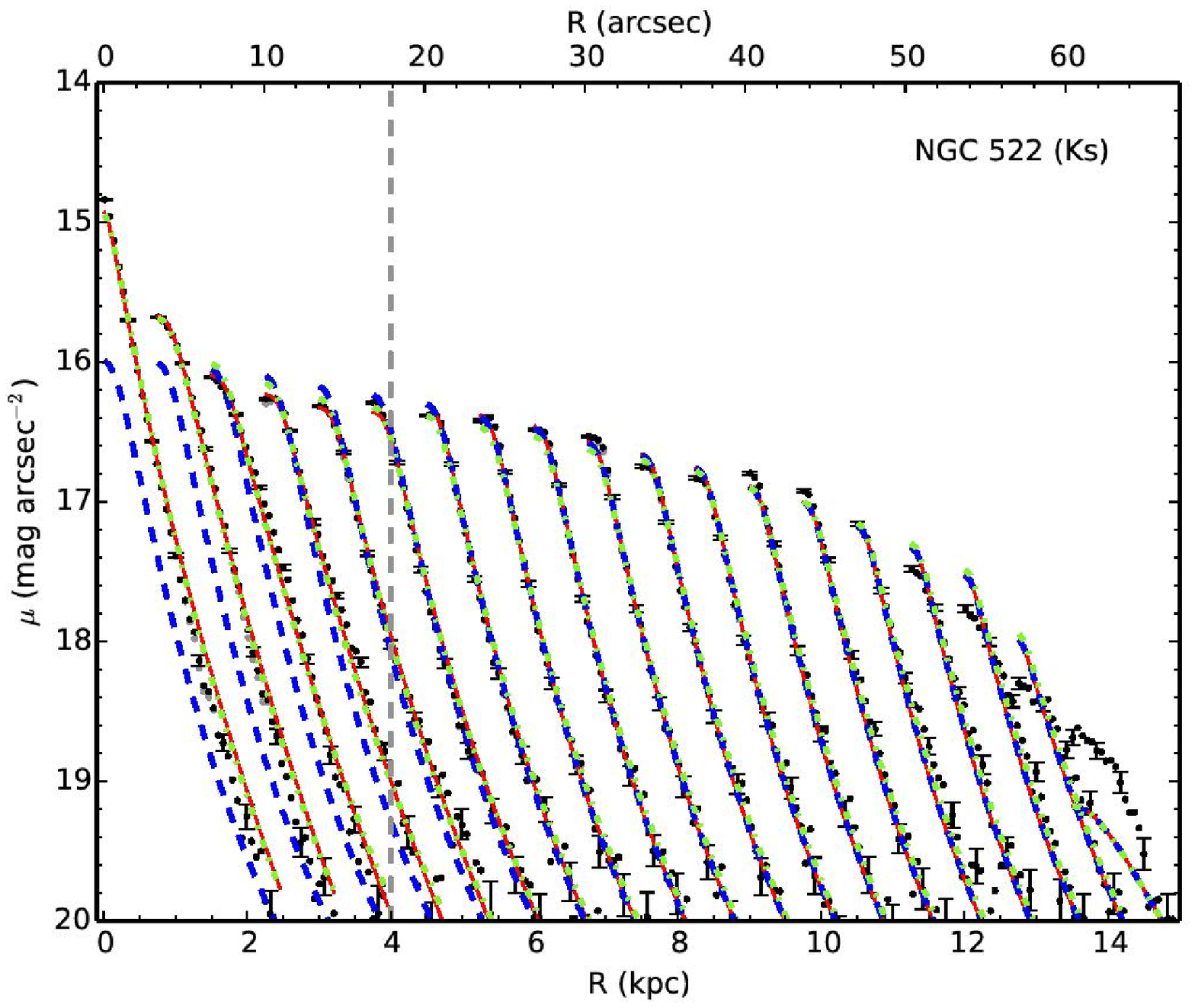}
\includegraphics[scale=0.7]{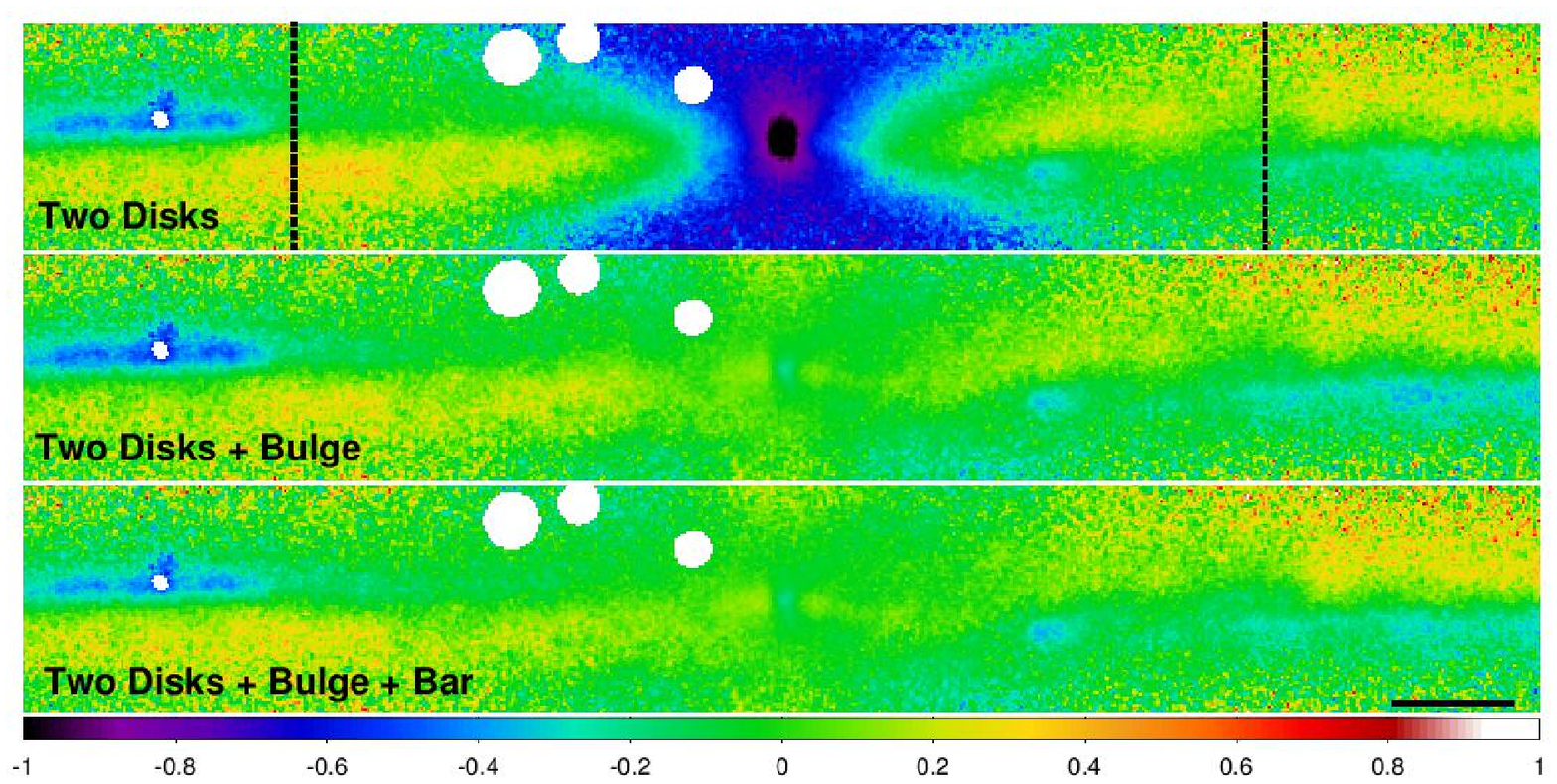}
\caption{Similar to Figure \ref{fig:ngc891fitplot}, but for
  NGC~522. From left to right in Table \ref{tab:ngc522bestfits}: blue
  dashed lines (two disk model fit outside of 4 kpc), green dot-dashed
  lines (two disk model with S\'{e}rsic bulge), and red solid lines
  (two disk model with S\'{e}rsic bulge and exponential bar). To improve
  S/N the data in the 1D profiles have all been binned to 1'' x 0\farcs3 ($R \times z$).}
\label{fig:ngc522fitplot}
\end{figure*}

To avoid NGC~4565's obvious inner truncation we first restrict our
models to fit outside of $R$=5 kpc. We find \citep[as do][]{Comeron11}
that only two disks are needed to model most of the light outside of
this radius, except for the contribution from the ring-like structure
visible in Figure \ref{fig:ngc4565attencorr} . We therefore chose to
change our models to ones with two disks and a ring, adding to our
fitting software a gaussian ring with the same functional description
as used in \citet{deLooze12}:
\begin{align}
I(R,z) =
\rho_{0}\int_{-\infty}^{\infty}\mathrm{exp}\bigg(&-\frac{(\sqrt{R^{2}+t^{2}\mathrm{sin}^{2}\,i}-R_{0})^{2}}{2\sigma_{0}^{2}}-\nonumber
\\ &\frac{|z\mathrm{sin}\,i-t\mathrm{cos}\,i|}{h_{z}}\bigg)dt,
\end{align}
where $R_{0}$ is the central ring radius and $\sigma_{0}$ is the
radial dispersion of the ring.  To help fit the thick disk, we used an
average of the thick disk scale-heights found in the 1D fits done for
NGC~4565 by \citet{Comeron11}.  Because their fits produced much
smaller values for the thick disk $h_{z}$ in their large radii bins
(14.5 kpc$\lesssim$R$\lesssim$23 kpc, regions outside of our WHIRC
coverage), we averaged only the values in their two bins nearest the
center of NGC~4565. (Note that all of the fits in \citealt{Comeron11}
are outside of the radii where the ring feature is present.) We leave
the scale-length and central surface brightness as free parameters, as
they are not reported in \citet{Comeron11}. The results of this fit
are shown as the blue dashed line in Figure \ref{fig:ngc4565fitplot}
and listed in the leftmost column of Table \ref{tab:ngc4565bestfits}.
This provides an adequate fit outside of $R=5$kpc, but deviates
substantially from the data within that radius.

While our fitting of NGC~4013 closely followed the same procedure as
we followed for NGC~891 in \citetalias{Schechtman-Rook13}, producing a
good fit to NGC~4565 at all radii was a significant challenge, due
largely to the unique morphology of this system. First, unlike in NGCs
891 and 4013, where the disk-only models clearly overpredict the
central surface brightness at small radii, in NGC~4565 the model
described above remains comparable or fainter than the peak surface
brightness at all radii. Therefore it is unclear whether a disk
truncation is actually needed, or if the ring merely creates the {\it
  appearance} of such a feature.

To improve our model fits at small radii and to investigate whether or
not our data support an inner disk truncation, we next held constant
the free parameters of the two disks from our previous best-fitting
model and added a nuclear disk and S\'{e}rsic bulge.\footnote{We
  attempted to use an exponential bar-like component instead of a
  bulge but it produced a poor fit, possibly because unlike NGCs 891
  and 4013, NGC~4565 has a very faint X pattern.} This model was fit
to the full radial range of data, and since the ring abutted our 5 kpc
cutoff from the previous fit we allowed it to vary freely. The results
of this fit are reported in the central column of Table
\ref{tab:ngc4565bestfits} and plotted in Figure
\ref{fig:ngc4565fitplot} with green dot-dashed lines.  This model is a
significant improvement at small radii from the initial model, but in
this radial range it over-predicts the amount of light at the smallest
and largest heights. From our experience fitting NGC 891 and 4013,
this indicates the need for an inner disk truncation (to remove model
light from the mid-plane) but without the redistribution of light into
a bar (to remove light at large height).

We then repeated this fit, adding in an outer truncation for the
nuclear disk and an inner truncation for the other two disks. This
third model is plotted with red solid lines in Figure
\ref{fig:ngc4565fitplot} and shown in the right column of Table
\ref{tab:ngc4565bestfits}. The addition of the truncation allows this
model to match the data at large heights over the entire radial
range. However, while significantly better than the model with no
truncation  (the corrected Akaike information criterion,
  AIC,\footnote{In the limit where the number of data points is much
    larger than the number of model parameters, ${\rm AIC} \sim \chi^2
    + 2k$, where $k$ represents the number of model parameters.} drops
  by $\sim$8\%; \citealt{Burnham02}) even this model is unable
to fully reproduce the the detailed shape of the light profiles near
the mid-plane. In particular, the offset peaks and central troughs of
the profiles between 2$\gtrsim$R$\gtrsim$5 kpc seen in Figure
\ref{fig:ngc4565fitplot} remain problematic.

One reason our models maybe failing to match the complex mid-plane
structure is that we are seeing an overabundance of light from the
inner edge of NGC~4565's ring due to dust scattering of bulge light.
To investigate this we looked at the profiles of the individual
quadrants of NGC~4565, under the assumption that this effect would
show up as a systematic offset in surface brightness between quadrants
showing the back and edges of the ring. However, we could find no
concrete evidence for this offset. In fact, it appears that two
quadrants diagonally opposite from each other (the upper left and
lower right quadrants in Figure \ref{fig:gri_jhk_sample}) show an
excess of flux. It is not clear why this should be, although if
NGC~4565 did have a weak bar at this position angle, that might
explain the observed behavior. While a full exploration of this effect
is beyond the present scope, it should not affect our analysis of the
broader disk parameters.

Our best-fitting model bulge has an effective radius of 1.77 kpc and
an axial ratio of $\sim$1.75. Additionally, it has a S\'{e}rsic index
of 0.71. This is close to an exponential radial light profile, but the
light distribution is much less oblate than the central light
concentrations we modeled as ``bars'' in the previous two galaxies NGC
891 and 4013.  The final disk components yield a thin disk with
$h_{z}$=350 pc, $\sim$90\% the height found at 3.6$\mu$m by
\citet{Comeron11} (averaged over all of their fits) and $\sim$75\% the
optical height fit by \citet{deLooze12}. In this case differences
between fitting functions would only serve to increase this disparity,
as adjusting from an exponential to a sech or sech$^{2}$ vertical
distribution would only decrease our $h_{z}$. A survey of other
literature fits indicates that overall our thin-disk $h_{z}$ is
anomalously small \citep{Wu02}, although most of these studies assume
that NGC~4565 is perfectly edge-on, which would bias fits to larger
values of $h_{z}$. However, the model of \citet{deLooze12} has
inclination comparable ($\sim$0.25$^{\circ}$ smaller) than ours. Given
the fact that most of the larger measurements for this quantity come
from optical data \citep[with the exception of][]{Rice96} and that our
results are close to those found at 3.6$\mu$m by \citet{Comeron11}, it
seems plausible that this discrepancy could be the result of
insufficient reddening corrections of the optical light profiles.

None of our models produce fits containing a (non-nuclear) disk
component with a scale-height consistent with a MW, NGC~4013, or
NGC~891-like super-thin disk. The stellar ring, the component
spatially associated with CO emission (and therefore star formation),
has a scale-height $\gtrsim$85\% of the thin disk. While our inability
to fit the offset peaks of the vertical profiles for
2$\gtrsim$R$\gtrsim$5 kpc could potentially leave room for a
super-thin component, given the quality of our fits at larger radii
such a component would likely be quite faint, with a smaller relative
$\ks$-band luminosity compared to the super-thin disks of NGCs 891 and 4013.

\subsubsection{NGC~522}
\label{sec:ngc522}

\begin{deluxetable*}{ccccc}
\tablewidth{0pt}
\tablecaption{NGC~1055 Best-fitting Models}
\tablehead{ & \multicolumn{3}{c}{Value} &  \\\cline{2-4}
  \colhead{Parameter$^{a}$} & \colhead{One Disk$^{b}$} & \colhead{Two Disks$^{b}$} & \colhead{Two Disks+Truncated Disk$^{b}$} & \colhead{Units}}
\startdata
$\mu_{0,tr^{c}}$  & \nodata& \nodata& 15.49$\pm$0.01& mag arcsec$^{-2}$\\
$h_{R,tr^{c}}$    & \nodata& \nodata& 3.38$\pm$0.09& kpc\\
$h_{z,tr^{c}}$    & \nodata& \nodata& 0.29$\pm$0.00& kpc\\
$L_{tr^{c}}$      & \nodata& \nodata& 9.00 $\times$ 10$^{9}$& $L_{\odot,K}$\\
$\mu_{0,1}$  & 11.17$\pm$0.00& 14.17$\pm$0.003& 14.05$\pm$0.02& mag arcsec$^{-2}$\\
$h_{R,1}$    & 1.80$\pm$0.00& 1.45$\pm$0.003& 1.60$\pm$0.01& kpc\\
$h_{z,1}$    & 0.01$^{d}$& 0.19$\pm$0.001& 0.10$\pm$0.00& kpc\\
$L_{1}$      & 3.34 $\times$ 10$^{10}$& 3.23 $\times$ 10$^{10}$& 2.10 $\times$ 10$^{10}$& $L_{\odot,K}$\\
$\mu_{0,2}$ & \nodata& 16.89$\pm$0.01& 16.02$\pm$0.04& mag arcsec$^{-2}$\\
$h_{R,2}$   & \nodata& 2.82$\pm$0.01& 10.00$^{e}$& kpc\\
$h_{z,2}$   & \nodata& 1.26$\pm$0.00& 0.15$\pm$0.01& kpc\\
$L_{2}$     & \nodata& 3.40 $\times$ 10$^{10}$& 3.20 $\times$ 10$^{10}$& $L_{\odot,K}$\\
$\mu_{0,bulge}$ & 12.48$\pm$0.01& 8.32$\pm$0.02& 10.38$\pm$0.05& mag arcsec$^{-2}$\\
$R_{e}$ & 5.93$\pm$0.01& 8.00$^{e}$& 8.00$^{e}$& kpc\\
$a/b$ & 1.83$\pm$0.00& 1.50$\pm$0.01& 1.44$\pm$0.00& kpc\\
$n^{f}$ & 3.19$\pm$0.01& 6.00$^{e}$& 4.67$\pm$0.02& \\
$L_{bulge}$ & 8.13 $\times$ 10$^{10}$& 3.94 $\times$ 10$^{10}$& 7.94 $\times$ 10$^{10}$& $L_{\odot,K}$\\
$i$ & 83.88$\pm$0.00& 86.21$\pm$0.00& 85.50$\pm$0.03& degrees\\
$R_{trunc,1}^{g}$ & \nodata& \nodata& 1.46$\pm$0.01& kpc\\
$R_{trunc,2}^{g}$ & \nodata& \nodata& 4.03$\pm$0.01& kpc\\
$L_{tot}$ & 1.15 $\times$ 10$^{11}$& 1.06 $\times$ 10$^{11}$& 1.41 $\times$ 10$^{11}$& $L_{\odot,K}$\\
$\chi_{\nu,total}^{2}\,^{h}$& 0.40& 0.39& 0.37&\nodata
\enddata
\tablenotetext{a}{$tr$ denotes parameters that we associate with the
  truncated disk.}
\tablenotetext{b}{Fit performed over regions of the galaxy with $R<6$
  kpc.}
\tablenotetext{c}{We call this disk the truncated disk rather than a
  super-thin disk, because it does not have a smaller scale-height
  than the nominal thin disk. See text for discussion.}
\tablenotetext{d}{This parameter minimized to the lower boundary of
  the allowed parameter space.}
\tablenotetext{e}{This parameter minimized to the upper boundary of
  the allowed parameter space.}
\tablenotetext{f}{Sersic index.}
\tablenotetext{g}{Truncations are for the truncated disk only, where
  truncation 1 is the inner truncation and truncation 2 is the outer truncation.}
\tablenotetext{h}{Reduced $\chi^{2}$. Since all models for NGC~1055 were fit
  over the same spatial regions, we only report the total $\chi_{\nu}^{2}$.}
\label{tab:ngc1055bestfits}
\end{deluxetable*}

\begin{figure*}
\centering
\includegraphics[scale=0.6]{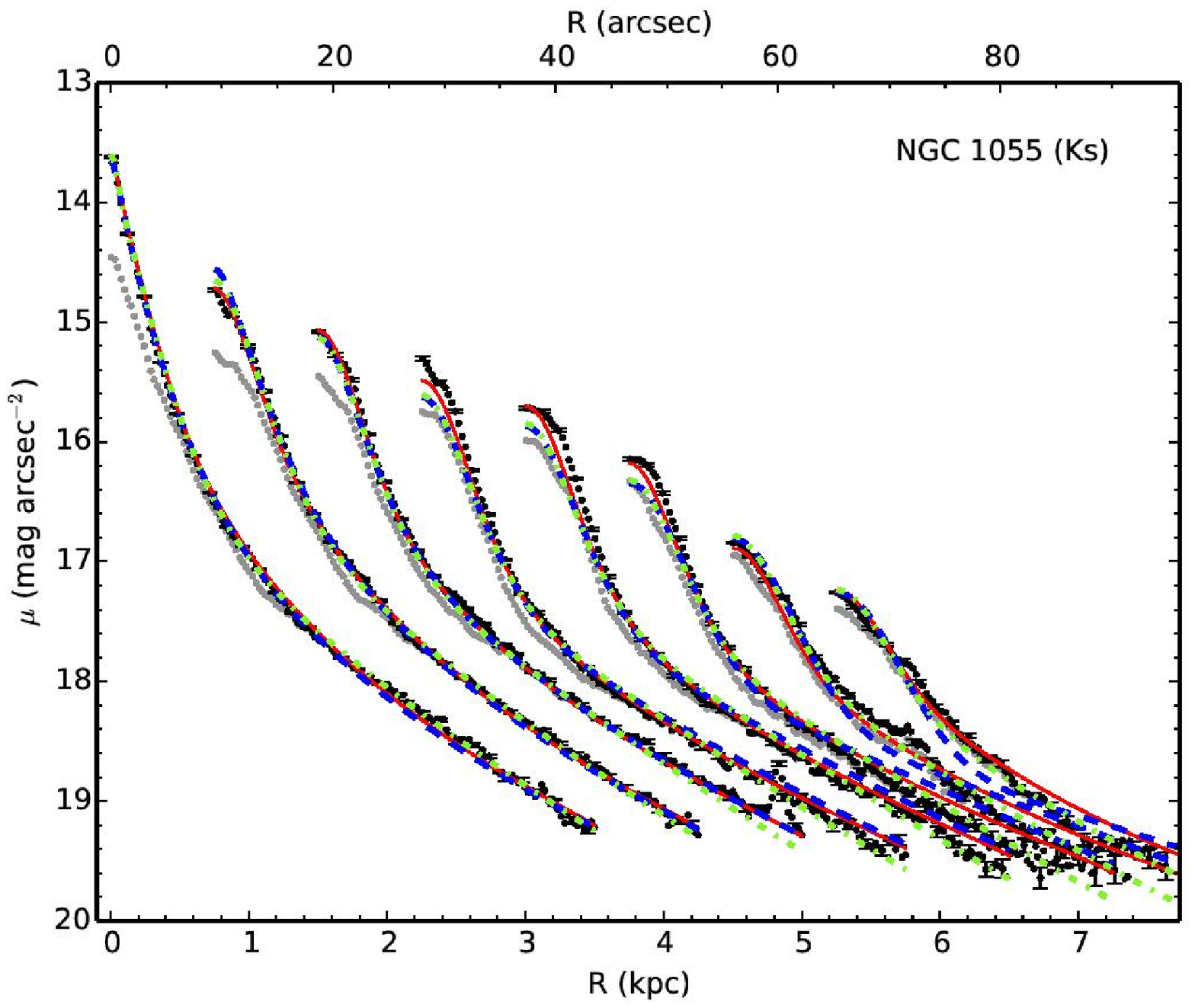}\\
\includegraphics[scale=0.6]{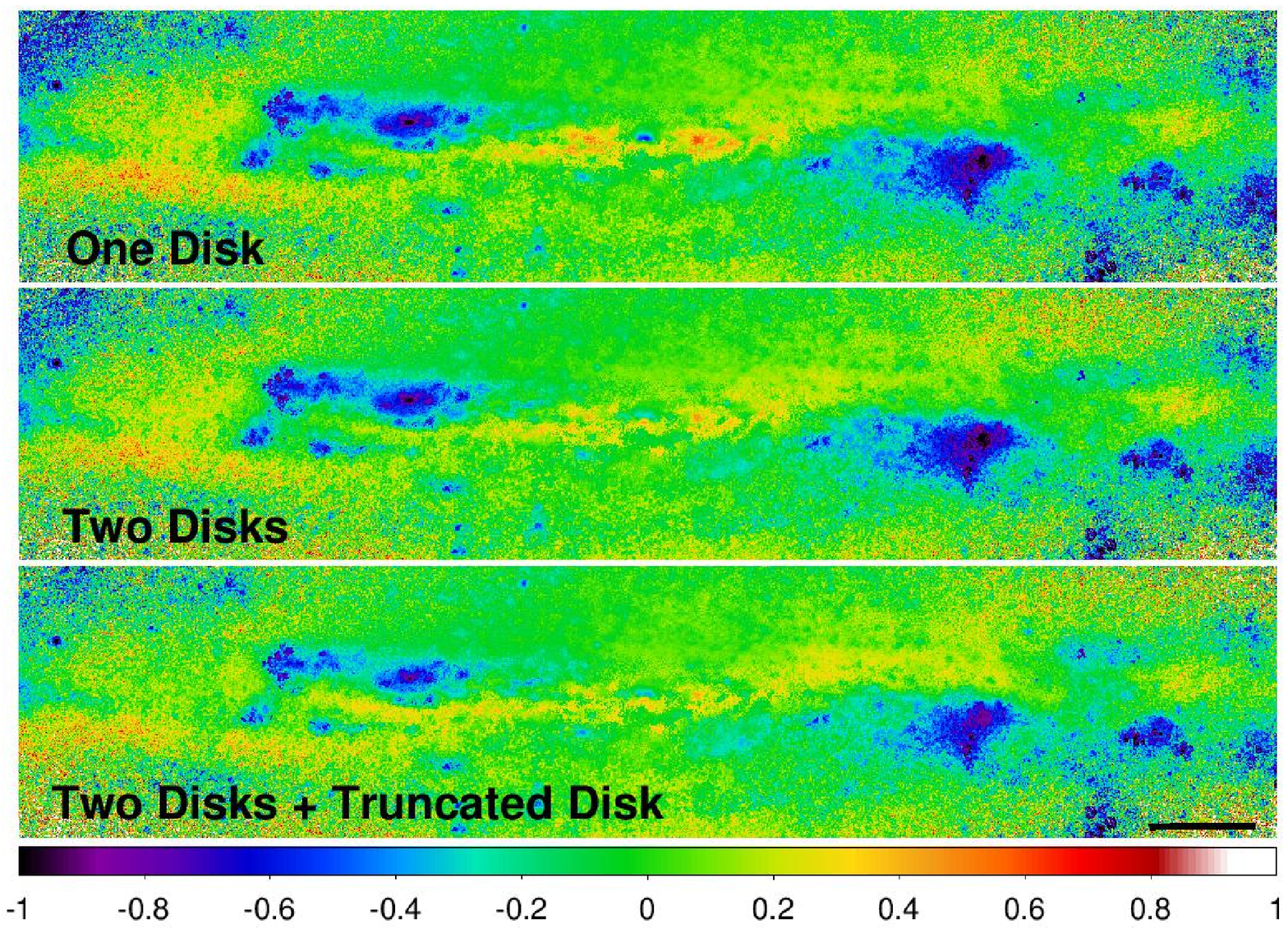}
\caption{Similar to Figure \ref{fig:ngc891fitplot}, but for
  NGC~1055. Here we show the best fitting models with a S\'{e}rsic
  bulge and: one disk (blue dashed lines), two disks (green dot-dashed
  lines), and three disks with one containing inner and outer
  truncations (red solid lines). To improve S/N the data in the 1D profiles have all been
  binned to 5'' x 0\farcs3 ($R \times z$). Details of the model
  parameters are in Table \ref{tab:ngc1055bestfits}.}
\label{fig:ngc1055fitplot}
\end{figure*}

\citet{Comeron11} produce several one-dimensional fits of NGC~522,
reporting thin and thick disk scale-heights of $\sim$200 and $\sim$830
pc, respectively. We begin by fitting our own version of that model,
restricted to $R>4$ kpc. Based on a visual inspection of the
surface brightness profiles (see Figure \ref{fig:ngc522fitplot}) the
outer part of the disk appears to be truncated, and so based on our
results for NGC 891 and 4013 we also include an outer truncation of
the thinner disk. Due to the relatively small scale-height found by
\citet{Comeron11} for NGC~522's thick disk, our data is sensitive
enough that we do not have to fix this parameter in our fits (as we
had done for previous galaxies). The results of this model are given
in the left panel of Table \ref{tab:ngc522bestfits} and shown as the
blue dashed lines in Figure \ref{fig:ngc522fitplot}.

NGC~522's surface-brightness profile is notable for several
reasons. First, the outer disk truncation at $\sim$13 kpc in radius
appears to be a downward break of the global radial surface brightness
profile rather than a truncation of a single disk component, as seen
in NGCs 891 and 4013.  Modeling such breaks in the radial slope is
beyond the scope of this paper. However, our simple outer truncation
provides a better fit than a model with no truncations (and will
therefore result in more accurate luminosity estimates), so we
continue to include this component in our other models.

NGC~522's disk is also quite flat in radius, with scale-lengths
roughly 1.5-2 times as large as those found for the other fast
rotators in our sample. This results in a scale-length to scale-height
ratio of $\sim$20 for the thin disk, significantly larger than found
in normal spirals \citep{Bershady10b}. It appears that NGC~522 is an
outlier in this regard, but note that our measurements may
be biased because of the distance-dependence of our attenuation
correction (Section \ref{sec:ngc522corr}). However, we would expect that
correcting for this bias would tend to decrease both $h_{z}$ \textit{and}
$h_{R}$, which could result in either a larger or smaller ratio of
$\frac{h_{R}}{h_{z}}$ depending on how exactly the data is biased. 

We see no evidence for a super-thin disk component in NGC~522,
although due to the uncertainty in our dust correction at this
distance we reiterate that we may be insensitive to such a
component. An inner disk truncation does appear to be warranted by the
data, although it is not nearly as obvious as for NGC 891 or 4013. A
visual inspection of Figure \ref{fig:gri_jhk_sample} shows a central
morphology closer to that of NGC~4565 than the other fast rotators
(chiefly in the lack of an obvious X-shape), so we choose to use a
S\'{e}rsic bulge instead of a bar to model NGC~522's inner light
profile.

It is unclear whether or not a separate disk (akin to either the bar
or nuclear disk components of NGCs 891, 4013, or 4565) inside of the
truncation is needed in this galaxy, so we first fit a model without
one (dot-dashed green lines in Figure \ref{fig:ngc522fitplot} and the
center column of Table \ref{tab:ngc522bestfits}). This model slightly
overpredicts the central surface brightness between $\sim$2 and
$\sim$4 kpc in radius. Including an inner disk produces an excellent
fit in this region (solid red lines in Figure \ref{fig:ngc522fitplot}
and rightmost column of Table \ref{tab:ngc522bestfits}). The
  change in the residual structure appears less dramatic, and indeed
  the AIC value decreases by only 3\%. This disk is quite extended,
with $h_{z}$ between the equivalent values found in the thin and thick
disks. This result is generally consistent with what we fit for bar
components in NGCs 891 and 4013, and therefore we adopt that
nomenclature for this component.

We note, however, that NGC~522's bar is much more extended in the
radial direction than in either NGC 891 or NGC 4013, and actually
minimizes to the largest value allowed in our fit (5 kpc). It is
possible that this is an indication that NGC~522's bar may be
physically different than the bars we find in our other fast rotators,
or possibly the conflation of two distinct morphological
components. Regardless of the correct interpretation for this
component, however, because it is truncated cleanly at 4.45 kpc in
radius it will have little effect on our estimation of this galaxy's
disk parameters (which are only present outside of this
radius). Assuming it is directly comparable, this bar component is
also much fainter (relative to the other components in the galaxy)
than in NGCs 891 and 4013, which is a reasonable result given the
comparative faintness of X-shaped bar features in NGC~522 (just barely
visible in Figure \ref{fig:gri_jhk_sample}).

\subsubsection{NGC~1055}
\label{sec:ngc1055}

NGC~1055 is morphologically very different from the other
fast-rotators in our sample, notably in its bright, extended bulge
(most visible in the SDSS image in Figure \ref{fig:gri_jhk_sample}),
but also in the large number of coherent dust features visible away
from the midplane. It is in a loose group with the extremely
fast-rotating (HyperLEDA gives $V_{rot}\sim$280 km sec$^{-1}$) spiral
Messier 77, and a recent interaction between these two systems could
explain the apparent disarray in NGC~1055's disk.

While its unique features make it an especially interesting object to
examine, they also conspire to make NGC~1055 tremendously difficult to
model. Due to the expansive nature of the bulge, we must abandon our
usual practice of first fitting just the disk components, and instead
we are compelled to include a S\'{e}rsic bulge in all of our models
and simultaneously constrain the bulge and disk parameters. Since we
are no longer ignoring the bulge contribution for any models, we do not
mask out the central region of our data for any of our
fits. Additionally, given the rapidly vanishing disk flux at larger
radii we choose to fit our models only to regions with radius less
than 6 kpc, as marked in Figure~\ref{fig:ngc1055attencorr}, which also
avoids portions of the outer disk which likely have a substantial
warp. 

In all of our models there is a clear preference for a very large
($R_{e}\gtrsim6$ kpc) bulge, the quantitative consequence of the
bright spheroidal component notably visible in the images. Our first
model, consisting of a bulge and single disk, is listed in the
leftmost column of Table \ref{tab:ngc1055bestfits} and shown with blue
dashed lines in Figure \ref{fig:ngc1055fitplot}. This model does a
reasonable job fitting the data at small radii, but underpredicts the
midplane surface brightness when 2.5$\gtrsim$R$\gtrsim$4.5 kpc and
poorly matches the extended light at $R\gtrsim$5 kpc. This model
predicts an extremely thin ($h_{z}\lesssim$10 pc) disk.

A model with a second disk (green dot-dashed line in Figure
\ref{fig:ngc1055fitplot} and the middle column of Table
\ref{tab:ngc1055bestfits}) does a better job reproducing the data at
large radii.  Its thinnest disk has $h_{z}=190$ pc, much thicker than
that of the single-disk model. The bulge parameters also change
significantly with the addition of a second disk, with larger central
surface brightness and effective radius. Despite these increases the
total bulge luminosity actually \textit{decreases}. This is due to the
S\'{e}rsic index $n$, which doubles compared to its value in the
single disk model -- while the bulge has a brighter central surface
brightness, the surface brightness decreases at a much faster rate,
resulting in a bulge that is fainter overall. The two disks have very
different thicknesses, with $\frac{h_{z,2}}{h_{z,1}}\sim6$. However,
the total luminosity of the two disks is very similar, an unusual
feature considering we generally find the thickest disk to be
significantly fainter than the thinner component(s).

The only region in Figure \ref{fig:ngc1055fitplot} that is still
poorly reproduced by the model with a bulge and two disks is near the
midplane, around 3-4 kpc in radius. It is not immediately clear what
is happening here; the excess only persists over $\sim$2 kpc, smaller
than what we associate with a disk component in NGC 4013 and 891, but
larger (and less peaked) than NGC~4565's ring.  Additionally, if this
excess is the sign of another disk, it would need to contain at least
an outer truncation, but given how well the two disk model fits the
midplane flux inside of 2 kpc this disk may also need to have an inner
truncation.  Indeed, when we add a third disk component we find the
fit prefers having both inner and outer truncations of roughly 1.5 and
4 kpc, respectively.  We list this model in the rightmost column of
Table \ref{tab:ngc1055bestfits}; it is represented as the red solid
curves in Figure \ref{fig:ngc1055fitplot}. This model still appears to
under-predict the mid-plane surface-bightness and over-predict the
light at large heights between radii of 2-5 kpc, which might indicate
a thinner truncated component would be preferable.

While not perfect, this third model does a fairly good job fitting
NGC~1055's entire light distribution. The AIC value for this
  model is 7\% lower than our initial mode, and 4\% lower than our
  second model. The scale-length of one of the disks minimizes to the
upper boundary of our parameter space (10 kpc), but given the overall
quality of the fit and the fact that increasing this parameter even
more would have a minimal effect on the radial regions we are able to
probe, increasing the range of our parameter space would have a
negligible impact on our results. The bulge contains over 50\% of the
total $\ks$-band luminosity, comparable to the single disk model, and
dominates the light profile along nearly every sightline. This model
is also notable for the sheer thinness of its disks. The
doubly-truncated component, with scale-height only 290 pc, is at least
twice as thick as the other two disks. Despite the limits of our
ability to model NGC~1055 given its disturbed morphology, the fits we
can make all point strongly to the presence of significant amounts of
starlight with $h_{z}\lesssim$200 pc.

\begin{deluxetable}{cccc}
\tablewidth{0pt}
\tablecaption{NGC~4244 Best-fitting Models}
\tablehead{ & \multicolumn{2}{c}{Value} &  \\ \cline{2-3}
  \colhead{Parameter$^{a}$} & \colhead{Disk Only$^{b}$} & 
\colhead{Disk+Nucleus$^{c}$} & \colhead{Units}}
\startdata
$\mu_{0,T}$ & 17.25$\pm$0.01 & 17.15$\pm$0.02 & mag arcsec$^{-2}$\\
$h_{R,T}$ & 1.80$\pm$0.01 & 1.84$\pm$0.01 & kpc\\
$h_{z,T}$ & 0.25$\pm$0.00 & 0.23$\pm$0.00 & kpc\\
$L_{T}$ & 3.09 $\times$ 10$^{9}$  & 3.19 $\times$ 10$^{9}$ & $L_{\odot,K}$ \\
$\mu_{0,nucleus}$ & \nodata & 19.05$\pm$0.02 & mag arcsec$^{-2}$\\
$R_{e}$ & \nodata & 0.46$\pm$0.00 & kpc\\
$a/b$ & \nodata & 1.21$\pm$0.02 & kpc\\
$n^d$ & \nodata & 0.53$\pm$0.02 & kpc\\
$L_{nucleus}$ & \nodata & 1.59 $\times$ 10$^{8}$& $L_{\odot,K}$ \\
$i$ & 86.53$\pm$0.05 & 85.81$\pm$0.10 & degrees\\
$L_{tot}$ & 3.09 $\times$ 10$^{9}$& 3.35 $\times$ 10$^{9}$& $L_{\odot,K}$\\
$\chi_{\nu,inner}^{2}\,^{e}$& 2.44& 2.17& \nodata\\
$\chi_{\nu,outer}^{2}\,^{e}$& 1.73*& 1.73& \nodata\\
$\chi_{\nu,total}^{2}\,^{e}$& 1.86& 1.81*& \nodata
\enddata
\tablenotetext{a}{$T$ denotes parameters that we associate with the
  thin disk.}
\tablenotetext{b}{Only fit to regions of the galaxy with $R>1$ kpc.}
\tablenotetext{c}{Fit performed over entire radial range of data.}
\tablenotetext{d}{Sersic index.}
\tablenotetext{e}{Reduced $\chi^{2}$. {\it inner} and {\it outer} correspond
  to the regions inside and outside of the 1 kpc fitting cutoff, while {\it
    total} is for both regions together. (*) denotes the $\chi_{\nu}^{2}$ that was
  minimized by each model.}
\label{tab:ngc4244bestfits}
\end{deluxetable}

\begin{figure*}
\centering
\includegraphics[scale=0.7]{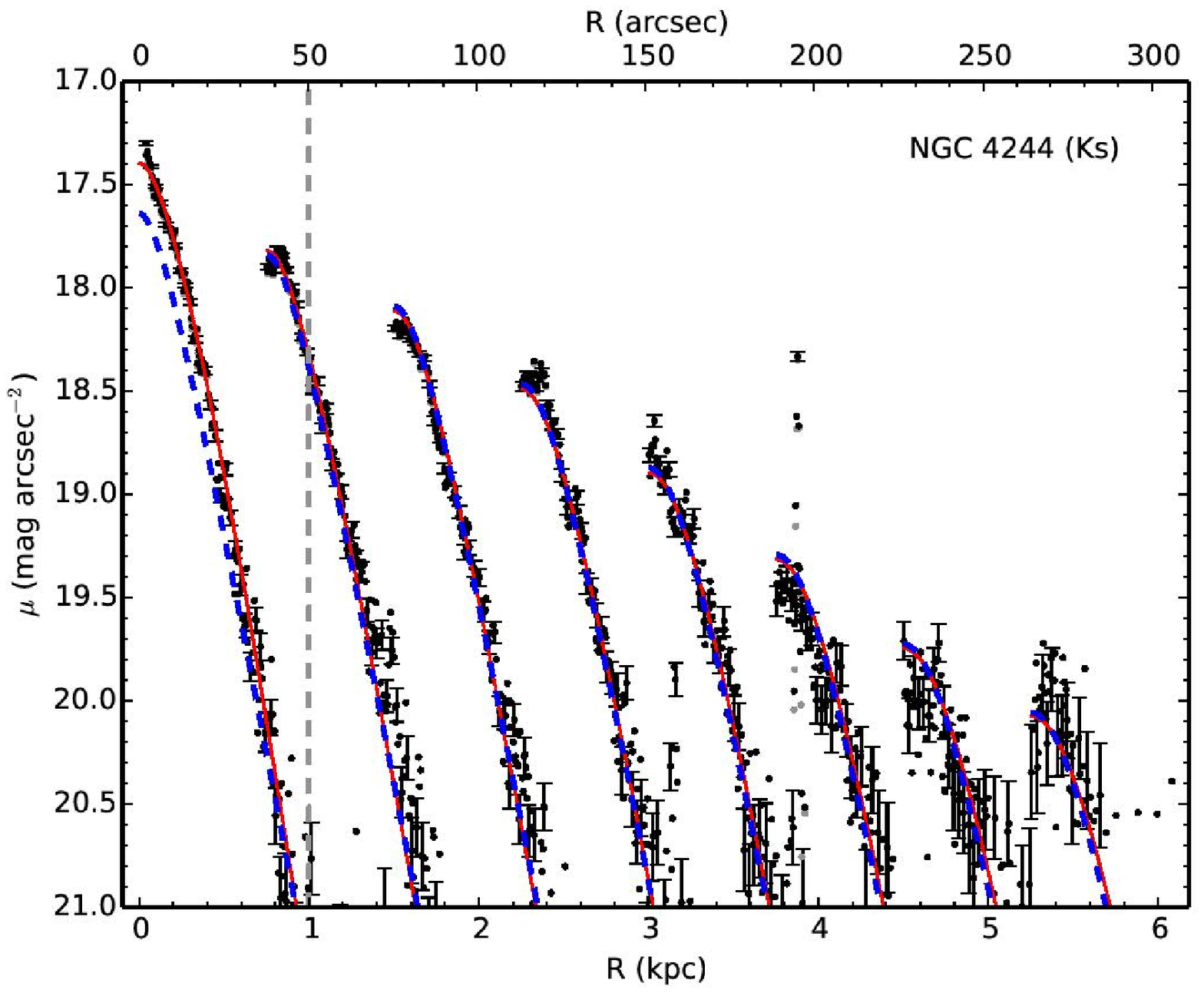}\\
\includegraphics[scale=0.7]{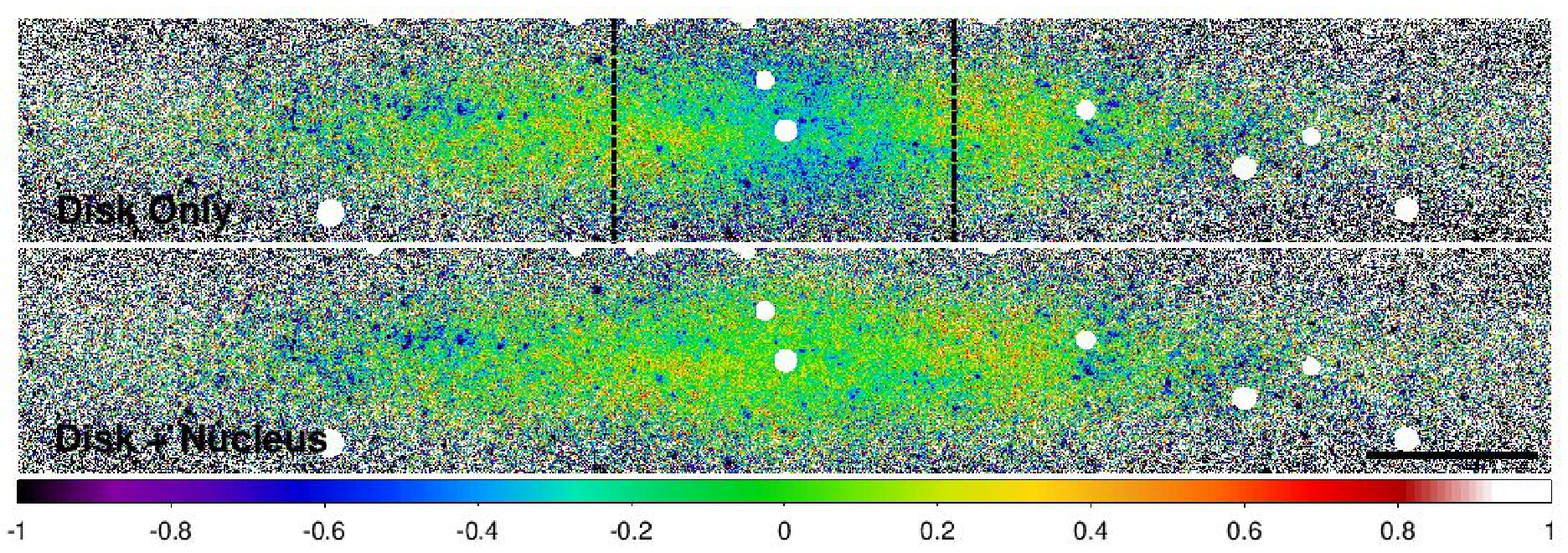}
\caption{Similar to Figure \ref{fig:ngc891fitplot}, but for
  NGC~4244. The blue dashed lines show the best fitting single-disk
  model, fit to all regions outside of 1 kpc in radius (denoted by the
  vertical gray dashed line). The solid red lines show the best
  fitting single disk + S\'{e}rsic bulge model, which was fit at all
  radii.  To improve S/N the data in the 1D profiles have all been binned to 10'' x
  0\farcs3 ($R \times z$). Details of the model parameters are in
  Table \ref{tab:ngc4244bestfits}.}
\label{fig:ngc4244fitplot}
\end{figure*}

\subsubsection{NGC~4244}
\label{sec:ngc4244}

Because this galaxy is essentially bulgeless \citep[save for a small
  nuclear star cluster;][]{Comeron11c} we only need to mask out the
innermost 1 kpc to avoid contamination.  A single disk is the only
component necessary to fit our NGC~4244 data outside of 1 kpc, as
shown as the blue dashed lines in Figure \ref{fig:ngc4244fitplot} and detailed
in the left column of Table
\ref{tab:ngc4244bestfits}. While there is currently an ongoing debate
about whether NGC~4244 has a thick disk (see \citealt{Comeron11c} for
a discussion), we note that our data is not sensitive enough to
measure such a disk even if it does exist. Our best-fitting disk is
almost identical to that found using optical data for NGC~4244
(\citealt{Fry99} and \citealt{Seth05}), further evidence that NGC~4244
is nearly dust-free. Our fitted inclination falls between the
literature measurements of \citet{Olling96} and
\citet{Zschaechner11}.

We next re-fit NGC~4244, adding a S\'{e}rsic bulge model to the single
disk to fit the nuclear star cluster.  Because of the unresolved
nature of the central star-cluster core and the PSF mismatch between
IRAC and WHIRC images, the attenuation correction fails at the very
center of NGC~4244. We therefore avoid fitting the inner 3.5 arcsec,
or $\sim$70 pc in radius. Since the fitted nucleus has a half-light
radius of 460 pc, the excluded light given the profile shape should
only be $\sim2$\% of the total luminosity of the nucleus component.
The best-fitting model for these additional free parameters
well-reproduces the observed surface brightness distribution, and is
shown alongside our initial model as red solid lines in Figure
\ref{fig:ngc4244fitplot} and in the right column of Table
\ref{tab:ngc4244bestfits}; the AIC value for this is reduced from
  our initial model by about 3\%. The resulting nucleus-to-disk
luminosity ratio is 0.05, similar to that found in the optical
\citep[0.02,][]{Fry99}.

\begin{deluxetable}{cccc}
\tablewidth{0pt}
\tablecaption{NGC~4144 Best-fitting Models}
\tablehead{ & \multicolumn{2}{c}{Value} &  \\ \cline{2-3}
  \colhead{Parameter$^{a}$} & \colhead{One Disk$^{b}$} &
\colhead{Disk+Bulge$^{c}$} & \colhead{Units}}
\startdata
$\mu_{0,T}$ & 17.78$\pm$0.01& 19.28$\pm$0.01& mag arcsec$^{-2}$\\
$h_{R,T}$ & 0.98$\pm$0.01& 2.19$\pm$0.02& kpc\\
$h_{z,T}$ & 0.35$\pm$0.00& 0.55$\pm$0.00& kpc\\
$L_{T}$ & 1.45 $\times$ 10$^{9}$& 1.28 $\times$ 10$^{9}$& $L_{\odot,K}$\\
$\mu_{0,bulge}$ & \nodata& 17.64$\pm$0.00& mag arcsec$^{-2}$\\
$R_{e}$ & \nodata& 1.07$\pm$0.01& kpc\\
$a/b$ & \nodata& 3.54$\pm$0.01& kpc\\
$n^{d}$ & \nodata& 0.90$\pm$0.00& kpc\\
$L_{bulge}$ & \nodata& 6.42$\times$ 10$^{8}$& $L_{\odot,K}$\\
$i$ & 85.18$\pm$0.09 & 88.32$\pm$0.00& degrees\\
$L_{tot}$ & 1.45 $\times$ 10$^{9}$& 1.92 $\times$ 10$^{9}$& $L_{\odot,K}$\\
$\chi_{\nu,inner}^{2}\,^{e}$& 0.66& 0.48& \nodata\\
$\chi_{\nu,outer}^{2}\,^{e}$& 0.48*& 0.49& \nodata\\
$\chi_{\nu,total}^{2}\,^{e}$& 0.58& 0.48*& \nodata
\enddata
\tablenotetext{a}{$T$ denotes parameters that we associate with the thin disk.}
\tablenotetext{b}{Only fit to regions of the galaxy with $R>1$ kpc.}
\tablenotetext{c}{Fit performed over entire radial range of data.}
\tablenotetext{d}{Sersic index.}
\tablenotetext{e}{Reduced $\chi^{2}$. {\it inner} and {\it outer} correspond
  to the regions inside and outside of the 1 kpc fitting cutoff, while {\it
    total} is for both regions together. (*) denotes the $\chi_{\nu}^{2}$ that was
  minimized by each model.}
\label{tab:ngc4144bestfits}
\end{deluxetable}

\subsubsection{NGC~4144}
\label{sec:ngc4144}

\begin{figure*}
\centering
\includegraphics[scale=0.6]{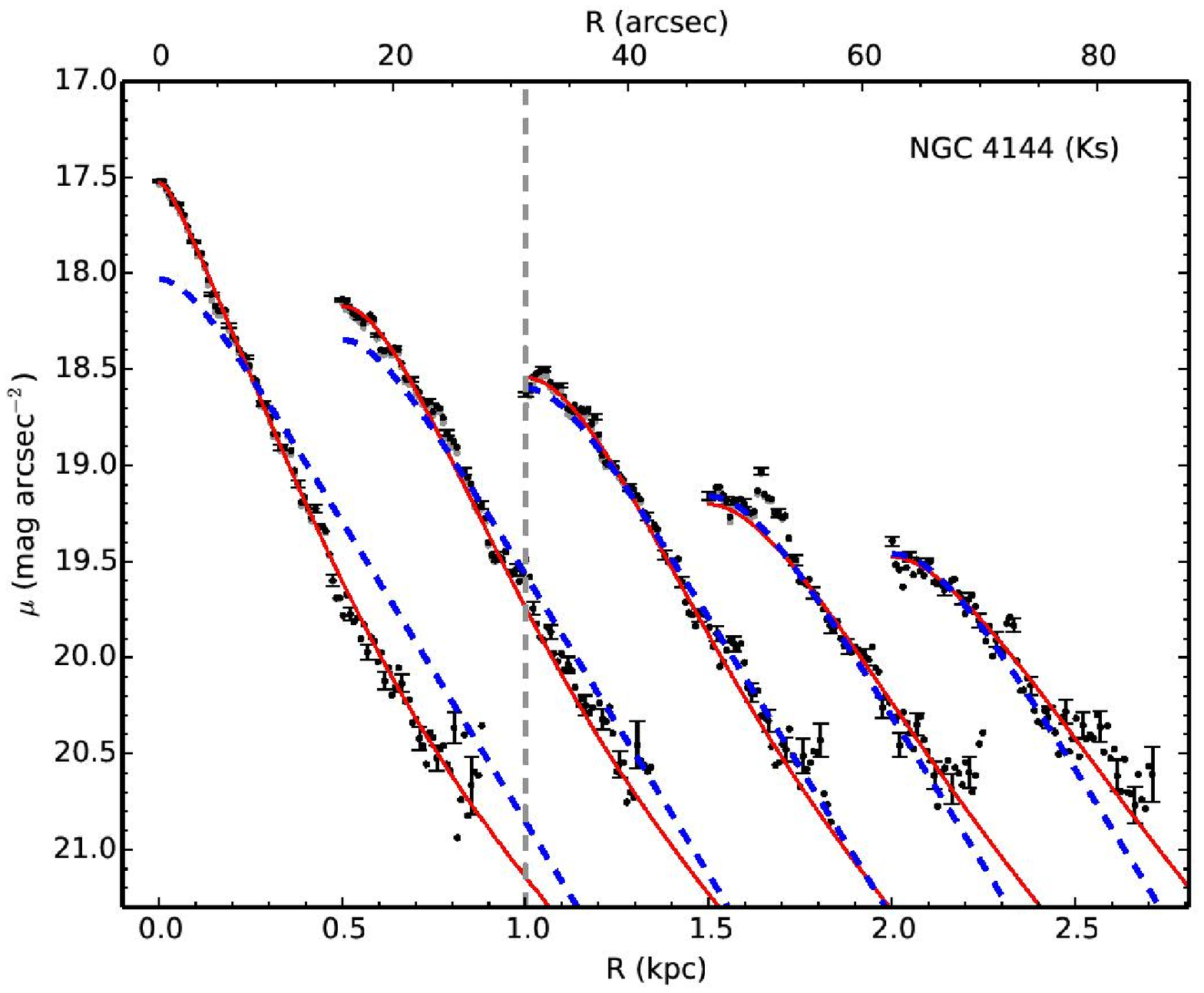}\\
\includegraphics[scale=0.6]{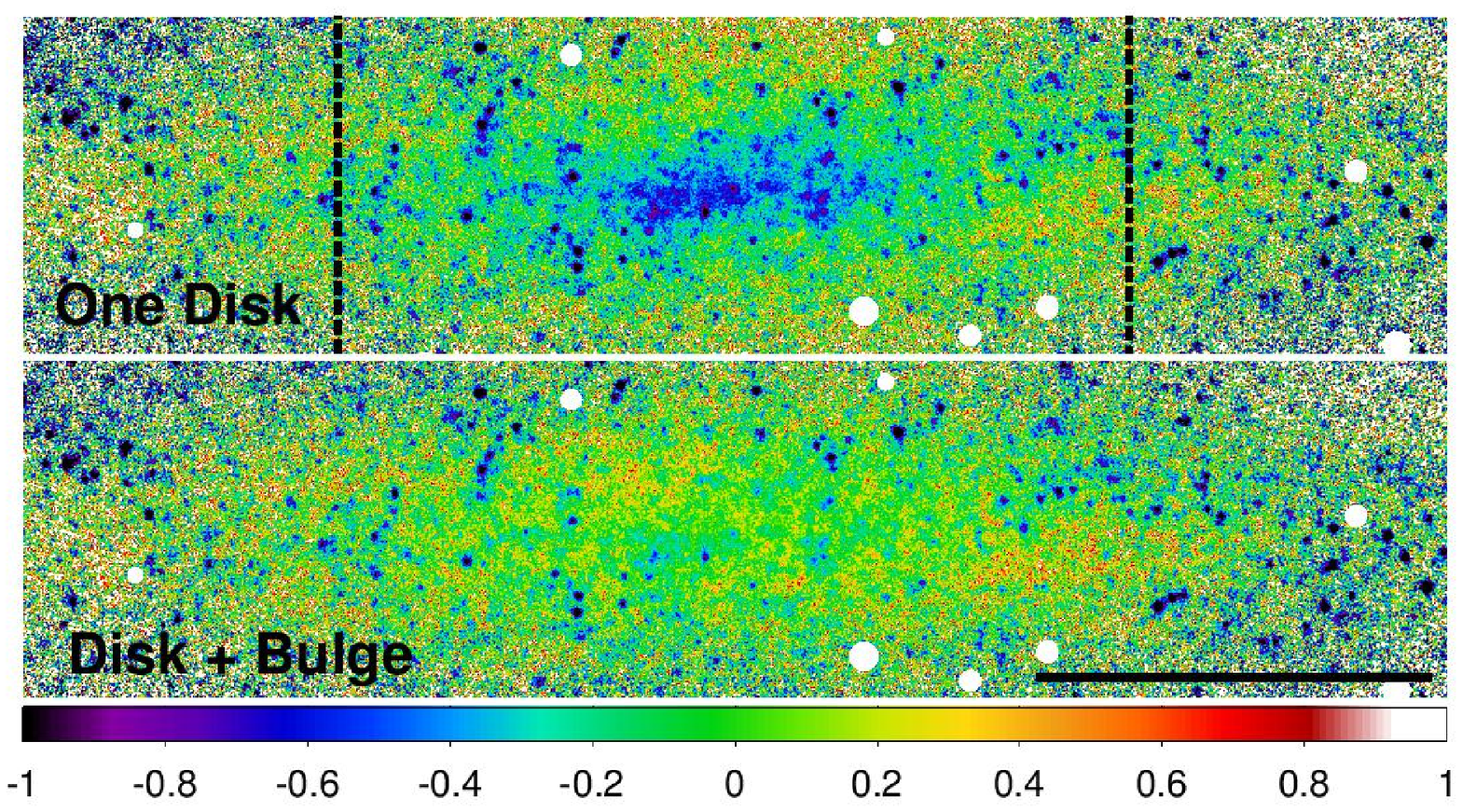}
\caption{Similar to Figure \ref{fig:ngc891fitplot}, but for
  NGC~4144. The blue dashed lines show the best fitting single-disk
  model, fit to all regions outside of 1 kpc in radius (denoted by the
  vertical gray dashed line). The solid red lines show the best
  fitting single disk + S\'{e}rsic bulge model, which was fit at all
  radii. To improve S/N the data in the 1D profiles have all been binned to 10'' x
  0\farcs3 ($R \times z$). Details of the model parameters are in
  Table \ref{tab:ngc4144bestfits}.}
\label{fig:ngc4144fitplot}
\end{figure*}

Modeling NGC~4144 is challenging because at large radii there is a
clear asymmetry in the light distribution, visible both in our WHIRC
and SDSS images (Figure \ref{fig:gri_jhk_sample}). This asymmetry
appears to increase at larger radii. Therefore we choose to perform
our fits restricted to be within a projected radius of 2 kpc from the
galaxy center. As it turns out, given the faintness of this galaxy
there is little data outside this area that would add significant
additional constraints on our models. Even with this restriction the
disk appears to flare, with larger radii having flatter vertical
profiles (Figure \ref{fig:ngc4144fitplot}).

A single disk fit, restricted to the region between 1 and 2 kpc in
radius and shown as the blue dashed lines in Figure
\ref{fig:ngc4144fitplot} and the leftmost column of Table
\ref{tab:ngc4144bestfits}, does a generally poor job of fitting the
data. At small radii this model underpredicts the light at small
heights and overpredicts the light at large heights.  Even at larger
radii, where the model is directly constrained, it underpredicts the
high-lattitude light. The telltale flattening of the radial midplane
profile we see in NGCs 891 and 4013 (e.g. the ridgeline of the
profiles in Figure \ref{fig:ngc4013fitplot}) does not exist here. This
indicates that an inner disk truncation is not a viable solution.

Based on a visual inspection NGC~4144 does not appear to have a
significant bulge component. Consequently, we next attempted to fit a
two-disk model to the entire radial range of the galaxy. Such a fit
results in low inclination ($\sim76^\circ$), a very thin inner disk
($h_{z}=30$ pc), and an exceptionally thick outer disk with $h_R/h_z
< 1$. The two-disk model does an excellent job fitting the mid-plane
light at small radii, but increasingly under-predicts the light near
the mid-plane at larger radii. However, without any argument about the
astrophysical plausibility of such an unusual disk, we can dismiss
this model simply because it predicts far too much light at large
scale-heights. For this reason we do not tabulate or plot the
fit. Nonetheless, the model results are informative regarding
limitation in modeling late-type, slow-rotating galaxies without
well-defined mid-plane dust-lanes, as follows.

Literature values for NGC~4144's inclination (e.g., 82 degrees in
\citealt{Rhee96} from LEDA, and 86.2 degrees from the current
measurements in HyperLEDA, both based on optical axial ratios and
assumed disk axial ratios) are comparable to what we found in our
single-disk model, but are significantly higher than what we found in
our two-disk model.  Visual inspection of archival HST
images\footnote{From the Hubble Legacy Archive at
  http://hla.stsci.edu/cgi-bin/display?image=HST\_9765\_19\_\linebreak[0]ACS\_WFC\_F814W\linebreak[4]\%2CHST\_\linebreak[0]9765\_19\_ACS\_WFC\_F606W.}
yields the impression that this galaxy could indeed be at an
inclination in the range of 70 to 80 degrees, based on the
intermittent ring of star-formation regions and patchy dust in the
nuclear regions.  However, without a clear morphological features such
as a coherent, thin dust-lane, the impression is inconclusive. What is
clear is that the thinness of the inner disk in the two-disk model is
only possibly in a model at relatively low inclination.  This reflects
the inherent degeneracy between separating intrinsic oblateness from
inclination angle.

Given the thickness of the second disk in the two-disk model, we next
turned to a disk plus bulge model. We found a single disk and a
S\'{e}rsic bulge were able to fit the data at all radii and heights
(shown as the red solid line in Figure \ref{fig:ngc4144fitplot})
better that the previous models, and with no failings. The AIC
  value for this model is 16\% lower than for our initial model. The
fitted parameters of this model are listed in the rightmost column of
Table \ref{tab:ngc4144bestfits}. The bulge is quite flattened
($a/b$=3.54) and has a S\'{e}rsic index near unity, making it similar,
but not identical to, an exponential disk with the same
oblateness. While the bulge has a brighter central surface brightness,
the disk component has twice the luminosity of the bulge due to its
more extended light distribution.

Based on this two-component model, NGC~4144's disk has a fairly large
scale-height of 550 pc, roughly twice that estimated from other
studies in the near-infrared. For example, \citealt{Seth05} find a
$\ks$-band scale-height of 228.5 pc, assuming an inclination of 83
degrees. This discrepancy could be due to contamination from the bulge
component, which contains a substantial amount of the total $\ks$-band
luminosity, but most likely reflects the degeneracy between
inclination and disk oblateness noted above. For example, in our disk
plus bulge model the best-fitting inclination is $\sim 88$ degrees. We
also note that some models fits with comparable $\chi^{2}$ to our
best-fitting model but have $h_{z}\sim100$ pc and inclinations closer
to 81 degrees. We would caution against interpretation of disk
scale-heights in this or other work until a better understanding of
the three-dimensional geometry of this non-asymmetric galaxy is
understood.

\section{Discussion}
\label{sec:discussion}

\subsection{Attenuation Profiles}
\label{sec:fastslowae}

\begin{figure}
\epsscale{1.3}
\plotone{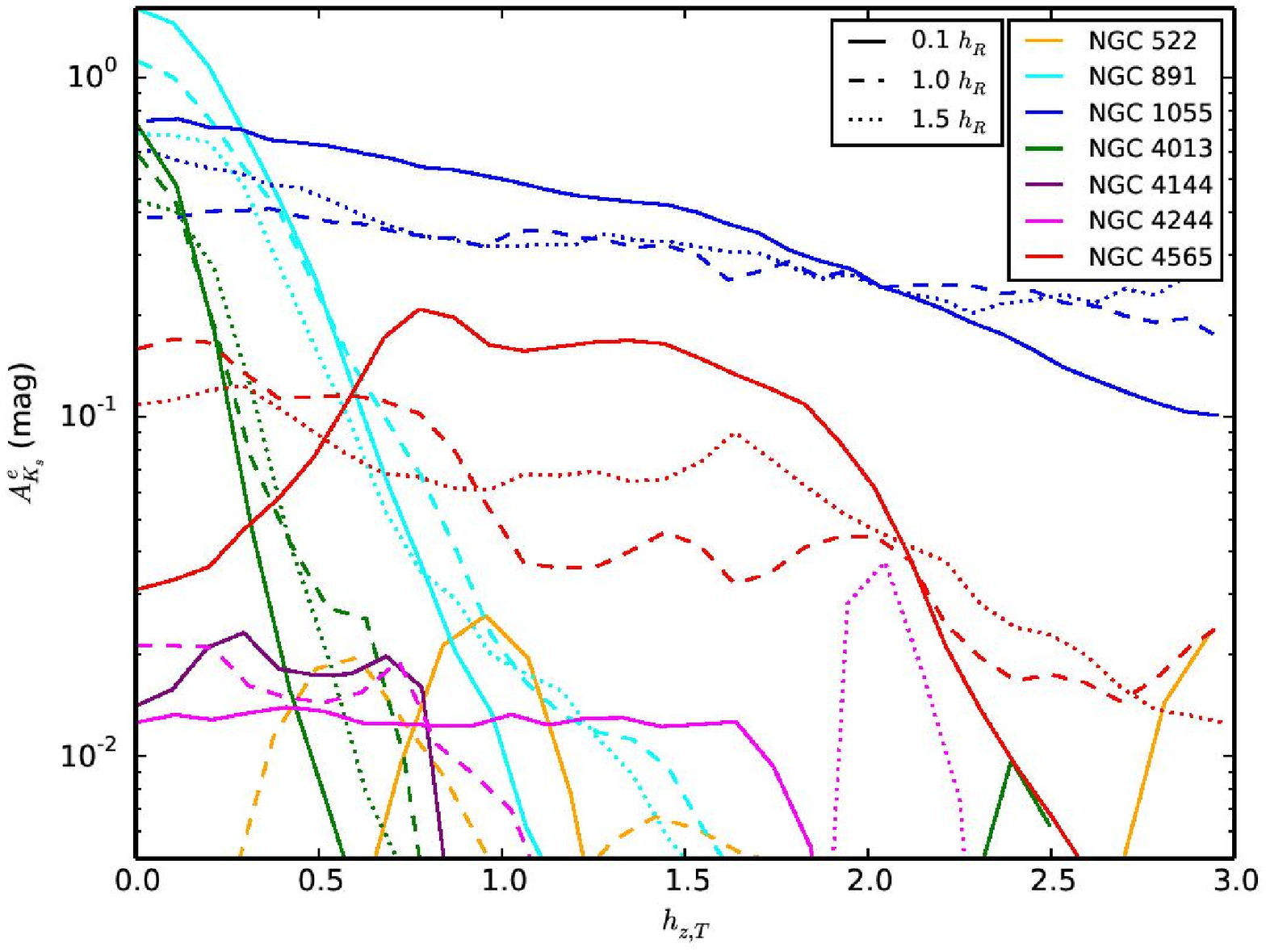}
\caption{Projected vertical $\ks$-band attenuation profiles as a
  function of best-fitting thin disk scale-height. Profiles are shown
  at multiple thin disk scale-lengths to facilitate comparison between
  the galaxies.}
\label{fig:allattenprofs}
\end{figure}

For the six galaxies analyzed in this work, an inspection of
Figures~\ref{fig:ngc891attencorr} through \ref{fig:ngc4144attencorr}
(excluding NGC~522) shows there is a clear distinction between the
fast-rotating systems and the slow-rotators. The fast-rotators contain
a fair amount of attenuation ($\gtrsim$0.6 mag near the midplane),
while the two slow-rotating spirals have almost zero $\ks$-band
attenuation. In Figure \ref{fig:allattenprofs} we compare the vertical
$\ks$-band attenuation profiles of our sample. The data have been
scaled by the scale lengths and heights of what we call the `thin'
disk in our best-fitting models (the rightmost fits in Tables
\ref{tab:ngc891bestfits}-\ref{tab:ngc4144bestfits}).\footnote{There is
  one exception to this: due to the odd behavior of the two brightest
  disks in our most complex NGC~1055 model, we choose to use the
  scale-height of disk 1 from the two disk model instead.}  While not
a perfect solution given the complexity of fitting some of these
galaxies and the fact that most are not pefectly edge on, this scaling
should allow for less biased comparisons than purely physical units
like kpc, as previously shown in Figures \ref{fig:ngc891attencorr} to
\ref{fig:ngc4144attencorr}.

Despite the rough distinction between fast and slow rotators, nearly
every galaxy in Figure \ref{fig:allattenprofs} has a visually distinct
attenuation profile {\it shape}. To first order, this is an
inclination effect. It is not a coincidence that the two nearly
perfectly edge-on galaxies NGC~891 and 4013 both have centrally-peaked
profiles with steep dropoffs from the midplane to nearly zero
attenuation by 1 $h_{z,T}$. Relative to NGC~4013, NGC~891 simply has
more attenuation.

For slighly less inclined systems, NGC~4565's more extended
attenuation profile shows how much more transparent massive spiral
galaxies become only $\sim$3$^{\circ}$ away from edge-on \citep[see
  Figure 17 of][]{Schechtman-Rook12}.  NGC~4565 has only a couple of
small regions with $\aeks\ge0.5$ mag.  Much of the dust in this galaxy
appears to be concentrated in a narrow ring, clearly visible in Figure
\ref{fig:ngc4565attencorr}, and evident in the attenuation profile at
0.1 $h_R$. This ring has been previously identified in molecular gas
\citep{Sofue94} as well as in dust emission
\citep{Kormendy10,deLooze12}. This dust morphology likely indicates
NGC~4565 is not as similar to the other fast-rotators as would be
assumed from the un-corrected images, and in fact it is the only
galaxy in our sample that is known to have an active nucleus (e.g.
\citealt{Chiaberge06}). This activity may be responsible for the
unusual {\it decrease} in attenuation at small radii near the
mid-plane.

NGC~1055, in contast, despite being at similar inclination as
NGC~4565, has much more dust attenuation even though it has only half
the neutral gas mass (cf. \citealt{Thilker07} and Dahlen et al. 2005).
NGC~1055 also has more attenuation at larger heights than any of the
other galaxies in our sample.  Since we find both galaxies have
similar inclination, NGC~1055's extended dust distribution is likely
to be due to significant reservoirs of extraplanar dust and not merely
the result of viewing this galaxy with $i<90^{\circ}$. This is
interesting because NGC~1055 also has disturbed morphology which may
indicate recent interactions are the cause of this extended dust
layer.

\begin{deluxetable*}{c cccc c cccc c cccc c cccc}
\tablewidth{0pt} \tablecaption{Luminosity Distribution Summary$^a$}
\tablehead{ & \multicolumn{4}{c}{Super-thin Disk} &&
  \multicolumn{4}{c}{Thin Disk} && \multicolumn{4}{c}{Thick Disk} &&
  \multicolumn{4}{c}{\%$L$} \\ \cline{2-5} \cline{7-10} \cline{12-15}
  \cline{17-20} \colhead{NGC} & \colhead{$\mu0$} & \colhead{$h_R$} &
  \colhead{$h_z$} & \colhead{\%$L$} && \colhead{$\mu0$} &
  \colhead{$h_R$} & \colhead{$h_z$} & \colhead{\%$L$} &&
  \colhead{$\mu0$} & \colhead{$h_R$} & \colhead{$h_z$} &
  \colhead{\%$L$} && \colhead{Ring} & \colhead{Nuclear Disk} &
  \colhead{Bar} & \colhead{Bulge} } 
\startdata 
891 & 14.4 & 3.4 & 0.16 & 22 && 15.9 & 5.7 & 0.47 & 43 && 18.5 & 4.8 &
1.44 & 10 && 0 & 6 & 19 & 0 \\
4013 & 14.5 & 2.3 & 0.21 & 26 && 16.0 & 2.8 & 0.60 & 28 && 19.0 & 4.0
& 2.96 & 14 && 0 & 3 & 29 & 0 \\
4565 & \nodata & \nodata & \nodata & 0 && 15.4 & 6.0 & 0.35 & 40 &&
18.9 & 5.4 & 2.23 & 9 && 18 & 4 & 0 & 29 \\
522 & \nodata & \nodata & \nodata & 0 && 15.5 & 7.7 & 0.38 & 40 &&
18.0 & 9.6 & 1.22 & 38 && 0 & 0 & 13 & 9 \\
1055 & 14.0 & 1.6 & 0.10 & 15 && 16.0 & 10.0 & 0.15 & 23 && 15.5 & 3.4
& 0.29 & 6 && 0 & 0 & 0 & 56 \\
4244 & 17.2 & 1.8 & 0.23 & 95 && \nodata & \nodata & \nodata & 0 &&
\nodata & \nodata & \nodata & 0 && 0 & 0 & 0 & 5 \\
4144 & \nodata & \nodata & \nodata & 0 && 19.3 & 2.2 & 0.55
& 67 && \nodata & \nodata & \nodata & 0 && 0 & 0 & 0 & 33 
\enddata
\tablenotetext{a}{Surface brightness values ($\mu0$) are $\ks$-band
  mag arcsec$^{-2}$, scale-lengths and scale-heights have units of
  kpc, and the luminosity percentage of each component (\%$L$) is
  relative to the total modeled luminosity in the $\ks$-band. All
  models shown here can be found in the right-most columns of Tables
  \ref{tab:ngc891bestfits}-\ref{tab:ngc4144bestfits}. Disks for
  NGC~1055 are arranged in order of increasing $h_{z}$.}
\label{tab:summary}
\end{deluxetable*}

\subsection{Vertical Color Gradients and Disk Heating}
\label{sec:colgrad}

The evidence from our analysis is that most of the galaxies in our
sample exhibit near-infrared $(J-\ks)$ color gradients in their
vertical light profiles, and in some galaxies these vertical gradients
change with radius.  Color gradients encode information about changes
in stellar populations. The existence of such gradients presents the
question of how they arise and if, for example, the gradients support a
picture of dynamical heating of the stellar disks. Correlation of
these gradients with the multi-component disk structure is informative
in this regard.

In general, color gradients are rather modest at large heights
($z>0.5$ kpc) in the fast rotators and at all heights in the slow
rotators; colors in these regions {\it at a given height} are
remarkably similar at all radii for any given galaxy. Distinguishing
between age and metallicity is premature without further
spectrophotometric diagnostics, but the smooth disk light distribution
and the lack of radial structure at $z>0.5$ kpc in fast rotators
(evident in
Figures~\ref{fig:ngc891attencorr}-\ref{fig:ngc522attencorr}) suggests
the stellar populations are dynamically relaxed, and hence
old. Stellar population synthesis models indicate that metallicity is
the primary driver of colors at these ages (Figure~\ref{fig:jkzage}),
but we cannot rule out a changing admixture of young and old
populations with height.

At smaller heights the fast-rotator color gradients have much more
structure, changing both with radius and between galaxies.  Moving
from large heights toward the mid-plane the gradients generally become
steeper between $0.2 {\rm kpc} < z < 0.5 {\rm kpc}$, and then flatten
at $z<0.2$ kpc, in some cases even getting bluer near the mid-plane
(NGC~891 and 4565). The reddest colors tend to be found near $z=0.2$
kpc, which is a factor of two above our physical resolution limit even
for NGC~522, our most distant source. The reddest colors in this
region are quite high, with $(J-\ks)>1$, which can only be matched in
stellar population synthesis models with the presence of a large
contribution from intermediate-age (0.2-2 Gyr) populations.

The two fast-rotating galaxies with super-thin disks (NGC~891 and
4013) also show the most dramatic changes in vertical gradients with
radius. In both galaxies, the reddest colors near $z=0.2$ kpc are
found in the inner regions ($R<3 $kpc), where the disk components are
truncated and their bars dominate the projected light. As a
consequence of their very red colors at intermediate heights, these
regions have the strongest gradients toward bluer colors at larger
heights ($z>0.2$ kpc). The mid-plane colors never get bluer than
$(J-\ks)=1$, while the reddest colors are between $1.2 < (J-\ks) <
1.4$. These very red colors are indicative only of near-infrared
star-light dominated by intermediate-age populations (see
Section~\ref{sec:galaxyapcolors} and Figure~\ref{fig:jkzage}).  This
suggests that the bars in these two galaxies are dynamically young and
that their creation rapidly shut down star-formation in the inner
regions. Bluer mid-plane colors, which fall below $(J-\ks)=1$ in
NGC~891, are found in the radial region where super-thin disks are
present in these galaxies, and indeed the gradients diminish above
$z=0.2$ kpc.

For the fast-rotating galaxies without major super-thin components, we
still note color gradients that correlate with their disk structure.
The small, super-thin nuclear disk in NGC~4565 does appear to lead to
bluer colors near the mid-plane at small radii. In NGC~522, vertical
color gradients are largest at intermediate radii ($6 {\rm kpc} < R < 8
{\rm kpc}$), where the thin disk contributes most to the total disk
luminosity.

It is also interesting that the one exception to the discussion of
vertical color gradients in fast-rotators is NCG~1055, which has
disturbed morphology indicative of a recent or on-going interaction.
It is well known that minor mergers heat disks in galaxy simulations
(e.g., \citealt{Walker96}). The clear lack of a vertical color
gradient in this galaxy is consistent with this merger picture, in
which any gradients that may have been present prior to the merger are
removed from the mixing effects of the relatively more impulsive disk
heating from a single merger event. The {\it absence} of a vertical
color gradient in this galaxy is also verification that the {\it
  presence} of vertical color gradients in our other fast rotators are
not artifacts of our attenuation correction. Despite the broader apparent vertical distribution of dust in this galaxy seen in
Figure~\ref{fig:allattenprofs}, due in part to its non-edge-on
orientation and its disturbed morphology, there is a
significant gradient in the attenuation with height. Indeed, the {\it
  uncorrected} colors do show a pronouced gradient
(Figure~\ref{fig:ngc1055attencorr}). The lack of a thick disk
component for this galaxy is somewhat puzzling in the context of a
minor merger or interaction. We note the model degeneracy between disk
thickness and inclination, discussed in the context of NGC~4144, may
be an issue here as well, although for NGC~1055 there are good
constraints on the thin inclination from the dust geometry.  Overall,
the disk system of this galaxy has near-infrared colors that are
consistent with either a very metal rich stellar population or one
dominated by intermediate-age stars. It may be that the disk here is
relatively young, having formed recently after a more major merger
that produced or contributed to the significant bulge component.

The picture emerging from this discussion is that the disk structure
parameterized in Section~\ref{sec:model} does indeed correlate with,
and offer insight on, the color gradients found in
Section~\ref{sec:galaxyapcolors}. The general behaviour is consistent
with a dynamical heating model where the mid-plane is dominated by the
youngest populations, intermediate age stars dominate the light at
intermediate heights, while older stellar populations dominate at
large heights. With additional spectrophotometric constraints it may
be possible to use vertical gradients in stellar populuations as
chronometers of relatively recent dynamical heating of disks.

\subsection{Disk Oblateness and Luminosity Ratios: Implications for Disk Maximality}
\label{sec:lumrat}

\begin{deluxetable*}{rllcccc}
\tablewidth{0pt} \tablecaption{Infrared AB Colors$^a$}
\tablehead{
\colhead{NGC} & \colhead{$L_K$} & 
\colhead{$M_{\rm HI}/L_K$} &
\colhead{$K-m_{24}$} &
\colhead{$K-m_{25}$} &
\colhead{$K-m_{60}$} &
\colhead{$m_{60}-m_{100}$} \\
\colhead{} & \colhead{($10^{10} L_\odot$)} & 
\colhead{($M_\odot$/L$_\odot$)} &
\colhead{} & \colhead{} & \colhead{} & \colhead{}
}
\startdata
891  & 13.3  & 0.044 & 0.312     & 0.40     & 2.89     & 1.02 \\
4013 & \! \ 8.80  & 0.049 & $\cdots$ & -0.08     & 2.30     & 1.34 \\
1055 & 14.1  & 0.040 & $\cdots$ & 0.57     & 2.89     & 1.11 \\ 
4565 & 15.3  & 0.098 & -0.86     & -0.83     & 0.82     & 1.64 \\
522  & \! \ 9.66  & 0.011 & $\cdots$ & $\cdots$ & 1.55     & 1.31 \\
4244 & \! \ 0.335 & 0.31  & -0.38     & $\cdots$ & $\cdots$ & $\cdots$ \\
4144 & \! \ 0.192 & 0.25  & -0.26     & $\cdots$ & $\cdots$ & $\cdots$
\enddata
\tablenotetext{a}{$L_K$ and $f_K$ are the best-fitting model {\it
    total} luminosity and apparent flux (adopting distances in
  Table~\ref{tab:obslog}) in the $\ks$ band from this study (see
  Tables \ref{tab:ngc891bestfits} - \ref{tab:ngc4144bestfits}).
  Remaining fluxes are from the literature (see text): $m_{24}$ is the
  $24\mu{\rm m}$ magnitude measured using the Spitzer MIPS instrument;
  $m_{25}$, $m_{60}$, $m_{100}$ are IRAS 25, 60 and 100 $\mu{\rm m}$
  magnitude measurements, respectively.}
\label{tab:fluxrat}
\end{deluxetable*}

To facilitate comparisons between galaxy disk structural parameters we
gather together the surface-brightness, scale-lengths, scale-heights
and relative $\ks$-band luminosity fractions from our best fitting
models from Tables \ref{tab:ngc891bestfits}-\ref{tab:ngc4144bestfits}
in Table \ref{tab:summary}. The Table suggests rough definitions of a
``super-thin disk'' having $h_z<250$pc, a ``thin disk'' having $300 <
h_z < 600$pc, and a ``thick disk'' having $h_z > 1$kpc for this galaxy
sample--not unreasonable in the context of the MW.

While the two slow rotators have very different distributions of light
between their disk and bulge components, overall their total
luminosities are fairly similar (within a factor of two).
Unsurprisingly, the fast rotators are all at least 25 times more
luminous than the slow rotators. These are all reflections of the
well-known scaling relation between luminosity and rotation speed.
NGC~1055 is especially interesting, as while it is one of the slower
fast-rotators in our sample it has a luminosity more consistent with
galaxies rotating $\gtrsim$30 km sec$^{-1}$ faster. This is largely
due to the light contribution from NGC~1055's prodigious bulge, which
accounts for over 50\% of the total $\ks$-band light in the galaxy.

All of the fast-rotating galaxies have multi-component disks. Excluding
NGC~1055's disturbed morphology, the other four (NGCs 891, 4013,
4565, and 522) show a general trend toward thicker components having
larger scale lengths, but with {\it decreasing} axial ratios
$h_R/h_z$. In other words, larger and thicker disks are less
flattened.  This may reflect the relative efficiency of disk heating
in the radial and vertical dimensions. The conjecture could be tested
by measuring the radial stellar velocity dispersion as a function of
scale height either directly or through a dynamical proxy such as
asymmetric drift between the tangential motion of gas and stars.

We focus now on just these four fast rotators since their disk
structural parameters are likely most relevant to the DiskMass Survey
calibration of $h_R/h_z$ (Bershady et al. 2010b). The disk components
in these galaxies, including nuclear disks and NGC~4565's ring,
contribute $75\pm5$ (we quote mean and full range throughout this
section) to their total $\ks$ luminosity. Of this, the
nuclear disks only contribute $3\pm3$\%.

We find $h_R/h_z$ values of $16^{+5}_{-6}$, $13^{+7}_{-9}$, and
$4^{+4}_{-2}$ for the super-thin, thin, and thick disks (respectively). In the context of the DMS, these values are still too
small to imply maximal disks based on the observed, face-on stellar
velocity dispersions (Bershady et al. 2011). Moreover, for the two
disks with super-thin components (NGCs 891 and 4013), the thin and
thick components have axial ratios $h_R/h_z$ of $6.5\pm2$, and
$2.5\pm1$. The small axial ratios for these two galaxies are a
reflection of the increased oblateness of the {\it thin} components in
the galaxies that are {\it without} super-thin components, which we
discuss in the next section.

Thin disks contribute $37^{+4}_{-9}$\% to the total $\ks$ luminosity
while thick disks contribute $18^{+20}_{-8}$\%, so it is possible to
characterize their contribution in luminosity roughly as $L_T \sim 2
L_{Th}$.  For the two galaxies that have super-thin disks (NGCs 891
and 4013), those components contribute $24\pm2$\% to the total $\ks$
luminosity. This translates into roughly $L_T + L_{Th} \sim 2 L_{ST}$
for these systems.

\subsection{Presence of Super-Thin Disks}
\label{sec:std}

\subsubsection{Slow-Rotators}

Given their well-known lack of narrow dust lanes \citep{Dalcanton04}
and the correlation between dust and star formation
\citep[e.g.][]{Boquien11}, it is perhaps unsurprising to find that
NGC~4244 and 4144 have little-to-no evidence for the existence of a
MW-like star-forming super-thin disk. However, NGC~4244's disk is
moderately thin in relative and absolute terms given its vertical
scale height of 230 pc and oblateness ($h_R/h_z = 8$). NGC~4144's
light profile is more complicated, but it seems clear that neither
galaxy contains the nested super-thin+thin+thick disk structure we
find in most of our fast-rotating galaxies. This does \textit{not} mean that slow
rotating galaxies cannot form very thin and flattened disks;
UGC~7321 for instance, a member of the class of ``super-thin''
low-surface brightness disk galaxies, has a stellar disk scale-height
of $\sim$150 pc but $V_{rot}<100$ km sec$^{-1}$ and no distinct dust
lane \citep{Matthews00}.

\subsubsection{Fast-Rotators}

The results for all of the fast-rotators are phenomenologically richer
than for their slow-rotating counterparts in this study.  As we noted
earlier, to first order the mass of the galaxy (its rotation speed)
correlates with the number of distinct components required to model
the light distribution. The distinct nature of the {\it model}
components may or may not reflect distinct {\it physical} components,
e.g., we could be approximating continuous distributions of age,
metallicity and structural parameters. However, this correlation hints
at the increase in complexity we might expect to see in the
hierarchical build-up of more massive systems that have assembled in a
weakly-disruptive fashion, i.e., in a way such that they have remained
dynamically cold by virtue of, say, the gas-richness or angular
momentum alignment of their merging events. In this context the
structural decomposition of these edge-on spiral galaxy light profiles
reads like an archaeological record of their mass assembly and
dynamical heating. We distinguish three groups, namely those with
regular morphology and super-thin components, (NGC 891 and 4013),
those with regular morphology but without super-thin components (NGC
4565 and 522), and those with disturbed morphology (NGC~1055).  No
doubt this grouping reflects the limited sample size of this study,
but it still enables some conclusions to be drawn about the nature of
super-thin disks in massive spiral galaxies.

Both NGC 891 and 4013 have similar relative luminosity in the
super-thin disk components, but NGC~4013 has a thin disk that has half
the flattening and two-thirds the relative luminosity as the thin disk
for NGC~891. The ``missing'' thin disk luminosity can be accounted for
if attributed to the prominently visible bar component of NGC~4013,
which has a comparable total luminosity to the super-thin and thin
disks. It is interesting to consider this in the context of the
relative thickness of the disks in NGC~4013 and 891.

As mentioned previously, \citet{Comeron11b} find that NGC~4013 has
disks with unusually large $h_{z}$. \citet{Comeron11b} choose to label
their disks as ``thin'', ``thick'', and ``extended''. However, given
that we do not find any significant additional very thin component we
designate the three disks as ``super-thin'', ``thin'', and ``thick''
(as outlined in Sections \ref{sec:ngc4013} and \ref{sec:lumrat}).
Based on our best fits the scale-heights of NGC~4013's two thinner
disks are $\sim$20\% larger than their apparent analogs in NGC~891;
NGC~4013's thickest disk is twice as vertically extended than its
putative counterpart in NGC~891.  NGC~4013's disk oblateness is significantly decreased; $h_R/h_z$ is roughly 2-3 times smaller
than in NGC~891 for all three disks.

It is unclear why NGC~4013's disks would be so much less oblate than
other fast rotating spirals. \citet{Comeron11} find other galaxies
with similar $V_{rot}$ but disks with more ``normal'' thicknesses
(e.g., PGC 013646).  The ratios of disk luminosity to dynamical mass
within $2.2h_R$ would argue in favor of NGC~4013 having a more maximal
disk than NGC~891, assuming similar $\ks$-band mass-to-light ratios.
All else being equal in the dynamical history of these disks, this
would argue in favor of NGC~4013 having thinner disks.  However, the
process (particularly the sources) of disk heating is not a well
understood phenomenon. Certainly the abundance of gas in this galaxy would be consistent with a quick reformation of a
super-thin disk after some relatively minor merger (e.g.,
\citealt{Puech12}), but the same could be said of NGC~891 as well. The
telling difference between these two galaxies is the prodigious
neutral hydrogen warp first detected by \citet{Bottema87}, and
what appears to be an associated giant stellar tidal stream
\citep{Martinez-Delgado09}.

In an absolute sense, NGC~4013's thick disk scale-height of $\sim$3
kpc is larger than what can be produced in simulations of minor
mergers \citep{Kazantzidis08}. However, disks made through minor
interactions tend to be flared at large radii \citep{Villalobos08}, so
it is possible that what we see in projection as a very thick disk is
just the flared or warped outer portions of a thinner, inner disk. In
contrast, \citet{Comeron11b} advocate that a disk this thick must have
formed at high redshift, when the galaxy was in the process of initial
buildup and had elevated levels of gas turbulence and heating via
giant star-forming clouds. This is consistent with the apparent
ubiquity of thick disks at high redshift, e.g., \citet{Elmegreen06}
and \citet{Forster_Schreiber09}, but it would imply such thick disks
should be fairly more common in the local universe, while \citet{Comeron11b}
report only 2 in 46 of their galaxies contain such a thick disk. Therefore
in addition they suggest the disk of NGC~4013 must have been further
heated by a single, significant merger event.  This would require
fine-tuning of the merger event since, for example, simulations
following multiple minor mergers fail to produce such thick disks
\citep{Kazantzidis08}. As dramatic as the giant stellar tidal stream
found by \citet{Martinez-Delgado09} appears, they estimate the
progenitor merger object was relatively low mass ($6\times10^8
M_\odot$).  

An alternative and perhaps simpler explanation is that the thickness
of the disks in NGC~4013 is a projection effect of a warp in the
stellar disk seen clearly in the gas disk. If the warp were
  oriented as the HI, we would expect to see the assymetry in the
  vertical light profile. Such an asymmetry is not apparent in our
  residual images (Figure \ref{fig:ngc4013fitplot}), but a more
  complicated, pretzel-like distribution of luminosity at lower
  surface-brightness at much larger radii is clearly visible in the
  deep optical images of \citet{Martinez-Delgado09}. This morphology
  may be consistent with the HI flare, but clearly there is additional
  structure in what appears to be tidal streams.  If the apparent
  thick disks seen in any galaxy are due to warps or flaring, it may be
  possible to probe this condition with integral-field spectroscopy
  assuming the systems is dynamically stable. In this case, the
  off-plane line-of-sight velocity distribution function would be
  skewed increasingly toward lower tangential velocities at larger
  heights since an increasing fraction of the stars would be situated
  away from the tangent point. Changing dust attenuation over the
  height and radius of the disk
  would complicate such a measurement, however, and would need to be
  carefully controlled for by any future investigations along these lines.

While the thickness of NGC~4013's disk components is difficult to
explain unless they are the projection of a warp or tidal disturbance,
the galaxy still requires three nested disks to fit its integrated
light profile like in NGC~891 or the MW. The remaining three
fast-rotating galaxies in our sample present further complexity.
While the two disks necessary to fit the data for NGC~4565 closely
resemble the thin and thick disks of the MW and NGC~891 in terms of
scale-heights and total luminosities, we see no strong evidence for a
super-thin component--even the stellar emission associated with the
CO-rich ring has $h_{z}\sim$300 pc. Given that this galaxy is only
slightly ($\sim$20\%) farther away than NGC~891, and that we recover
components with $h_{z}\le200$ pc in galaxies even more distant than
NGC~4565, this is not a resolution issue. It's possible that, while
outwardly similar to the MW and NGC~891, NGC~4565 is just not forming
many new stars. The minimal amounts of extraplanar dust
(\citealt{Howk99}, unlike in NGCs 4013 and 891; see
\citealt{Rueff13,Howk97}) would corroborate this argument.

The need for a ring to fit NGC~4565's light profile is also unusual,
insofar in that none of the other six galaxies in this work required
such a component. \citet{Buta90} notes that prominent rings are
preferentially observed in early-type disk galaxies. While both
NGC~891 and NGC~4565 are classified as Sb in HyperLEDA, given
that NGC~4565's central light concentration is better fit with a bulge
model, it could actually be an earlier type. Another possibility is
that this feature is a pseudo-ring caused by the projection of the bar
or the spiral arms, which \citet{Buta90} indicates is more common for
later-type spirals.

Like NGC~4565, NGC~522 also does not appear to require a super-thin
disk. Given its large distance we have noted our concerns for the
accuracy of the attenuation-corrected vertical profile. The lower
physical resolution we have for this galaxy would tend to reduce the
contrast in the correction between regions of high and low attenuation
when these regions are spatially proximate and unresolved, as we would
expect near the disk mid-plane.  However, the disk structural
parameters for these two galaxies are remarkably similar. In particular,
their thin-disk components are nearly identical in size and thickness,
and both contribute 40\% to the total $\ks$-band luminosity,
comparable to NGC~891's thin-disk but twice as flat in terms of
$h_R/h_z$. (Their thin-disk oblateness is comparable to the super-thin
oblateness of NGC 891 and 4013.) Their thick disks are also not
dissimilar to that of NGC~891, although NGC~522 stands out at having
the largest thick-disk luminosity (and scale-length) in both absolute
and relative terms. What also differentiates both NGC~4565 and NGC~522
from NGC 891 and 4013 is the diminshed strength of nuclear disks, and
stronger bulge versus bar components.  One might speculate that
nuclear and super-thin disks are both manifestations of
relatively recent gas accretion. The distinction between bulge and bar
seems less clear, particularly given the low S\'{e}rsic indices we
find ($n\sim1$) for these systems, with axial ratios not very
different from those of the bars in NGC 891 and 4013.

Finally, while NGC~1055 is not directly comparable to any of the more
isolated fast rotators we have studied, it is interesting to note that
despite clear signs of interaction and containing a bulge component
more luminous than any disk by a factor of $>$2, this galaxy manages
to have disk components with scale-height $<$200 pc. In fact, roughly
38\% of its total $\ks$-band luminosity is in such super-thin
components, and all of the disk components have scale-heights below
300 pc. Either the very thin disks are sufficiently self-gravitating
or gas-rich and dissapative to be robust against tidal disruption, or
they have reformed rapidly. The large amount of attenuation in this
galaxy indicates the presence of abundant HI (estimated at 10$^{9.56}$
M$_{\odot}$ by \citealt{Thilker07}).

\begin{figure*}
\centering
\subfigure{
  \includegraphics[scale=0.44]{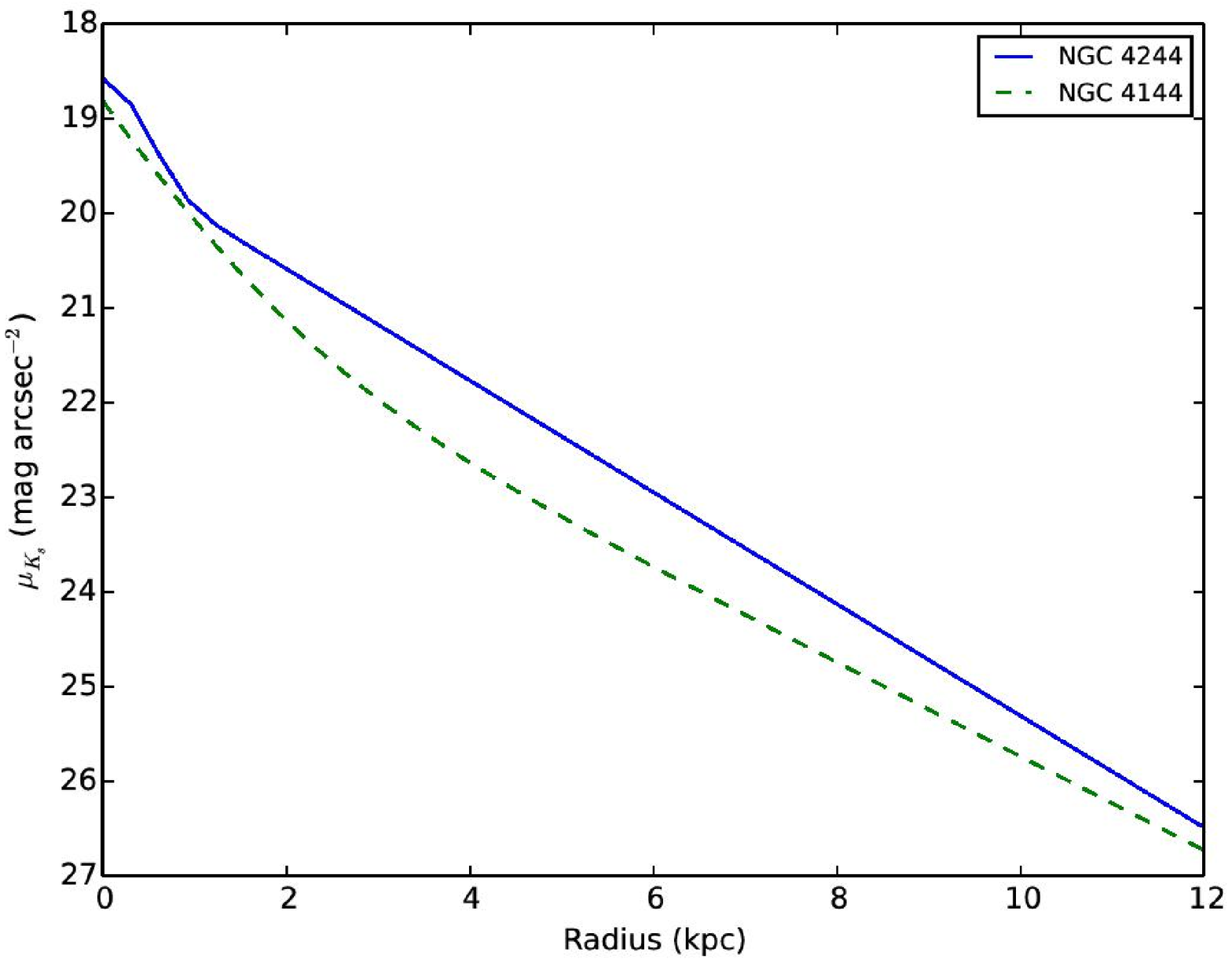}
}
\subfigure{
  \includegraphics[scale=0.44]{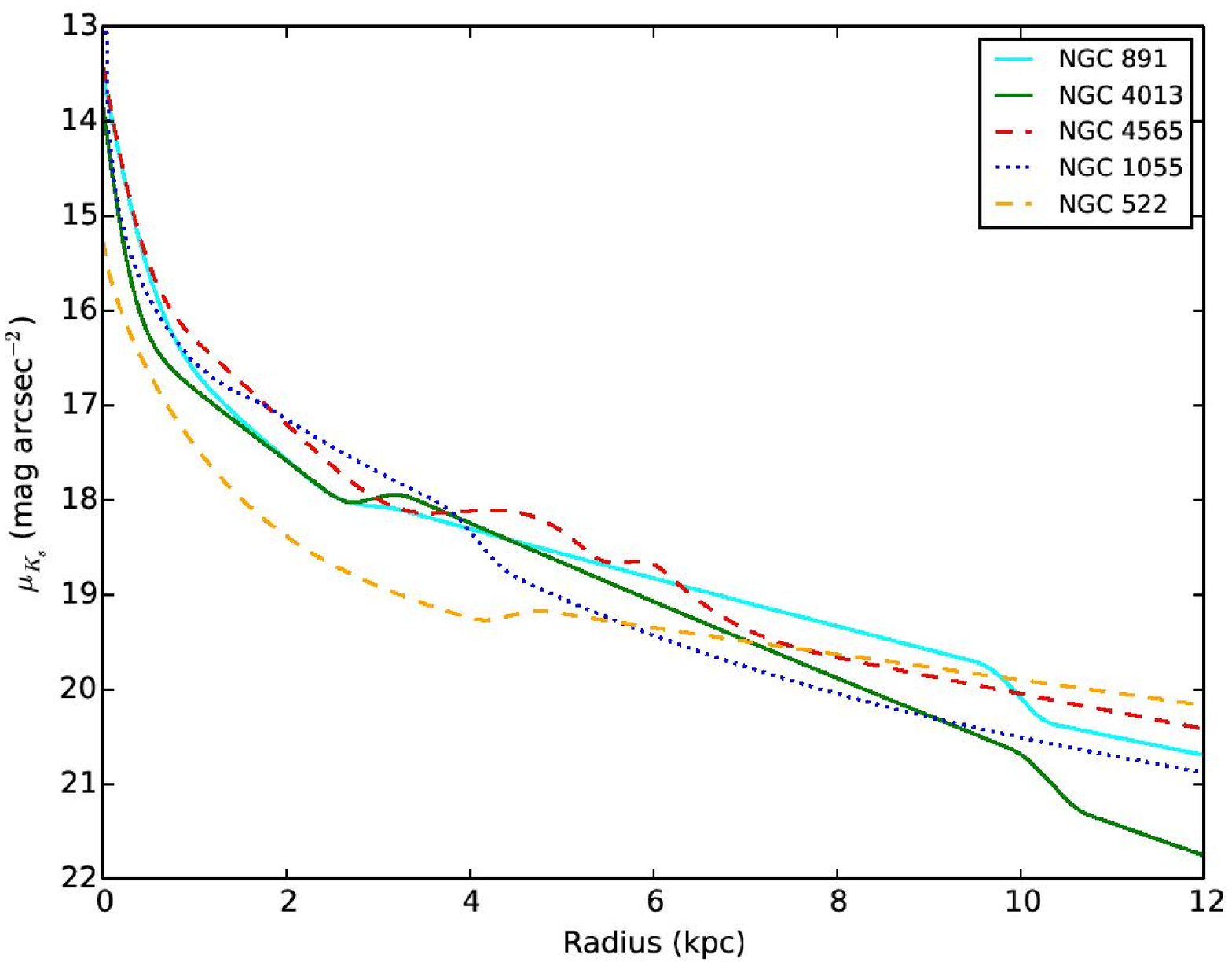}
}
\caption{Predicted face-on radial surface brightness profiles for the
  galaxies in our sample. Left: slow rotators. Right: fast
  rotators. The line styles of the fast rotators correspond to
  galaxies with super-thin disks (solid lines), galaxies without
  (dashed lines), and galaxies with disturbed morphologies (dotted
  lines).}
\label{fig:faceonprofiles}
\end{figure*}

\subsubsection{Infrared Colors as a Diagnostic}

An exasperating outcome of our modeling is the roughly even split
between fast-rotating spirals with and without super-thin {\it
  stellar} disks that appear, based on their optical and
near-infrared images and colors, to be rather similar. Only with
attenuation corrected high angular resolution near-infared images
and 2D profile fitting is the presence of the super-thin components
revealed. This is rather unsatisfactory, particularly in the context of
dynamical studies of galaxies that are not viewed edge-on, such as the
DiskMass Survey.

However, with the supposition that the super-thin disks are
star-forming regions, it is reasonable to expect that either a {\it
  large} gas to stellar mass ratio or a {\it small} near to
mid/far-infrared luminosity ratio would indicate the presence of a
super-thin disk. To test this hypothesis we have compiled our total
$\ks$-band model luminosities along with literature values for atomic
hydrogen mass from HyperLEDA, as well as Spitzer and IRAS mid- and
far-infared luminosities from
NED\footnote{http://nedwww.ipac.caltech.edu/.} in
Table~\ref{tab:fluxrat}. Colors are in the AB system. Wherever possible we used the IRAS values
from \citet{Sanders03}.  Unfortunately, molecular gas-mass
estimates remain surprisingly hard to come by in the literature
or in on-line databases, but we note the correlation
between mid-IR 24$\mu$m and and CO fluxes used by \citet{Westfall11} based on data in \citet{Leroy08}.

Inspection of the tabulated colors indicates that the atomic gas to
stellar mass ratio (technically, $M_{HI}/L_K$) is {\it not} a good
diagnostic of the presence of a super-thin disk, most likely because
atomic gas is a necessary but not sufficient condition for
star-formation. In contrast, the near- to mid/far-infared colors
{\it do} seem to correlate with the presence of a super-thin disk in
the expected sense that a relative enhancement of mid- and far-infared
flux (from dust warmed by star-formation) accompanies the presence of
a super-thin stellar disk. Further corroborating this astrophysical
picture is the indication that the 60 to 100 $\mu$m color appears to be
slightly 
enhanced when the near- to mid/far-infrared is depressed, i.e., when
there is more dust emission per unit stellar emissivity, the dust
emission appears to be somewhat warmer.  As best we can reckon given
our limited statistics and extant literature data, $K-m_{60} >
1.86_{-0.31}^{+0.44}$ is the threshold for the presence of a significant
super-thin disk component in fast-rotating spiral galaxies.

\subsection{Face-on Profiles of Truncated Galaxies}

The model fits for the slow rotating galaxies in our sample contain
disk components that do not require radial truncations. From this fact
we can infer these galaxies should have a classical Type I face-on
radial surface brightness profiles \citep{Freeman70}.  We have
confirmed this by re-projecting the models to a face-on orientation
and plotting them in the left panel of Figure
\ref{fig:faceonprofiles}).  It is also clear from inspection of
Table~\ref{tab:summary} that these slow-rotating disks have central
disk surface-brightness 4 to 10 times lower than that of a a Freeman
disk (Freeman 1970) which would have $\mu_0(K) = 18.15\pm0.5$ for
colors in the plausible range of $3 < (B-K) < 4$.

The necessity of an inner disk truncation for our fast rotators
indicates that these galaxies may have more complex face-on light
distributions. Of interest is the possibility that the inner disk
truncation may give rise to Freeman Type II profiles, where there is
an apparent down-turn of the light profile at intermediate radius
where the dominance of the disk and bulge (and/or bar) transitions.
Using an inner disk truncation to produce Type II profiles has been
employed by \citet{Kormendy77} and \citet{Baggett98}. The application
here is different because our analysis is based on fitting the edge-on
light distributions where the radial truncation, if it exists, is seen
in projection. However, by virtue of this projection we are able to
capture the vertical dependence of disk trunction as we have done in
the preceeding sections. Specifically, the truncation radii of
different thickness disk components are not always the same.

The right panel of Figure \ref{fig:faceonprofiles} shows the face-on
radial profiles predicted by our models of NGCs 891, 4013, 4565, 522,
and 1055. These disks all have central surface brightness values
within a factor of 2 of the Freeman value, much brighter than the
slow-rotators. Since the truncations we used in our models are abrupt
in radius they produce unrealistic discontinuities in the face-on
profiles. (In projection to edge-on this abruptness is naturally
smoothed.)  We therefore convolve the regions within 1.0 kpc of any
truncations using a uniform kernel of 0.25 kpc.  To ensure continuity,
points near the edges of the smoothed region are constructed of a
weighted average between the smoothed and unsmoothed profiles, with
more weight being given to the unsmoothed data closer to the edge of
the smoothing region. We use the Meijer G functional form described by
\citet{Baes11} to reproject our fitted bulges.

Two of the five galaxies exhibit light profile morphology that would
clearly be classified as Type II (inner break). NGC 4013 and 522 show
dips in their light profile around their innermost truncations near 3
kpc and 4.5 kpc respectively.  NGC~891's inner truncation is less
obvious, although it too produces a lower-amplitude break near 2.7 kpc
where its disks truncate; it might also be classified as a
Type II profile. NGC~4565 is more difficult to identify, largely due
to the bump at R$\sim$5.5 kpc caused by its ring. If the flux excess
caused by the ring was considered to be small-scale structure and
masked from the fit \citep[as is often done, e.g.][]{Baggett98}
NGC~4565 would almost certainly be classified as having a Type I
profile. Despite its disturbed morphology NGC~1055's face-on
appearance is that of a Type I spiral, although the outer truncation
in the profile is clearly visible (around 4 kpc). Our results suggest
that inner Type II profiles are at least sometimes caused by
truncations in the disk luminosity distributions.

\section{Summary}
\label{sec:conclusions}

We have expanded our study of the intrinsic NIR light distribution of
spiral galaxies, begun in \citetalias{Schechtman-Rook13} for NGC~891,
to a sample of seven spiral galaxies at a variety of rotation speeds
ranging from 69 to 245 km s$^{-1}$. This work makes advances in three
broad areas. First, on a technical level, we have generalized our
method for estimating and correcting for dust attenuation in the
near-infrared. This method is now applicable to a wide range of dust
morphologies, and requires no assumptions about the intrinsic infrared
colors of stellar populations. We have also dramatically increased the
complexity of our model two-dimensional light-profiles, and the
dynamic range over which we can probe model parameters.  Second, on a
descriptive level, we have characterized the multiplicity of disk
structure, and the trends this multiplicity exhibits with rotation
speed--at least within the limitations of the present sample. Third,
we have found a correlation between disk structure, as manifest in our
modeling, and the attenuation-corrected NIR color gradients in height
and radius. This correlation points to an astrophysical explanation
for the origins of this structure in disk heating.

At a technical level, we continue to utilize 3D radiative transfer
models to estimate the attenuation due to dust (both emission and
absorption), and in this work we further develop these models for a
more realistic representation of the dust and stellar distributions in
slower-rotating spirals. To compute the attenuation accurately for the
entire sample we also improved upon the correction scheme of
\citetalias{Schechtman-Rook13}, where we assumed a constant color of
$(\ks-4.5\mu{\rm m}) = 0$ for the unattenuated star-light. While this is
a suitable approximation for NGC~891, in this study we find that
sample galaxies have a small but non-negligible range of IR colors
($-0.1 < (\ks-4.5\mu{\rm m}) < 0.5$) in regions where there is little
dust attenuation.

In this study we incorporate SSP models with our estimated
\textit{differential} attenuation in the $J,H,\ks,3.6\mu{\rm m}$ and
$4.5\mu{\rm m}$ bands, simultaneously constraining the age,
metallicity, and attenuation for every binned pixel. To produce
accurate attenuation corrections in regions of low S/N, we employed
adaptive binning based on the Voronoi tesselation.  While the
estimates of age and metallicity are uncertain and quite degenerate,
our purpose here has been to estimate the attenuation, which we find
is {\it not} particularly covariant with other model parameters that
produce suitably good fits to the multi-band data.  Specifically, we
estimate the accuracy of our attenuation corrections is $0.1$ mag in
the $\ks$ band; comparable accuracy is achieved for the color excess
$E(J-\ks)$ for $\aeks<0.5$ mag. This enables us to construct relibale
two-dimensional NIR surface-brightness and color profiles.

We have fit two-dimensional models of galaxy light profiles to our
attenuation-corrected images using an improved Levenberg-Marquardt
minimization algorithm that increases flexibility relative to our
previous application in \citetalias{Schechtman-Rook13}. In the current
analysis we are able to fit a broader variety of analytic functions
(including rings and generalized S\'{e}rsic profiles), as well as
inclination. The increased sophistication of the models was required
in order to accurately reproduce the features clearly seen in the
attenuation-corrected galaxies, and to account for the fact that few
of the galaxies in our sample are perfectly edge on. The need
  for additional model complexity is illustrated in one-dimensional
  comparison of surface-bright profiles and two-dimensional residual
  maps. These visual assessments are corroborated and quantified by
  significant decreases in $\chi^2$ values with additional model
  parameters.  In all cases, the presence of non-axisymmetric
  structure in the observed galaxies light distributions results in
  significant residuals with respect to even our most complex, but
  always axisymmetric, models. The non-axismmetric residuals utimately
  limit our ability to reliably constrain additional model complexity.
  For the galaxies analyzed in this study we appear to be above this
  limit, but approaching it for the three cases of NGC~522, NGC~1055,
  and NGC~4244.

We obtained dramatic increases in computational power by optimizing
our software to use local, distributed computing resources. This was
critical for enabling us to probe a wider range of models, thereby
increasing the robustness of our conclusions since we could fully
explore some of the inherent degeneracies in our parameters (e.g.,
disk oblateness and inclination for NGC~4144). With the exception
  of NGC~4144, inclination is well constrained in our models, and the
  remaining model parameters for disk scale-length, height, and
  luminosity are also well constrained even in the context of
  three-component disks.

At a descriptive level we find that in such a detailed examination all
the galaxies in our sample have unique qualities. Even arguably the
two most similar galaxies, NGCs 891 and 4013, have fairly different
scale-heights and bar-to-disk luminosity ratios. A general trend is
the need for fewer structural components to accurately model slow
rotating spirals (lower-mass disks are simpler).

In fast rotators the most common features seem to be the presence of
inner disk truncations.  The commonality of these truncations may be a
key to understanding why some face-on spiral galaxies appear to have
inner breaks in their radial profiles. When reprojected to a face-on
projection, our best-fitting models for several of our fast rotating
galaxies yield radial surface-brightness profiles with breaks similar
to classic Type-II morphology \citep{Freeman70}. Whether or not the
bars in these instances act to \textit{radially} redistribute stars to
produce the break, it is clear they are responsible for redistributing
the stars into a thicker \textit{vertical} distribution at small radii
($<1h_R$).

Our slow rotators both have Type-I profiles and appear to be without
bars, from which we conclude that Magellanic irregulars are absent
from our sample; their disks generally appear to have $h_{z}$
comparable to the thin disks in fast rotating spirals.  With the
exception of NGC~4144, the vertical disk structure is well constrained
or all galaxies in our sample, with systematic errors in inclination
below 1 deg for all galaxies in our sample exception for NGC~4144.
Given the lack of well-defined, thin dust-lane and apparent
irregularity in the star light for NGC~4144, the inclination estimate
is highly covariant with scale-height. We caution that estimates of
NGC~4144's disk scale height here or in the literature are highly uncertain (to
within factors of 5).

Roughly half of our sample of fast-rotating disks contain a super-thin
($h_z < 250$ pc) disk component.  This demonstrates that super-thin
disks are not ubiquitously significant components of massive,
star-forming spiral galaxies, but they are common. It also serves to
confirm that NGC~891 and the MW are not alone in their nested
multi-component disk structure, in which thinner disk components are
also radially smaller. In particular, NGC~4013 appears to have
super-thin, thin and thick components with similar luminosity ratios
to NGC~891. 

However, each of NGC~4013's disks are more oblate than their putative
counterparts in NGC~891, being both thicker and radially shorter. The extreme
thickness of NGC~4013's thickest disk is quite difficult to explain in
the theoretical paradigm of disk heating through minor mergers; this
is either a clue that the nested disk structure we see in our own
galaxy may arise from a variety of merger histories or alternatively
that the decreased oblateness of NGC~4013's disks are due to
projection of a warp in the stellar disk seen clearly in the neutral
gas. Given the giant, stellar tidal stream known to exist in this
system, apparently due to a low-mass progenitor, a third alternative
is that our theoretical understanding of disk heating from minor
mergers is incomplete. As a further example, NGC~1055 also has
super-thin disk components; what is unusual about this galaxy is its
disturbed morphology and pronounced lack of a {\it thick} disk. While
the complexity of its light distribution precludes a definitive
characterization of its intrinsic disk structure, our results indicate
very thin disks may survive or be able to reform efficently after
merger events in gas rich systems \citep[e.g.][]{Hopkins09}.

Neither apparent optical and near-infared morphology or gas richness
of galaxies in our sample provide a good predictor of a super-thin
disk component.  For example, despite frequently being named as a good
analog to the MW \citep[e.g.][]{vanderKruit84,Shaw89,Jones89}, no
super-thin component is necessary to fit NGC~4565's vertical light
profile. However, for the fast rotators, we do find that an
enhancement of mid- and far-infared flux relative to the $\ks$ band
does correlate with the presence of a super-thin disk component.  This
is consistent with super-thin disks being associated with young
stellar populations and star-formation. In future work we intend
  to use this criterion to estimate the statistical frequency of
  super-thin disks using larger galaxy samples, and also apply this
  result to the DiskMass Survey for an improved dynamical analysis.

The two fast-rotating galaxies without super-thin disks (NGCs 4565 and
522) do have the thinnest and most oblate ``thin'' disk components,
which may be indicative of recent, but perhaps not on-going (or
vigorous) star-formation. NGC~4565 is known to be one of the few
massive galaxies with little extraplanar dust, a feature which usually
signifies outflows from young, massive stars \citep{Howk99}. A
ring-like {\it stellar} feature with a radius of $\sim$4.5 kpc is
clearly seen in NGC~4565 from even a cursory visual inspection of the
attenuation-corrected NIR image; the ring has a thickness comparable
to that of NGC~4565's thin disk. NGC~4565 it is the only galaxy in our
sample requiring such a feature. It is also the only galaxy in our
sample with a known active nucleus, which plausibly may be responsible
for the relatively diminished current star-formation and absence of a
super-thin disk.

With the exception of the one morphologically disturbed galaxy in our
sample (NGC~1055), all show distinctive, attenuation-corrected
near-infared ($J-\ks$) color gradients with height above their
mid-planes. The trends are significant with respect to our estimated
systematic errors, and correlate with the indpendently estimated disk
components, e.g., super-thin, thin, etc.  The general trends are to
bluer $J-\ks$ colors at larger heights, which could reflect both
gradients in age and metallicity.  Our fast rotating galaxies have
strong color gradients at small heights, with $J-\ks$ increasing by
0.2-0.4 mag from the high-latitude value of $\sim$1, and peaking near
$z\sim0.2$ kpc. These near-mid-plane colors imply the existence of a
significant population of intermediate-age AGB (carbon) stars. In
some galaxies, notably NGC~891, the $J-\ks$ color also becomes
significantly bluer below $z\sim0.2$ (where NGC~891's super-thin disk
dominates the light profile).

These results paint a picture consistent with a disk heating scenario
in which stars form in a dynamically-cold super-thin gas layer, and
diffuse at some rate to less oblate (thicker, dynamically warmer)
distributions. The rate of heating and its discreteness (i.e., whether
heating is a continuous or episodic process) are of general interest
in understanding how disk galaxies evolve. Determining whether
  super-thin disks are transient structures--rapidly destroyed by
  mergers to form the thin and thick disks, or are slowly and
  continuously evolving components--requires both a larger statistical
  sample of galaxies and better estimates of stellar ages. The
results here suggest that with suitable additional spectrophotometric
measurements it may be possible to utilize stellar chronometers to
measure the heating rate of disks. At the same time it is also clear
that edge-on spiral galaxies contain a significant amount of
information on the stellar populations of these systems by virtue of
these vertical gradients.  In contrast to post star-burst systems used
to probe the relative importance of AGB stars in the integrated
stellar populations of galaxies (e.g., \citealt{Kriek10},
\citealt{Zibetti13}), where the young component may contribute only a
small fraction of the near-infrared light, here we are utilizing the
unique geometry of disk vertical structure to probe a significant
dynamic range in mean stellar population age.

Overall this work has shown that while the MW's vertical structure is
clearly not unique, the local universe contains a wide range of disk
structure even for intermed-type, fast-rotating systems. The limiting
factor in obtaining a statistical understanding of this structure is
the relative small number of galaxies which are close enough to edge
on and nearby enough to be adequately resolved. Better resolution
imaging data, particularly in the mid-infrared, should be possible
with the {\it James Webb Space Telescope} (JWST) NIRCAM
instrument.\footnote{http://jwst.nasa.gov/nircam.html.}  With the
advent of JWST, a much increased volume will become available for this
kind of study, yielding the larger sample sizes necessary to probe the
diversity of disk structure and potentially the fossil record of disk
heating.

\acknowledgements{This research was directly supported by the
  U.S. National Science Foundation (NSF) AST-1009471.  MAB
  acknowledges the generous hospitality of the Institute for Cosmology
  and Gravitation (University of Portsmouth) and financial support of
  the Leverhulme Foundation. The authors are grateful to John
  MacLachlan for sharing his NGC~4244 data, and Arthur Eigenbrot for his
  significant assistance with data collection. Computational support
  came from the UW-Madison Department of Computer Sciences Center For
  High Throughput Computing (CHTC), supported by the Wisconsin Alumni
  Research Foundation, the NSF and the U.S. Department of Energy
  Office of Science as part of the Open Science Grid. We also made use
  of SDSS and
  2MASS\footnote{http://www.ipac.caltech.edu/2mass/releases/allsky/},
  databases and archives.}

\bibliography{citations}

\appendix
\section{NGC~4244 Attenuation Correction}
\label{sec:ngc4244attencorr}

\begin{figure}
\plotone{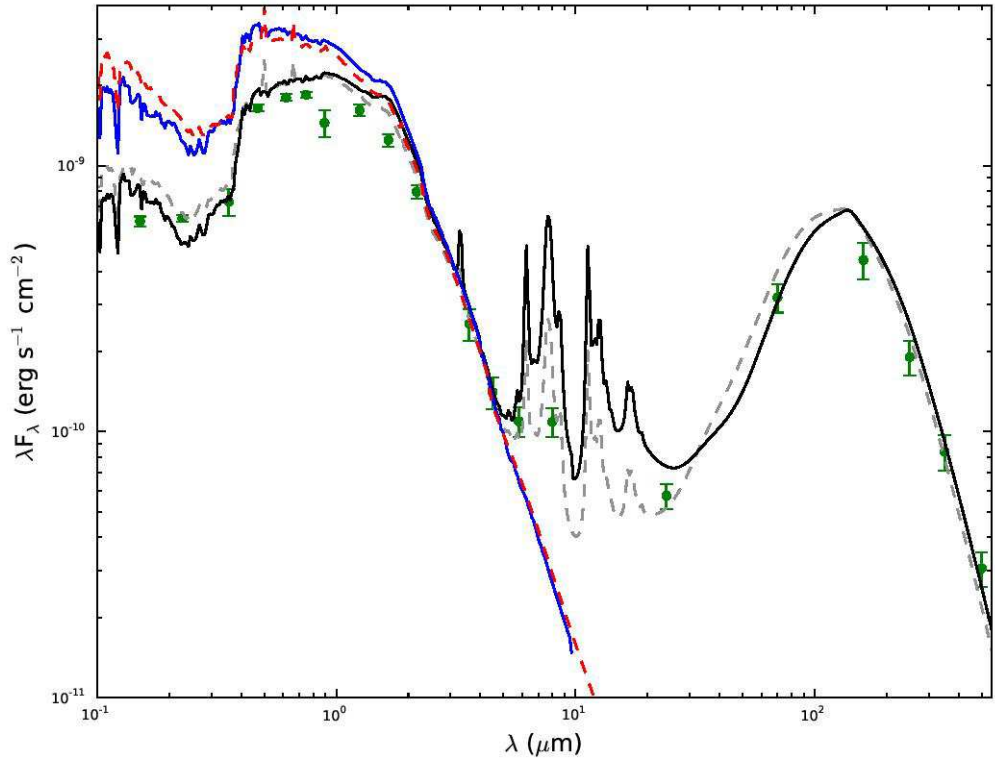}
\caption{Model SEDs for NGC~4244. Green points show data and are the
  same as in \citet{MacLachlan11}. The blue solid line shows the input
  SED using the PEGASE stellar library, while the black solid line
  shows the corresponding HYPERION-processed output SED. The red and
  gray dashed lines show the input and output SEDs (respectively) from
  the GALEV evolutionary synthesis model (including the PAH reduction
  discussed in the text).}
\label{fig:ngc4244sed}
\end{figure}

Multiple authors have fit RT models to NGC~4244, two notable recent
efforts being that of \citet{MacLachlan11} and \citet{Holwerda12}. We
use the results from these works as starting points for our models,
which are run using HYPERION \citep{Robitaille11} . To obtain input
SEDs we use the method of \citet{Bianchi08}, splitting a PEGASE
\citep{Fioc97} synthesis model of a spiral galaxy into high-mass and
low-mass components (as a proxy for young and old stars,
respectively). For NGC~4244 we find that a Sc galaxy model at only 3
Gyr after formation provides the best fit to the data, an indication
of the relative youth of this galaxy's stellar
population. Additionally, unlike for NGC~891, we find that a lower
dust density threshold for the addition of young stars to a grid cell
is not required to reproduce the SED (shown in Figure
\ref{fig:ngc4244sed}). A list of the relevant parameters used in our
fit is shown in Table \ref{tab:ngc4244rtmodel}.

To compute the attenuation correction we use the same technique as in
\citetalias{Schechtman-Rook13}, producing model images with and
without dust and taking the ratio. The ratio image is compared to a
$(\ks-4.5\mu{\rm m})$ image on a pixel-by-pixel basis. We then fit this
relationship. NGC~891 required a piecewise fit, where pixels with
$(\ks-4.5\mu{\rm m})<1.3$ were well fit by a fourth order
polynomial and pixels with redder colors followed a linear trend. For
NGC~4244, since there is much less attenuation, there are no pixels
with $(\ks-4.5\mu{\rm m})>1.3$, so we just fit the fourth-order
polynomial (note that these fits are for the dust distribution called
Model A in \citetalias{Schechtman-Rook13}). The spectra used for this
modeling have intrinsic colors of $(\ks-4.5\mu{\rm m}) \sim 0$.  To make
this fully general for any intrinsic $(\ks-4.5\mu{\rm m})$ color,
however, we formulate these relations in terms of the color excess
$E(\ks-4.5\mu{\rm m})$, as discussed in the text:
\begin{align}
\aej = &-0.03+1.12\times E(\ks-4.5\mu{\rm m})+7.42\times E(\ks-4.5\mu{\rm m})^2- \nonumber \\
       &26.44\times E(\ks-4.5\mu{\rm m})^3+29.82\times E(\ks-4.5\mu{\rm m})^4
\end{align}
\begin{align}
\aeh = &-0.01+0.67\times E(\ks-4.5\mu{\rm m})+4.92\times E(\ks-4.5\mu{\rm m})^2- \nonumber \\
       &16.41\times E(\ks-4.5\mu{\rm m})^3+18.35\times E(\ks-4.5\mu{\rm m})^4
\end{align}
\begin{align}
\aeks = &-0.01+0.40\times E(\ks-4.5\mu{\rm m})+3.24\times E(\ks-4.5\mu{\rm m})^2- \nonumber \\
        &10.27\times E(\ks-4.5\mu{\rm m})^3+11.33\times E(\ks-4.5\mu{\rm m})^4.
\end{align}

\begin{figure}
\plotone{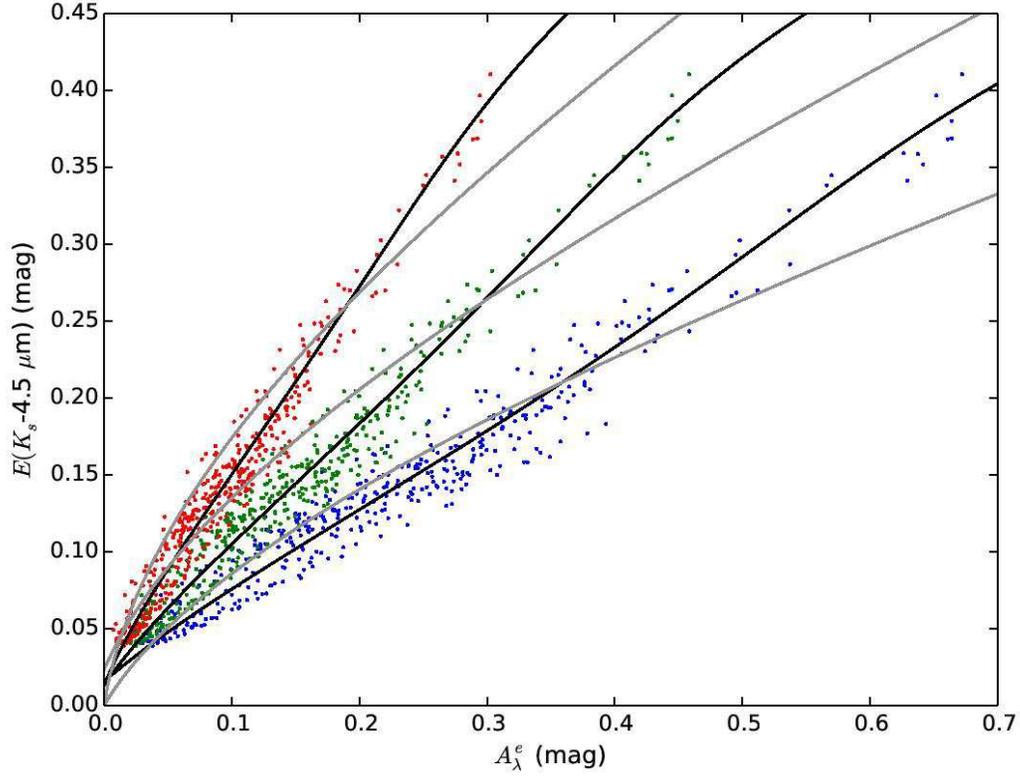}
\caption{$\aeff$ as a function of $\ks$-4.5$\mu$m color. Blue, green,
  and red points show individual J, H, and $\ks$ model pixels, while
  black lines show the best-fit lines to those pixels. Gray lines show
  the best fits for NGC~891 using dust Model B in
  \citetalias{Schechtman-Rook13}. The differences between the two
  models (especially at low attenuation) are mainly due to the fact
  that, even though the same stellar library was used for both fits,
  the input SED for NGC~4244 is much younger than NGC~891. }
\label{fig:ngc4244attencorrdemo}
\end{figure}

Note that some of the higher-order coefficients on these fits are very
large. This is because the models have $\sim$zero pixels with
$(\ks-4.5\mu{\rm m})\gtrsim 0.4$ mag.  Therefore these fits should
{\it not} be used to correct any data with $(\ks-4.5\mu{\rm m})$
colors redder than this, as the correction function will be
erroneously dominated by these large high-order terms. A graphical
representation of these corrections is overlaid on top of the model
pixels used in the fit in Figure \ref{fig:ngc4244attencorrdemo} along
with the fit used in \citetalias{Schechtman-Rook13}\footnote{Note that
  there is a typo in Appendix C of \citetalias{Schechtman-Rook13}. The
  equation for $\mathrm{A}_{\mathrm{H}}^{\mathrm{e}}$ of Model B when
  $(\ks -4.5\mu{\rm m}) > 1.3$ should read $1.75(\ks - 4.5\mu{\rm m})
  + 0.22$.}. We also show $E(J-\ks)$ as a function of $\aeks$ for our
attenuation corrections as well as the foreground screen model of
\citet{Cardelli89} in Figure \ref{fig:modelejks}.

While the PEGASE models provide a good fit, we also investigated the
GALEV \citep{Kotulla09} models used by \citet[][private
  communication]{MacLachlan11}. We found that with the GALEV spectra the only
way to produce a model that properly fit NGC~4244's SED around 5
$\mu$m was to decrease the fraction of PAHs in the model by a factor
of $\sim$4. The SED of that model is shown alongside our PEGASE model
in Figure \ref{fig:ngc4244sed}.

A dust distribution different from that found in the MW is not
necessarily a problem, as low-mass spiral galaxies are known to have
different metallicities than their high-mass counterparts
\citep{vandenBergh90,Dalcanton04} which would naturally result in a
different grain distribution. However, while the shape of the
attenuation corrections are very similar, the GALEV models also
predict different intrinsic stellar $(\ks-4.5\mu{\rm m})$
colors. Because our new attenuation correction method only relies on
the shape of the attenuation correction and uses SSPs to compute the
attenuation zeropoints, the use of either input SEDs will therefore
produce very similar results. To maximize consistency between our
methodology between our galaxies, we choose to use the correction
resulting from the PEGASE model. Ultimately, however, we show in
Section \ref{sec:fastslowae} that the slow-rotating galaxies have
negligible $\ks$-band attenuation over their entire disks, rendering
the choice of attenuation correction largely academic in this work.

\begin{deluxetable}{c c c}
\tabletypesize{\footnotesize}
\tablewidth{0pt}
\tablecaption{NGC~4244 RT Model Parameters}
\tablehead{\colhead{Parameter} & \colhead{Value} & \colhead{Units}}
\startdata
Thin disk bolometric luminosity & 2.5$\times 10^{9}$& L$_{\odot}$\\
Star-forming disk bolometric luminosity & 1.3$\times 10^{9}$&L$_{\odot}$\\
Thin disk scale-length & 1.9&kpc\\
Star-forming disk scale-length & 3.4&kpc\\
Thin disk scale-height & 0.2&kpc\\
Star-forming disk scale-height & 0.2&kpc\\
Dust disk central density & 1.1$\times 10^{-26}$&g cm$^{-3}$\\
Dust disk scale-length & 3.4&kpc\\
Dust disk scale-height & 0.2&kpc\\
Dust clumping fraction & 0.5& \\
Number of largest-scale clumps  &130 &
\enddata
\label{tab:ngc4244rtmodel}
\end{deluxetable}

\end{document}